\documentclass[final,5p,times,one]{elsarticle}

\usepackage{graphicx}
\usepackage{amssymb}
\usepackage{amsmath}
\usepackage{hyperref}
\usepackage{enumitem}
\usepackage[capitalize]{cleveref}

\usepackage{lineno}
\usepackage{colortbl}
\usepackage[normalem]{ulem} 

\definecolor{applegreen}{rgb}{0.55, 0.71, 0.0}
\definecolor{dodgerblue}{rgb}{0.12, 0.56, 1.0}
\definecolor{electricindigo}{rgb}{0.44, 0.0, 1.0}
\definecolor{electricviolet}{rgb}{0.56, 0.0, 1.0}
\definecolor{darkviolet}{rgb}{0.58, 0.0, 0.83}
\definecolor{majorelleblue}{rgb}{0.38, 0.31, 0.86}
\definecolor{mauve}{rgb}{0.88, 0.69, 1.0}
\definecolor{lavendergray}{rgb}{0.77, 0.76, 0.82}
\definecolor{lavenderpurple}{rgb}{0.59, 0.48, 0.71}
\definecolor{deepcarmine}{rgb}{0.66, 0.13, 0.24}
\definecolor{mediumspringgreen}{rgb}{0.0, 0.98, 0.6}

\def\ergs{{\rm erg\,\,s^{-1}}}
\def\cm2{{\rm cm^{-2}}}
\def\ergcms{{\rm erg\,cm^{-2}\,s^{-1}}}

\usepackage{xspace}

\newcommand{\un}[1]{\,\ensuremath{\rm{#1}}}

\newcommand{\degrees}{$^{\circ}$}

\newcommand{\msun}{\,$M_{\odot}$\xspace}

\newcommand{\ergpersec}{\un{ergs\;s^{-1}}\xspace}

\newcommand{\xray}{X-ray\xspace}
\newcommand{\xrays}{X-rays\xspace}

\newcommand{\asm}{\textit{\small{ASM}\/}\xspace}

\newcommand{\integral}{\textit{\small{INTEGRAL}}\xspace}

\newcommand{\rxte}{\textit{\small{RXTE}\/}\xspace}
\newcommand{\chandra}{{Chandra}\xspace}
\newcommand{\xmm}{{\textit{XMM}-\textit{Newton}}\xspace}
\newcommand{\swift}{\textit{Swift}\xspace}
\newcommand{\suzaku}{{Suzaku}\xspace}
\newcommand{\nustar}{{\em NuSTAR}\xspace}

\newcommand{\jemx}{{JEM-X}\xspace}

\newcommand{\grs}{GRS\,1915$+$105\xspace}

\newcommand{\gx}{GX 339$-$4}
\newcommand{\GA}{1E~1740}
\newcommand{\cyg}{Cyg~X-1}
\newcommand{\swJ}{{SwJ1745}}

\newcommand{\xrbs}{{XRB}s\xspace}

\journal{Journal Name}

\begin{document}

\begin{frontmatter}


\title{The \textit{INTEGRAL} view on Black Hole X-ray Binaries}



\author[1,2]{S.\,E. Motta}
\author[3]{J. Rodriguez}
\author[4]{E. Jourdain}
\author[5]{M. Del Santo}
\author[6]{G. Belanger}
\author[7]{F. Cangemi}
\author[8]{V. Grinberg}
\author[9]{J. J. E. Kajava}
\author[10]{E. Kuulkers}
\author[4]{J. Malzac}
\author[11,12]{K. Pottschmidt}
\author[4]{J.P. Roques}
\author[6]{C. Sanchez-Fernandez}
\author[13]{J. Wilms}

\address[1]{INAF--Osservatorio Astronomico di Brera, via E.\,Bianchi 46, 23807 Merate (LC), Italy}
\address[2]{Department of Physics, Astrophysics, University of Oxford, Denys Wilkinson Building, Keble Road, OX1 3RH Oxford, UK}
\address[3]{D\'ept. of Astrophysics,  lab AIM, CEA/IRFU, CNRS/INSU, Universit\'e Paris-Saclay \& Universit\'e de Paris, Orme des Merisiers F-91191, Gif-sur-Yvette, France }
\address[4]{IRAP, Université de Toulouse, CNRS, CNES, UPS, (Toulouse), France\\}
\address[5]{INAF, Istituto di Astrofisica Spaziale e Fisica Cosmica, via Ugo La Malfa 153, 90146 Palermo, Italy}
\address[6]{European Space Astronomy Centre (ESA/ESAC), Science Operations Department, Villanueva de la Ca\~nada (Madrid), Spain;}
\address[7]{CNRS, Laboratoire de Physique Nucléaire et des Hautes Energies (LPNHE), Sorbonne Université, 4 place Jussieu, Paris 75005, France}
\address[8]{Institut f\"ur Astronomie und Astrophysik, Universität T\"ubingen, Sand 1, 72076 T\"ubingen, Germany}
\address[9]{Department of Physics and Astronomy, FI-20014 University of Turku, Finland}
\address[10]{European Space Agency (ESA), European Space Research and Technology Centre (ESTEC), Keplerlaan 1, 2201 AZ Noordwijk, The Netherlands}
\address[11]{CRESST, Department of Physics and Center for Space Science and Technology, UMBC, Baltimore, MD 210250, USA}
\address[12]{ NASA Goddard Space Flight Center, Code 661, Greenbelt, MD 20771, USA}
\address[13]{Dr.~Karl Remeis-Sternwarte \& Erlangen Centre for Astroparticle Physics, Friedrich-Alexander-Universit\"at Erlangen-N\"urnberg, Sternwartstr.~7, 96049 Bamberg, Germany}

\begin{abstract}

\noindent \integral is an ESA mission in fundamental astrophysics that was launched in October 2002. It has been in orbit for over 18 years, during which it has been observing the high-energy sky with a set of instruments specifically designed to probe the emission from hard \xray and soft $\gamma$-ray sources. 
This paper is devoted to the subject of black hole binaries, which are among the most important sources that populate the high-energy sky. We present a review of the scientific literature based on \integral data, which has significantly advanced our knowledge in the field of relativistic astrophysics. 
We briefly summarise the state-of-the-art of the study of black hole binaries, with a particular focus on the topics closer to the \integral science. We then give an overview of the results obtained by \integral and by other observatories on a number of sources of importance in the field. Finally, we review the main results obtained over the past 18 years on all the black hole binaries that \integral has observed.
We conclude with a summary of the main contributions of \integral to the field, and on the future perspectives. 

\end{abstract}

\begin{keyword}
black hole physics \sep X-rays: general \sep gamma-rays: general \sep accretion disc \sep jets \sep binary stars
\end{keyword}
\end{frontmatter}


\section{Introduction}

\xray binaries (XRBs) are binary systems consisting of an ordinary star and a compact object, which can be a stellar-mass black hole (BH) or a neutron star (NS). Those containing a BH are referred to as black hole binaries (BHBs). The compact object is formed by the collapse of the core of a massive star at the end of its (relatively quick) evolution. The gas from the companion star is stripped away and forms an accretion disc around the compact object, where the gas is heated up through viscosity to millions of degrees, causing the gas to emit \xrays.
The presence of an accretion disc, coupled with the deep gravitational potential well around the compact object, gives rise to different kinds of outflows, streams of matter launched from different regions of the accretion flow. Outflows are able to carry away from the system a significant fraction of the matter and energy, which could have potentially fallen onto the compact object, so that XRBs act to heat their environment rather than behaving only as sinks. Outflows can take the form of relativistic, transient or persistent collimated jets, but also warm or cold, poorly collimated winds (and essentially anything in between), and are observed in different portions of the electromagnetic spectrum, from the radio wavelengths up to the \xrays. 

With their fast evolution, which takes place on humanly accessible time-scales, BHBs, together with NS binaries \cite{Papitto2020}, constitute the best laboratories where to study the extremes of physics in real time: around collapsed objects one can probe the behaviour of matter in the presence of extremely curved space-time, which is at the base of the accretion process, but also the physics underlying the particle acceleration behind the launch of relativistic jets and winds.

\smallskip 

There are approximately 60 known BHBs in our Galaxy. Of these, around 20 host black holes dynamically confirmed via spectro-photometric observations \cite{Casares2014}. Another $\sim$40 systems are referred to as 
BH candidates: they share a strong phenomenology with the confirmed black holes (e.g. \cite{vanderKlis2005,Munoz-Darias2014}). In the following, we will not differentiate between candidate and confirmed BHs, and we shall refer to these systems collectively as the BHBs.
The BH discovery rate is currently limited by the X-ray outburst rate of BHBs (the number of persistent systems being so dramatically lower than that of transient ones), which is relatively small - on average 2 outbursts per year - due to the typically low duty cycles of these systems. Most transient BHBs are \xray  bright for only a small fraction of the time that is on the order of a few percent, and most of the time, they are in a quiescence state, i.e. very \xray dim or (more typically) invisible to observers from Earth \cite{Campana2001}. Therefore, the total of $\sim$60 known BHBs in the Galaxy is necessarily only a minute fraction of the actual Galactic BHB population, which is thought to be greater than $\sim 10^4$ \cite{Yungelson2006}. In turn, the Galactic BHB population represents an equally minute fraction of the total number of stellar mass BHs in the Milky Way, estimated to be in the range of 10$^8$–10$^9$ \cite{vandenHeuvel1992,Brown1994}.


\subsection{History}\label{sec:history}

The first BHB, Cygnus X-1 (Cyg X-1), was discovered in 1964 \cite{bowyer:65a,Giacconi1967}, at the dawn of \xray astronomy, via sounding rockets carrying \xray sensitive instruments. Cyg X-1 became the first candidate for systems hosting a BH when optical observations revealed its binary nature, thus allowing an estimate of the mass of the compact object \cite{Bolton1972,Webster1972}. 
Cyg X-1 is one of the few known persistent BHB systems in the Galaxy, which are relatively rare in the local Universe as it was established decades after its discovery. 
We now know that most BHBs  are transients, and thus only visible during rare \textit{outbursts}.
A~0620$-$00 was the first transient system which was also classified as a BHB \citep{McClintock1986b}, with a lower limit to the mass of the compact object of 3.2 \msun \cite{Elvis1975}. 
A few additional transient BHBs were later discovered, including two \xray sources in the Large Magellanic Cloud (LMC X-1 and LMC X-3), which were afterwards classified as BH candidates, and GX~339$-$4, one of the most extensively studied BHBs  \cite{Markert1973}. 

With the launch of the Ginga satellite and its all-sky monitor, the number of known BHBs started to increase. Among these, a notable case is that of GS~023+338 in 1989 \cite{Makino1989}, which was quickly connected with the optical source V404 Cygni, and also linked to an optical transient observed over 50 years earlier \cite{Hurst1989}. 
\citet{Casares1992} obtained the first determination of the system's mass function ($f$(M) = 6.26 $\pm$ 0.31 \msun), making GS 2023+338 what is considered the first dynamically-confirmed BHB (even though \citet{Bolton1972} already claimed such record for Cyg X-1). 

This era of early observations began to elucidate some of the basic properties of BHBs, namely that they vary on relatively-short timescales, and show significant flux changes, which led to
the definition of a BHB's high and low states \cite{Tananbaum1972}. Such states are largely based on observations of Cyg X-1, which were initially performed in a limited X-ray band only extending up to 10 keV. 
More detailed studies based on several observations---in particular of the transient GX 1124$-$683 (also known as Nova Muscae 1991), GX 339$-$4 and Cyg X-1---showed that the complex behaviour of BHBs could not be reduced to two states, and additional states were defined, based on the complex spectral and timing properties of these systems  \cite[see, e.g., ][]{Miyamoto1993, Homan2006, McClintock2009,Done2007}. 
In particular, the first spectral studies based on the modelling of the accretion disc emission opened the way to techniques aimed at measuring the inner disc radius \cite{Mitsuda1984}. 

The hard X-ray/soft $\gamma$-ray sky has been notably explored in the 1980' by the {\it{CGRO}} and \textit{Granat} observatories, which both carried instruments covering a large energy domain, from a few keV to a few MeV (and even more in the case of {\it{CGRO}}/EGRET).
The study of accreting BHBs took a jump forward with the launch of the Rossi \xray Timing Explorer (\rxte) satellite in 1995. \rxte carried an All Sky monitor (\asm) that allowed a lot more outbursts of old and new transients to be detected. \rxte also carried two high time-resolution instruments with large effective areas, which could operate in a very flexible way while accumulating high signal-to-noise data for timing and spectral studies in the 2-200 keV energy band.
The \textbf{INTErnational Gamma-Ray Astrophysics Laboratory} (\integral) was launched in 2002, just few years after the \xmm and \chandra \xray satellites (both launched in 1999).
It contributed significantly to the field of BH research by improving the knowledge of the high-energy emission from accreting XRBs, thanks to its sensitive high-energy instruments. 
These four observatories brought on an explosion of information, which led to new insights into the emission properties of BHBs that resulted in a much clearer understanding of the process of accretion onto compact objects \cite{vanderKlis1996a, vanderKlis1997, McClintock2006,  Done2007, Belloni2016}. 

\smallskip 

Since their discovery, stellar-mass accreting BHs in Galactic binaries have been studied in comparison with the super massive black holes (SMBH) in Active Galactic Nuclei (AGN), with which they share a number of fundamental properties, reflecting the fundamentally similar physics underlying their emission \cite{Gallo2003,Merloni2003,Falcke2004, Plotkin2012, Gultekin2019}. Of course, Galactic BHBs have also been largely compared with other types of accreting binaries, first and foremost NS XRBs. Such systems in many respects show the same properties and characteristics of their BH counterparts, and thus provide the best control sample to test which of the fundamental processes governing the inflow/outflow of matter and energy around all accreting objects may be unique to BHs.

Another type of systems typically compared with BHBs is that of the enigmatic Ultra-Luminous \xray (ULX) sources, a class of very bright, unresolved, extra-galactic \xray sources, generally located at some distance from the centre of their host galaxies, and currently believed to be powered by accreting compact objects \cite{Kaaret2017}.
The compact object hosted in these systems have been proposed to be either a stellar mass BHs or (more likely) NS accreting either significantly above the Eddington accretion rate or non-isotropically (see e.g. \cite{Israel2017}); or an intermediate mass black holes (IMBHs) accreting at much lower accretion rates \cite{Earnshaw2016}. 

A final type of BH-powered system that is being compared to BHBs especially in the late years is that of the tidal disruption events (TDEs), where a star is gravitationally captured by a (typically quiescent) SMBH and is partially or completely torn apart by tidal forces \cite{Rees1988, Guillochon2015,Holoien2016}. When such an encounter takes place, a relatively short-lived accretion disc forms around the SMBH from the stellar matter of the captured star.
Such an accretion disc would then power the production of radio jets in many ways similar to those produced in BHBs and in the AGN \cite{Komossa2015}. This further strengthens the idea that the accretion/jet generation process occurs according to the same principles in \textit{many} systems that share one key property: the combination of a deep gravitational potential well and an accretion disc.

\subsection{In this paper}

This review focuses on the systems identified as BHBs (candidates or dynamically confirmed), with an emphasis on the contribution that \integral has made to improve our understanding of these systems.
We will restrict ourselves to the science of these systems, while the details relating to the \integral mission and its instruments are reviewed in \citet[and reference therein]{kuulkers2021},
and the data reduction is discussed in the \integral user manuals\footnote{See \url{https://www.isdc.unige.ch/integral/analysis}.}

\integral is an ESA medium size mission dedicated to the spectroscopy and imaging of celestial hard \xray/soft $\gamma$-ray sources, in the energy range 15 keV to 10 MeV with concurrent source monitoring in the X-ray (3-35 keV) and optical (V-band, 550 nm) energy ranges. 
The \integral payload consists of the two main soft $\gamma$-ray instruments,  both using a coded mask imaging technique, with a field of view of $\sim 30^{\circ}$: SPI, a spectrometer that operates in the 18 keV--8 MeV energy range with an energy resolution of 2.2 keV (FWHM) at 1.33 MeV; and the imager IBIS, with a spatial resolution of 12 arcmin, operating between 15 keV and 10 MeV. 
\integral carries also two monitors, JEM-X (formed by two coded-mask units), sensitive to the X-rays (3--35 keV) with a field of view of 5\degrees, and the optical monitor OMC, equipped with a Johnson V-filter to cover the wave-length in the 500--600 nm range. All instruments are co-aligned, and thus are able to cover simultaneously a very broad energy range. 

\integral has permitted several important leaps forward in the understanding of BHBs and various aspects  of accretion-ejection mechanisms. This progress has not been due to specific instrumental capabilities - even if at the time of launch in 2002, SPI featured the best spectral resolution and IBIS the best spatial resolution above 20 keV, compared to the other X-ray missions operational back then. Instead, advances have been made possible by the combination of many unique properties of the \integral mission, such as: a very wide field of view of 30\degrees\ by 30\degrees\ (900 square degrees); the ability to perform long uninterrupted observations of up to 3 full days; and a high sensitivity above 20 keV.
These capabilities are also  what made \integral ideal for the study of BHBs. Specifically, early stages of outbursts in the hard state could be detected serendipitously, so that simultaneous multi-wavelength observing campaigns could be promptly organised, and surveys and monitoring observations --- especially of the Galactic 
plane and Centre ---  could be planned as core programs and encouraged as guest observers 
programs. All the above allowed for spectral transitions to be observed and probed in detail, and important connections between the X-rays (accretion flow)/soft $\gamma$-rays ('hard tails') and the radio (jets) emission to be investigated. 

This paper is organised as follows: \cref{sec:states} describes the typical outburst evolution of BHBs, the states and transitions that characterise them and their properties. 
\cref{sec:discjet} looks at the different types of outflows that can be observed from BHBs, and reviews the current understanding of the connection that exists between accretion and outflows. 
\cref{sec:high-energy} examines the properties of the hard X/soft $\gamma$-ray emission probed by \integral, and briefly reviews the interpretations and models that have been put forward since the beginning of the mission. 
\cref{sec:sources} details the main results obtained from \integral observations of BHBs.
We first focus on a few sources of historical importance, which were most relevant for the understanding of BHBs in general, and to the understanding of which \integral particularly contributed (\cref{sec:GRS1915} to \cref{sec:annihilator}). Then we will briefly review the sources that have been discovered thanks to \integral (\cref{sec:INTdiscoveries}), and finally all the sources that \integral observed since its launch (\cref{sec:otherBHs} and \cref{sec:notbh}).

\section{Outbursts}\label{sec:states}

In BHBs, the \xray emission arises from the inner portion of the accretion flow, and possibly from the base of the jets (observed primarily in radio), which may or may not coincide with part of the accretion flow itself \cite{Markoff2010}. While a few BHBs are persistent sources, like the archetypal BH system Cyg X-1, (which is a persistent accretor and shows a luminosity consistently above $10^{37} erg s^{-1}$), most of them are of transient nature. With a recurrence period that varies between several months and decades \cite{Campana2001}, the accretion rate onto the central objects increases by orders of magnitude and the sources enter their {\bf outburst} phases, which are interleaved by (typically much longer) {\bf quiescence} phases. During outbursts, the \xray luminosity of XRBs increases, peaks and then decays, roughly tracing the changes of accretion rate onto the BH. 

While active, most BHBs show significant luminosity variations that take them from the low quiescent luminosity of the order of $L \sim 10^{30-31}\,\ergs$ \cite{Reynolds2014}), to values of the order of $L \sim 10^{38-39}\,\ergs$ or more, and through marked spectral and fast time-variability changes. 
For most BHBs, outbursts---historically referred to as \xray novae---last between a few weeks and a few months (although a few systems have shown decades-long outbursts, like GRS 1915+105), and generally recur on the timescale of years to decades \cite{Tanaka1995a}.

\subsection{Accretion states}

The different properties observed during the outbursts of transient BHBs are generally associated with different accretion states, but even restricting ourselves to the two original accretion states identified in Cyg X-1 \cite{Tananbaum1972}, the (low) hard and (high) soft states, the details of the emission are rather complex \citep[e.g.,][]{Gilfanov2010}, and have led to a number of state classifications \cite{zdziarski04_states, Homan2005,McClintock2006}. In this review, we will arbitrarily follow the classification originally proposed by \citet{Homan2005} (see also \cite{Belloni2011,Belloni2016}). 

\begin{figure}
\centering
\includegraphics[width=0.45\textwidth]{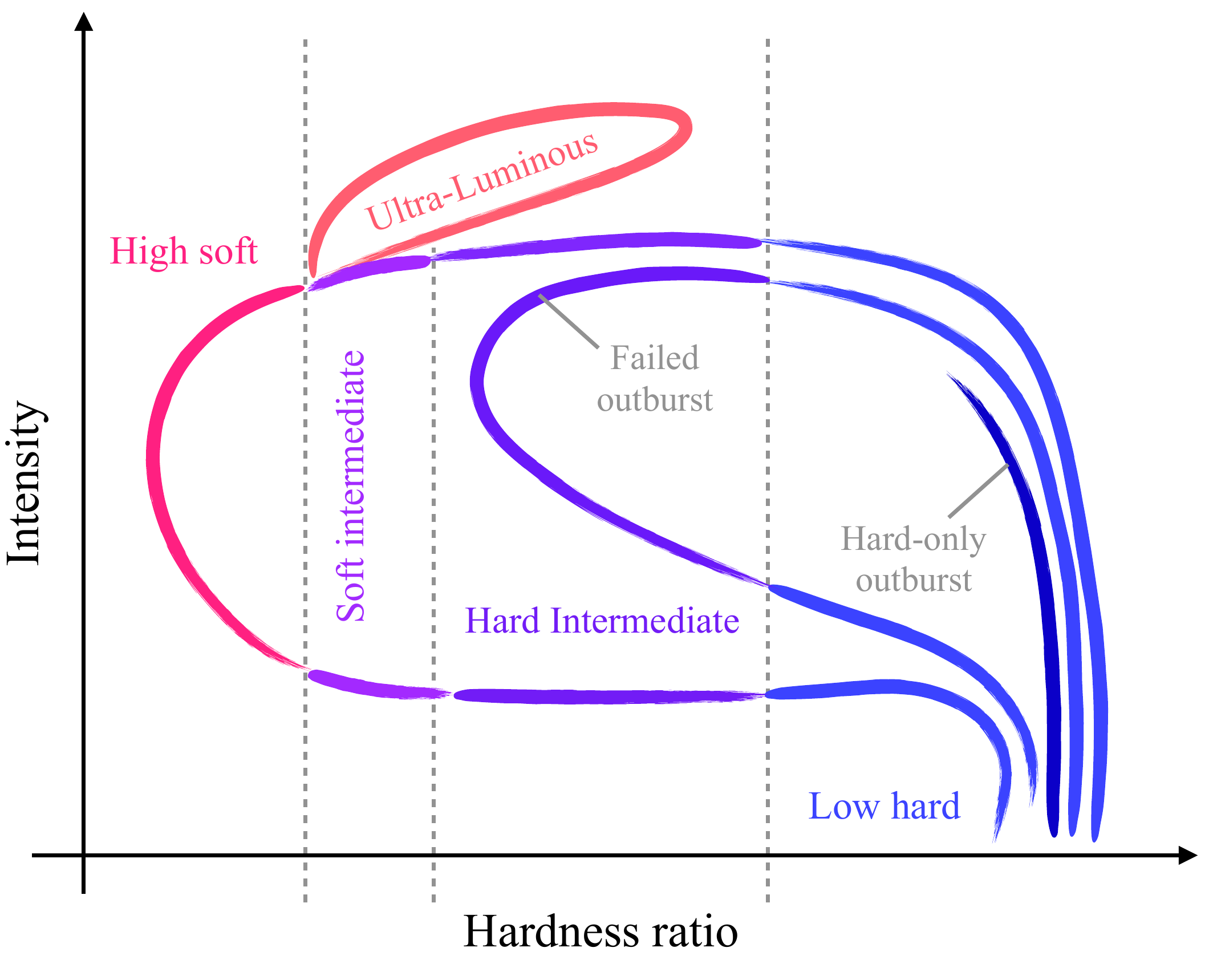}
\caption{A sketch of a typical Hardness Intensity Diagram (HID). All the accretion states sampled during an outburst---save for the ultra-luminous state---occupy 4 different vertical bands in the HID. The name of each each state is marked on the plot. During an outburst, the ``q-shaped'' loop is travelled anti-clockwise from the bottom right corner. All or only some of accretion states are sampled. Hard-only and failed outbursts are also marked on the figure. As noted in the text, the qualifications ``low'' and ``high'' attached to the hard and soft state denominations have an historical origin and and do not hold a physical meaning. }\label{fig:HID}
\end{figure}

The various physical components that contribute to the energy spectra, as well as the various fast time-variability properties of the emission from BHBs, vary quickly (in $\sim$hours or faster) and in a correlated fashion, in ways that vary in different states and/or across certain state transitions. Outbursts of different sources, and even different outbursts from the same systems, show time evolution that can differ considerably.
Fortunately, these variations follow repeating cyclical patterns of behaviour that become apparent in a so-called \xray \textbf{hardness-intensity diagram}, or \textbf{HID} \cite{Belloni2016} (see \cref{fig:HID}), where accretion states and transitions can be identified as specific regions of a q-shaped track. 
HIDs have been commonly used to describe the behaviour of accreting stellar mass BHs for over two decades now \cite{Belloni2016}, but they have been used also in the case of NS binaries \cite{Munoz-Darias2014}, and even in AGN \cite{Koerding2006, Svoboda2017}.
The HID is the \xray equivalent of the Hertzsprung–Russell diagram used in stellar astrophysics, but instead of being populated by several sources at different evolutionary phases spanning several million years, the HID shows the same source evolving quickly (over weeks to years) through different states. 
An HID shows the \xray intensity in units of (instrumental) count rate, luminosity, or flux as a function of a hardness ratio, calculated as a ratio of count rates (or luminosity, or fluxes) measured in two energy bands (hard over soft is typically used).

The general evolution of a transient BHB during an ordinary (as opposed to failed, see below) outburst is roughly always the same: a ``q-shaped'' track is travelled counterclockwise in the HID from the bottom right, which corresponds to the initial phase of an outburst, just above the quiescence, which is in general not included as most \xray telescopes are not sensitive enough to detect sources at such low luminosity. 
The \textbf{(Low) Hard State (LHS)}, which is observed at the beginning and at the end of an outburst, can be roughly identified with the right-hand vertical branch; the \textbf{(High) Soft State (HSS)}, generally including or following closely the peak of the outburst, corresponds to the left-hand side vertical branch\footnote{Note that the right and left-hand side branches are not exactly vertical. This is an effect of the logarithmic scale typically used to display HIDs, which minimises the significant softening observed as sources brighten. } of the q-loop. We note that the qualifications ``low'' and ``high'' attached to the hard and soft state denominations are a legacy of the original state classification based on the first observations of Cyg X-1, and thus have an historical rather than physical meaning.

The transition between the two main states takes place at two different flux levels: at high flux, as the source moves from the LHS to the HSS, and at low flux, when it transitions back from the HSS to the LHS completing an hysteresis cycle \cite{Miyamoto1995}. Along the two horizontal branches that complete the q-track, lie the Intermediate States, which are often split in two states - the \textbf{Hard Intermediate State (HIMS)} and \textbf{Soft Intermediate State (SIMS)}. These two intermediate states can be distinguished from each other thanks to their different timing properties \cite{Belloni2016}. In a few sources a fifth state has been observed, the so-called \textbf{Ultra-Luminous State (ULS)}, or anomalous state \cite{Motta2012}. Such a state shares some characteristics with the two intermediate states, and is characterised by a high luminosity, which can  reach or even exceed the Eddington Limit 
\cite{Uttley2015}. 

Over the decades of \xray monitoring of several BHBs with, a general pattern has emerged, according to which a source moves around the hardness-luminosity plane following the sequence LHS-HIMS-SIMS-HSS-SIMS-HIMS-LHS. Occasional brief excursions back and forth from and to the HIMS and SIMS have been observed in a few systems, as well as transitions all the way back to one of the main states (LHS or HSS) in rare cases (see \cite{Fender2004} and, e.g, \cite{Motta2009a,Motta2012} for GX 339$-$4 and GRS 1655$-$40, \cite{Brocksopp2002} for XTE J1859+226). 

Despite the general pattern followed by many transient BHBs in outbursts, a relatively small number of unusual outbursts has been observed over the past decades, in which sources do not show all the main accretion states.
A number of sources never leave the LHS during the outburst \cite{Brocksopp2004}, and show hard-state-only outbursts. Other sources proceed to an intermediate state before returning to the HS and quiescence, without, however, completing a full transition to the HSS and back (\cite{Capitanio2009,Ferrigno2012,Soleri2013}, see Sec. \ref{sec:H1743}, \ref{sec:MAXIJ1836}, \ref{sec:Swift J1753}). Such outbursts are generally referred to as \textit{failed} outbursts, even though the same term is also sometimes improperly used in the literature to indicate hard-state-only outbursts (see below). 
It is worth noting that some sources have undergone both canonical outbursts and failed outbursts \cite{Sturner2005}. Since most of the failed outbursts are under-luminous, it has been argued that the lack of a transition to the HSS might be connected with a premature decrease of the mass accretion rate, as during the 2008 outburst of H1743$-$322 \cite{Capitanio2009a} (see Sec. \ref{sec:H1743}).

Weaker outbursts can follow the initial outbursts by some weeks to a few months \citep[e.g.,][]{Kajava2019}. These events, referred to as re-flares or re-brightenings, are very similar to the hard-state-only outbursts mentioned above, so much so that it is not clear whether these two types of outbursts are physically distinct. During re-flares, a source temporarily rises out of quiescence to a higher luminosity, but no transition to a softer state is observed, so that the system remains in the hard state, thus only travelling 
up and down the right-hand side vertical branch of the HID. Analogous 
secondary outbursts are also seen in systems hosting a white dwarf (see, 
e.g., \cite{Kuulkers1996}) or a neutron star (e.g., 
\cite{Lewis2010,Patruno2016}), suggesting a disc instability origin, or 
reflecting an increase in the mass released from the companion star, perhaps 
induced by \xray heating of its outer layers during the primary outburst 
(e.g., \cite{Tanaka1995a, Lasota2001}).

\bigskip

The accretion states are defined based on specific properties, which emerge both from the analysis of the \xray spectra of the source considered, as well as on their fast-time variability properties. The latter play a crucial role especially when it comes to the definition of the two intermediate states (see Section \ref{sec:timing}). In the following sections, we will give a brief overview of the main spectral and fast-time variability properties typically observed in transient  BHBs. 

\subsection{Spectral properties of BHBs}\label{sec:spectralprop}

The typical X-ray ($\sim$0.1 --200 keV) spectrum of a BHB is characterised by two main components (unlike the more complex spectra from accreting NSs), which play a different role in the different accretion states. 
A geometrically thin, optically thick disc \cite{Shakura1973}  truncated at a certain radius - the inner disc truncation radius - generally larger than the innermost stable circular orbit (ISCO, \cite{Bardeen1972}), is responsible for black-body-like emission below $\sim$6 keV, and takes the form of a multi-colour disc black body in the spectrum \cite{Shakura1973}. The emission from such a disc peaks at a few keV and is scarcely variable on time-scales of hours and shorter, thus does not contribute significantly to the overall variability of BHBs on such time-scales. 

A cloud of electrons located close to the BH is believed to be primarily responsible for the non-thermal emission due to Comptonisation of disc seed photons, appearing typically above 5--10 keV \cite{Sunyaev1980, Haardt1994}. 
Such a cloud of electrons has been historically referred to as the corona, in analogy with the solar corona, although it is now clear that the BH coronae have little to do with stellar coronae, and a lot to do with the accretion flow itself, which the corona is most likely part of. The geometry of the corona is still debated, but it seems that at least part of it coincides with to what is generally referred to as \textit{hot flow}. This is the central part of the accretion flow,  which puffs up into a relatively geometrically thick, optically thin accretion flow. The hot flow is believed to replace or sandwich the razor-thin Shakura-Sunyaev disc that exists further out \cite{Shapiro1976,Narayan1994,Esin1997,Done2007,Marcel2018a,Marcel2018,Marcel2019}, so that the accretion flow from the inner disc truncation radius inwards is best described by a geometrically thick disc. The electrons in the hot flow (which are thermalised) up-scatter the disc photons (or seed-photons) giving rise to hard X-ray emission above 6~keV. Such photons produce a power law-like component in the \xray spectra featuring a high-energy cut-off that reflects the temperature of the Comptonising electrons \cite{Eardley1975,Sunyaev1980, Zdziarski2004, Joinet2005, Done2007}. Such non-thermal emission, unlike the disc emission, is highly variable, and it is responsible for the marked fast time variability typical of accreting BHBs. 

In the soft state, where the electrons have cooled down by the increasing amount of seed photons being up-scattered, this hard \xray component does not show any cutoff below MeV energies. Therefore, a small population of electron with a non-thermal/powerlaw distribution, has been proposed to be responsible for the up-scattering of the seed photons \cite{Coppi1999, Poutanen1996, Gierlinski1999,McConnell2002, Zdziarski2002}. 

Besides the emission from photons Comptonised by thermally-distributed electrons, an additional non-thermal component may also be present, depending on the accretion state a source is sampling. This type of emission appears as a second power law like tail, which does not show any cut-off and extends up to the MeV region \cite{Bouchet2003}. The physical origin of this  high-energy tail is not clear, even though it has been explained in terms of a contribution of the jets, or as the result of the Comptonisation of the seed photons on a population of non-thermal electrons. This topic has been largely investigated by \integral, and more details on the issue of the origin of this soft $\gamma$-ray emission in BHBs is treated in more detail below, in \cref{sec:high-energy}.   

A third component - the reflection - can appear in the spectra as well, superimposed to the spectral continuum determined by the aforementioned components. The reflection emission is believed to arise from the hot flow emission reflecting off the geometrically thin disc, and takes the form of a number of lines around 6.4 keV (including the Fe K$\alpha$ line and a number of lines attributed to several Fe ionization states) and a high-energy hump peaking at around 30 keV \cite{Garcia2013}. In some sources, where the presence of a local absorber significantly affects the X-ray emission (see, e.g. V404 Cyg, Sec. \ref{sec:v404cyg}, \cite{Motta2017}), what is called \textit{reflection} includes the radiation reprocessed in the local absorber (including up- and down-scattering in a cool, possibly in-homogeneous, nearly optically-thick medium), and should be  more correctly referred to as \textit{reprocessed} emission \citep[e.g.,][and Sec. \ref{sec:V4042015}]{Motta2017b}. 

The contributions of all these components vary in the different accretion  states, and determines the overall spectral shape in the \xray band ($\sim$1--200~keV) and fast time-variability properties of BHB in outburst. The accretion disc dominates the \xray emission in the HSS, which is thus characterised by a soft \xray spectrum and very little to no thermal (i.e. Comptonised) emission in the $\sim$1--200 keV energy band, and very limited fast-time variability ($< 5\%$). The non-thermal emission from the hot flow dominates the LHS, which shows hard spectra extending up to several tens and sometimes even hundreds of keV, with little to no contribution of a thermal disc-blackbody, and a large fraction of fast time-variability (up to 30\%). In the HIMS and SIMS (and during the ULS, when sampled), both the thermal and non thermal emissions are present, with the contribution of each component smoothly decreasing/increasing in the spectrum depending on the outburst phase: as a source moves from the LHS to the HSS, the inner disc truncation radius decreases, the hot flow emission decreases, and the thermal emission rises. The opposite trend is observed when a source moves from the soft to the hard state, with the disc emission progressively decreasing and the hot-flow emission increasing, as a consequence of an increase in the inner disc truncation radius \cite{Done2007}. These trends drive changes in the fast time variability, which decreases along the LHS to HSS transition (from above 20\% to 10\% in the HIMS, and between 5\% and 10\% in the SIMS), and increases along the opposite direction. This is due to both the decrease/increase of the number of non-thermal and variable photons, and the increase/decrease of thermal/non-variable photons which dilute the integrated variability.

\subsection{Fast time variability properties of BHBs}\label{sec:timing}

Black hole \xray binaries have been known to be variable on short timescales since the very first observations of Cygnus X-1 \cite{Tananbaum1972}. When the available \xray instrumentation became sufficiently sophisticated to allow detailed variability studies on short (sub-second) time-scales, the definitions of states and transitions were refined to include fast timing properties \cite{vanderKlis2005, McClintock2006, Belloni2016}.

While the long-term (days to years) variability of transient BHBs in outburst is most commonly described through the HID, the shorter time-scales cannot as easily be appreciated by the same means, and even the inspection of a source's light curve can be unpractical. Hence, the fast (i.e.,  sub-minute) time variability in accretion binaries is best studied in the Fourier domain, and the Power Density Spectrum (PDS) - the modulus squared of the Fourier transform of the light curve - is a tool commonly employed to study the emission from \xray binaries. 
The PDS from BHBs typically displays several different features, ranging from various types of broad-band noise spanning several decades in frequency (i.e. essentially scale-free), to much more narrow features, called quasi-periodic oscillations (QPOs). 

QPOs have been discovered in the late '70s in the BHB GX 339$-$4 \cite{Samimi1979}, and have been observed since then not only in virtually all BH and NS \xray binaries, but also in a number of ULXs and AGN \cite{Kaaret2017, Alston2016}.  QPOs reflect the oscillations of the accretion flow in the strong field regime, and yield accurate centroid frequencies linked to the motion of matter and/or accretion-related timescales. Thus, the study of variability in general and QPOs in particular provides a way to explore the accretion flow around BHs in ways which are inaccessible via the spectral analysis alone. The association of QPOs with specific spectral states and transitions indicates that they could be a key ingredient in understanding the details of the physical conditions underlying the accretion states.
Furthermore, being produced in the vicinity of relativistic objects such as BHs and NSs, QPOs are expected to carry information about the condition of matter in the strong field regime, thus providing powerful probes of the predictions of the Theory of General Relativity.

Most of the components found in a typical PDS, and in particular QPOs, have been studied in great detail by fitting the Poisson noise subtracted PDS of many observations with various empirical models, the most commonly used consisting of a sum of Lorentzian functions. From Parseval’s theorem, the PDS can be normalised such that its integral over all positive frequencies is equal to the square of the fractional root mean square deviation (or \textit{rms}) of the corresponding time series (this results in a PDS fractional rms normalisation). The rms essentially quantifies how variable a given time series is in a given Fourier frequency range. The typical frequency range sampled by the recent \xray instruments is from a few mHz to a few kHz (ms to ks). 

QPOs in BHBs are normally divided into two large groups, based on the frequency range where they are usually detected: the low frequency QPOs and the high-frequency QPOs. The former in BHBs have traditionally been classified into three types: A, B and C. 

Type-C QPOs  are by far the most common type of QPO in BH systems. Type-C QPOs can be detected in all accretion states (perhaps except the SIMS), although they are most prominent in the HIMS and at the bright end of the LHS, where the overall variability is large - up to 30\% rms in the 0.1--100 Hz band. The centroid frequency of type-C QPOs is highly sensitive to the state of the source, rising from a few mHz in the LHS at low luminosities to $\sim$10 Hz in the HIMS, and up to $\sim$30 Hz in the ULS. Type-C QPOs have also been observed at optical (e.g. \cite{Motch1983,Imamura1990,Gandhi2010}), ultraviolet \cite{Hynes2003a} and infrared \cite{Kalamkar2016} wavelengths.
Type-B QPOs  have been detected in a large number of BHBs and they appear during the SIMS, which is in fact \textit{defined} by the presence of a Type-B QPO \cite{Belloni2016}, and by a rms level of $\sim$5-10\% in the 0.1-100 Hz band. They are characterised by a relatively high amplitude (up to $\sim$5\% rms) and narrow ($Q \gtrsim 6$) peak, and generally show centroid frequency at 5-6 Hz (even though type-B QPOs with significantly lower centroid frequencies have been also reported \cite{Motta2011}). Being associated to a short-lived accretion state, type-B QPOs are comparatively much less frequent than Type-C QPOs. The same is true for type-A QPOs, which are the least common type of LF QPO in BHBs: the entire \rxte\ archive only contains $\sim 10$ significant Type-A QPO detections.

High Frequency QPOs (HFQPOs) are rare features in BHBs (but much more common in NS XRBs, \citep[e.g.,][]{Motta2017}), and to date most of the confirmed detections have been made by \rxte\ (but see also \cite{Belloni2019}). The first HFQPO was detected in 1997 in GRS 1915+105 (at $\sim 67$ Hz \cite{Morgan1997}) which is the source where the vast majority of HFQPOs come from. Other sources who have been claimed to show HFQPOs - all at frequencies significantly higher than in GRS 1915+105 - are XTE J1550$-$564, GRO J1655$-$40, XTE J1859+226, H 1743$-$322 , GX 339$-$4, XTE J1752$-$223, 4U 1630$-$47, IGR J17091$-$3624 (which also shows a high-frequency QPO at $\sim$67Hz, \cite{Altamirano2012}). However, most detections have been proven to be not statistically significant, leaving only a handful of detections from two sources (XTE J1550$-$564, GRO J1655-40, see \cite{Belloni2012}).


Many models have been proposed for both LFQPOs and HFQPOs, some of which have been developed for years and largely tested against data, while others are little more than ideas mentioned in a few papers at most. Dwelling on the details of the theoretical models for QPOs is beyond the scope of this work, and therefore we refer the interested reader to \cite{Ingram2020} and references therein for a recent review on the topic. Here we will only mention that the only type of QPO that has a relatively widely accepted explanation  is the type-C.Type-C QPOs  have been ascribed to the Lense-Thirring precession of particles within the accretion flow, a General relativity effect that arises from the frame dragging occurring in the vicinity of spinning massive compact objects \cite{Lense1918, Ingram2009,Motta2018}.

\section{The accretion-ejection connection.}\label{sec:discjet}

The '90s saw a leap forward in the knowledge of accreting BHBs thanks to a remarkable increase in the \xray data available. These years were also marked by the discovery of relativistic jets from the BHBs GRS 1915+105 \cite{Mirabel1994} and 1E1740.7--2942 \cite{Mirabel1992}, to which followed the discovery of more jets from a number of other \xray binaries hosting either BHs (see e.g. \cite{Hannikainen2000, Hannikainen2001,Corbel2005}) or NSs \cite{Fomalont2001,Migliari2006}. BHBs began to be referred to as \textit{micro-quasars}, in analogy to a jetted sub-class of Active Galactic Nuclei (AGN) - the quasars - where radio jets had been already observed frequently in radio for at least a few decades. Such a discovery opened a new path to the study of accretion and its connection with the generation of jets, and of outflows in general, which is often referred to as to \textit{disc-jet coupling\footnote{Even if this term has been initially coined to describe the connection that exists between the accretion disc and the generation of jets, it is now more broadly used to describe the connection between the accretion flow and all the outflows that the disc is capable of generating. }}, or accretion-outflow connection. 

Similarly to what happened for the radio jets, which were observed much earlier around SMBHs than around stellar mass BHs, powerful winds, relativistic and not - already observed in a number of AGN - were also discovered initially only in BHBs \cite{Ponti2012} and later in NS systems as well \cite{Ponti2014}. Unlike jets, which are believed to be mildly to highly collimated and approximately orthogonal to the accretion disc, winds appeared to be flowing in an almost equatorial direction with respect to the accretion disc, and are thus detected mainly in sources observed at high inclination (i.e. close to edge-on)  \cite{Ponti2012}. Probably due to the low number of simultaneous multi-band observations available, it initially seemed that disc winds and jets were mutually exclusive, with the first being launched in the radio-free HSS, where the radio activity was generally thought to be quenched. It must be noted that in the past observations were carried out mainly in the \xray band, which prevented to fully grasp the contribution of some components of accreting systems (such as jets and winds), which can be energetically very important. 
It has now become clear that the full picture is significantly more complex than that available only a decade ago, and new observations, more and more often carried out simultaneously in several energy bands (primarily in the soft and hard \xrays and radio, but also in optical, infra-red, and sub-mm band) are continuously unveiling unknown properties of jetted accreting systems that highlight how the connection between the accretion process and the generation of outflows is far from being fully understood.

The general picture which started to emerge in the early '00s remains overall true. During the LHS, steady, compact (i.e., generally not resolved in high-resolution radio images) and mildly relativistic jets dominate the radio emission. Simultaneous \xray and radio observations of BHBs in the LHS have evidenced the existence of a non-linear correlation between the \xray and the radio luminosity, originally observed in GX 339$-$4 \cite{Hannikainen1998} and then found in many other BHBs by several authors in the last two decades. This correlation was initially considered universal \cite{Gallo2003} - which led to a number of works on the fundamental plane of black hole activity \citep[e.g.,][]{Merloni2003,Falcke2004, Plotkin2012, Gultekin2019}. More recently it became clear that most transient BHBs are considerably less luminous in the radio band than GX 339$-$4 \cite{Coriat2011, Motta2017}, and populate a second track in the radio-\xray plane, which effectively includes the majority of the known BHBs \cite{Motta2018b}.

During the intermediate states, and in particular close to the HIMS-SIMS transition, powerful relativistic radio ejections are observed, which are often resolved as synchrotron emitting blobs in radio Very Long Baseline Interferometry (VLBI) high-resolution images. Jets---both the steady compact jets, and the transient relativistic one---are believed to transport an enormous amount of (kinetic) energy, which is transferred to the BH surrounding. Observational evidence exists  that the power output of low-luminosity – i.e. the overwhelming majority of – stellar mass BHs is dominated by the kinetic energy of ‘dark’ outflows, whose key signature is the eventual energising of the ambient medium \cite{Gallo2005}.

In the soft states, highly ionised winds - the presence of which is determined via the detection of characteristic \xray absorption lines - are launched on a quasi-equatorial direction at up to relativistic velocities. Winds are thought transport a huge amount of matter away from the accretion disc, possibly comparable with the amount of mass accreting onto the black hole, and perhaps enough to trigger a return to the hard state \cite{Munoz-Darias2016}.
Occasionally, cold winds (i.e., consistent with being not ionised), which appear to co-exist with the radio jets (either the compact steady jets, or the discrete, relativistic ones), have also been observed in the optical band during the hard and hard intermediate states \citep{Munoz-Darias2016}. Such finding was among the first pieces of observational evidence to cast doubts on the idea according to which jets and winds are associated with different accretion states, and therefore cannot co-exist.

A different sort of cold outflow has been observed in a limited number of systems, all believed to erratically or persistently accrete at or above the Eddington limit: a non-homogeneous, optically-thick mildly relativistic outflow which arises from the inner part of the accretion flow as a result of radiation pressure forces, which develop at high accretion rates. In a rather spectacular way, V404 Cyg displayed the best example of such an outflow observed to date (see Sec. \ref{sec:v404cyg}), but other sources show a similar property, albeit in a less extreme way (see e.g.  \cite{Miller2020,Neilsen2020,Koljonen2021,Motta2021} for the case of GRS 1915+105, \cite{Revnivtsev2002} for V4641Sgr, and \cite{Hare2020, Munoz-Darias2020} for  Swift J1858.6$–$0814). Interestingly, again similarly to the case of both jets and winds, outflows of this sort have been already observed - although not frequently - around SMBHs, in those AGN which are generically referred to as obscured (see e.g., \cite{Motta2017} for a discussion on the topic). 

\section{The high-energy emission from BHBs in the \integral era}\label{sec:high-energy}

Understanding the mechanisms of production of high-energy photons by XRBs is crucial for understanding the physical processes at work in the  regions surrounding the central compact objects. The properties and evolution of the two main components constituting the energy spectra of BHBs - the cool, geometrically thin disc, and the hotter, geometrically thicker hot flow (see Sec. \ref{sec:spectralprop}) - explain well the spectral shape evolution from the sub-keV region up to $\sim$100--200 keV. As already mentioned above, the $\sim$10--200 keV spectrum is well understood in terms of Comptonisation of soft photons from the disc by the Maxwellian-distributed electron population in the hot flow \cite{Zdziarski2004}. 

As soon as the signal to noise ratio became high enough thanks to improved instrumental sensitivity, it also became clear that the BHB emission above $\sim$100--200 keV deviates from the spectral shape expected in the presence of a thermal Comptonisation continuum.
The earliest observations of Cyg X-1 (see Sec. \ref{sec:CygX1}) in its hard state, revealed the first hints of spectral hardening above a few hundreds of keV (see \cite{Ling1987} and references therein) as well as the presence of possible annihilation line features \citep{Nolan1983}. 
More robust results followed at the beginning of the ‘90s, with the launch of space-bound missions covering a broader energy domain, which featured improved sensitivity that started to shed light on the soft $\gamma$-ray emission from BHBs. Instruments such as OSSE \&  COMPTEL on {\it{CGRO}}, and \textit{SIGMA} on \textit{Granat}, which observed the sky from 15 keV up to few MeV, really opened the way to the  exploration of the high-energy emission from BHBs. 

A soft $\gamma$-ray tail, in excess of a thermal Comptonisation and in the form of a simple power law, was confirmed in Cyg X-1, both in the soft and in the hard states \cite{McConnell1994, McConnell2002}. A similar component was also tentatively detected in the transient BHB GX 339$-$4 \cite{Wardzinski2002} and in Nova Persei (GRO J0422+32), a bright \xray transient discovered in 1992 by \textit{CGRO}/BATSE \citep{Grove1998,Jourdain1994, Ling2003}. Such findings have emphasised the need to explore the high-energy sky above a few hundreds keV. 

Thanks to \integral, more and more observations covering the high-energies have become available over the years, revealing a more complex configuration. The \integral mission was specifically designed to investigate in greater detail the  hard \xray/soft $\gamma$-ray domain, in particular thanks to two instruments - IBIS and SPI - featuring high sensitivity in the energy band from 20 keV  to a few MeV. 
\integral observations of Cyg X-1 assessed the presence in the hard states of a spectral component extending to the MeV, beyond the Comptonised emission observed in all the known BHBs \cite{Bouchet2003, Cadolle2006, Malzac2008,Jourdain2012b,Del_Santo_2013a}. 
Thereafter, the presence of such a soft $\gamma$-ray tail was revealed by \integral in several BHBs, when in hard or hard-intermediate states (e.g. GX 339$-$4 \cite{DelSanto2008},  GRS 1758-258, \cite{Pottschmidt2008}, 1E 1740 \cite{Bouchet2009}, V404 Cyg \cite{Roques2015}, Swift J174510.8--262411 \cite{DelSanto2016}, MAXI J1820+070 \cite{Roques2019}, GRS 1716--249 \cite{Bassi2020}) and also in a few systems harbouring a neutron star (e.g., GS 1826$-$238, \cite{Rodi2016}).
After the launch of \integral, it started to become clear that if the presence of the second  component (referred to in the literature as the ``hard tail'' or the ``soft $\gamma$-ray tail'')  is omitted in the model description of a BHB broad-band spectrum, the temperature of the Comptonising medium may be affected and will be overestimated. Furthermore, variations in the soft $\gamma$-ray component may mimic an evolution of the apparent temperature and/or optical depth of the Comptonised continuum. This is particularly important for instruments with limited sensitivity in the corresponding energy domain (i.e., above 200 keV), which would not be able to properly probe the soft $\gamma$-ray tail and its variations.

Various works were dedicated to the study of the variability of this high-energy emission and to its correlation with other spectral components (see the individual source sections below), and evidence was found that the soft $\gamma$-ray tail evolves independently from the thermal Comptonisation emission \cite{Droulans2010}. However, none of these works has allowed the validity of one particular scenario over the others to be confirmed \cite[e.g.,][]{Joinet2005,Malzac2006,Pepe_2015a,Jourdain2017}. 
Various hypotheses on the nature of the soft $\gamma$-ray tail have been put forward: a second Comptonisation region (or a gradient of temperature in one single emission region), a pair plasma \cite[e.g.,][]{Bouchet2009, Nolan1983} or - probably the most popular one - an hybrid thermal/non-thermal electron population \cite[e.g.,][]{Coppi1999, Poutanen1998}. It has also been suggested that the high-energy tail could be at least partially explained in terms of a jet contribution to the \xray emission (see also  \cref{sec:discjet}), but the level of this contribution is the subject of intense controversies. 

\integral detected a strong polarisation of the soft $\gamma$-rays (0.4--2 MeV) from Cyg X-1 (\cite{Laurent2011, Jourdain2012}, see  \cref{sec:CygX1}), which indicates that the jet synchrotron emission likely dominates at these energies, in particular in the hard states \cite{rodrigue15_cyg}. Polarisation measurements on other BHBs proved to be difficult mainly due to low number statistics limitation, as polarisation measurements based on IBIS and SPI data require the accumulation of an important number of photons above $\sim$100 keV, which is often challenging in sources not as bright as Cyg X-1. Therefore, the vast majority of the polarisation measurements from XRBs made by \integral\ come from observations of Cyg X-1, even though tentative polarisation detections exist for two extreme transients (V404 Cyg, \cite{Laurent2016} and MAXI J1348$-$630, [Cangemi, F, PhD thesis, 2020]). 

At low energies (i.e. below a few tens of keV), the measured polarisation is limited ($<$10 percent in the case of Cyg X-1, \cite{Long1980,Chauvin2018}), which seems to favour Comptonisation in an extended hot flow for the bulk of the observed \xray emission below $\sim$200 kev \cite{Chauvin2018, Chauvin2019}. However, some authors have pointed out that Comptonisation, or hybrid Comptonisation, may also arise in the jets \citep[e.g.,][]{Markoff2005,Kylafis2008}, but for this component to dominate the \xray emission, the energy budget requires the ejected plasma to be dominated by electron-positron pairs \cite{Malzac2009,Lucchini2021}.
A highly polarised emission in the soft $\gamma$-rays, instead, points at the presence of a synchrotron emission arising from an ordered medium, most probably a jet. It seems therefore possible that the (polarised) soft $\gamma$-ray emission arises from the same population of electrons that generate the synchrotron emission in radio, and that the jet produces most of the photons observed above 300--400 keV. The fact that the polarisation fraction increases with energy (from less than 20\%  below $\sim$230 keV, to greater than 75\% above $\sim$400 keV) is consistent with a scenario where the jet contributes marginally to the emission below $\sim$100 keV - where instead inverse Compton is responsible for the bulk of the emitted photons - and dominates above $\sim$400 keV in a form of a power law-like spectrum with photon index $\sim$1.6 \cite{Laurent2011, Jourdain2012}.

Theoretical studies performed by \citet{Zdziarski2014} demonstrated that jet models can in fact explain the soft $\gamma$-ray-energy emission assuming continuous particle acceleration along the jet. However, standard jet models usually require rather extreme parameters in order to reproduce the observed MeV spectrum of \xray binaries \citep[e.g.,][]{Zdziarski2014,Bassi2020,Kantzas2021}, such as a very hard electron energy distribution (electron index $\sim1.5$) and strong constraints on the magnetic field. Therefore, on the one hand, the spectrum observed above the high-energy cut-off is more easily interpreted in terms of inverse Compton emission from a small population of non-thermal particles located near the BH \citep[e.g.,][]{Poutanen1998,Malzac2009a,Poutanen2014}. On the other hand, the problem in this scenario is that the degree of polarisation expected from these hybrid thermal/non-thermal Comptonisation models would be much weaker than that observed, e.g., by \integral in Cyg X-1 above 300 keV \cite{Beheshtipour2017}. 

Even though more and better data are necessary to further investigate the nature of the soft $\gamma$-ray tails in BHBs in order to solve the above discrepancy, the high degree of polarisation observed in the very hard tail makes a jet origin quite plausible.


\section{Galactic Black Holes}\label{sec:sources}

\begin{table*}[htb]
    \centering
    \begin{tabular}{c c c}
    \hline
    System & Coordinates & References \\
           & J2000       & (based on \integral\ results) \\ 
    \hline
    \hline
GRS 1915+105        & 19 15 11.55 +10 56 44.90   & \cite{Droulans2009} \cite{Fuchs2003} \cite{hannikainen05} \cite{hannikainen03}  \cite{rodrigue08_1915a}    \cite{rodriguez2008} \cite{Rodriguez2019}\\    
Cygnus X-1         & 19 58 21.67 +35 12 05.78   & \cite{Bazzano2003} \cite{Bouchet2003}  \cite{Cabanac_2011a}  \cite{Cadolle2006} \cite{Cangemi_2021a} \cite{Del_Santo_2013a} \cite{Jourdain2014} \cite{Jourdain2012} \\
    & &  \cite{Jourdain2012b}   \cite{Laurent2011} \cite{Lubinski2020a}  \cite{Malzac2006} \cite{Pottschmidt_2003a} \cite{Pottschmidt_2006a} \cite{rodrigue15_cyg}
\\
V404 Cyg            & 20 24 03.82 +33 52 01.90   &       \cite{Alfonso2018} \cite{Ferrigno2015} \cite{Jourdain2017} \cite{Kajava2018} \cite{Kuulkers2015a} \cite{Kuulkers2015b} \cite{Motta2017a} \cite{Motta2017b}  \\
&       &  \cite{Munoz-Darias2016} \cite{Natalucci2015} \cite{Rodi2017} \cite{Rodriguez2015} \cite{Roques2015} \cite{SanchezFernandez2017} \cite{Siegert2016}             \\
GX 339$-$4          & 17 02 49.38 -48 47 23.16   &  \cite{Belloni2006} \cite{Caballero-Garc'ia2009} \cite{CadolleBel2011}  \cite{DelSanto2009} \cite{DelSanto2008}    \cite{Joinet2007} \\
1E 1740.7$-$2942      & 17 43 54.83 -29 44 42.60   &   \cite{Bouchet2009} \cite{Bosch-ramon2006} \cite{Decesare2011} \cite{Delsanto2005} \cite{Delsanto2008p} \cite{Castro2014}    \\
\hline
IGR J17091$-$3624   &  	17 09 07.61	-36 24 25.70  & \cite{Capitanio2012} \cite{Kuulkers2003} \cite{Lutovinov2003} \cite{Rodriguez2011}\\
IGR J17098$-$3628   &  17 09 45.93	-36 27 57.30  &  \cite{Capitanio2009a} \cite{Chen2008} \cite{Grebenev2005a} \cite{Grebenev2005}    \cite{Prat2009a}\\
IGR J17177$-$3656   &  17 17 42.62 -36 56 04.5   &   \cite{Frankowski2011} \cite{Paizis2011a}\\
IGR J17464$-$3213   &  17 46 15.60	-32 14 00.86  &  \cite{Capitanio2009a} \cite{Capitanio2005}  \cite{Joinet2005} \cite{Parmar2003}   \cite{Revnivtsev2003}\\
IGR J17454$-$2919   &  17 45 27.68	-29 19 53.45  &  \cite{Chenevez2014b} \cite{Chenevez2014a} \cite{Paizis2015b} \cite{Paizis2015}\\
IGR J17497$-$2821   &  17 49 38.04	-28 21 17.50  & \cite{Kuulkers2006}   \cite{Soldi2006} \\
\hline
GRO J1655$-$40      & 16 54 00.14 -39 50 44.90  &  \cite{Caballero-Garcia2007} \cite{DiazTrigo2007} \cite{Joinet2005} \cite{Shaposhnikov2007} \\
GRS 1716$-$249      & 17 19 36.93 -25 01 03.43  &  \cite{Bassi2020}\\
GRS 1758$-$258      & 18 01 12.40 -25 44 36.10  &  \cite{Pottschmidt2006}  \cite{Pottschmidt2008}\\
GS 1354$-$64        & 13 58 09.70 -64 44 05.80  &  \cite{Pahari2017}\\
MAXI J1828$-$249    & 18 28 58.07 -25 01 45.88  & \citep{Filippova2014} \citep{Filippova2013}  \cite{Grebenev2016} \\
MAXI J1836$-$194    & 18 35 43.45 -19 19 10.48  & \cite{Grebenev2013} \\
Swift J1745$-$26    & 17 45 10.85 -26 24 12.60  & \cite{DelSanto2016} \cite{Kalemci2014}\\
Swift J1753.5$-$0127& 17 53 28.29 -01 27 06.22	& \cite{Cadolle2007}  \cite{Kajava2016} \cite{Rodi2015}\\
XTE J1550$-$564     & 15 50 58.70 -56 28 35.20  & \cite{Sturner2005} \\
XTE J1720$-$318     & 17 19 58.99 -31 45 01.11  & \cite{Cadolle2004} \\
XTE J1752$-$223     & 17 52 15.09 -22 20 32.36  & \cite{Chun2013} \\
XTE J1817$-$330     & 18 17 43.53 -33 01 07.80  &  \cite{Sala2007}\\
XTE J1818$-$245     & 18 18 24.43 -24 32 17.96  &  \cite{Cadolle2009}\\
MAXI J1348$-$630    & 13 48 12.79 -63 16 28.48  & \cite{Cangemi2019b} \cite{Cangemi2019a}  \cite{Lepingwell2019}\\
MAXI J1631$-$479    & 16 31 14.22 -47 48 23.44  & \cite{Fiocchi2020} \cite{Onori2019} \\
MAXI J1820+070      & 18 20 21.94 +07 11 07.19	& \cite{Roques2019} \\
\hline
SS433               & 19 11 49.56 +04 58 57.82  & \cite{Cherepashchuk2020}  \cite{Cherepashchuk2005} \cite{Cherepashchuk2013} \cite{Cherepashchuk2009} \cite{Cherepashchuk2003}   \\
Cygnus X$-$3        & 20 32 25.78 +40 57 27.90  & \cite{Beckmann2007} \cite{Cangemi2021} \cite{Hjalmarsdotter2008} \cite{Szostek2008}\\
    \hline
    \end{tabular}
    \caption{A list of the black hole \xray binaries in this work. }
    \label{tab:transient BHBable}
\end{table*}

In this section, we will briefly review the results that have been obtained thanks to \integral \ on BHBs. 
With no pretence of being exhaustive, we aim to give an overall view of the contribution of \integral to the field, focusing on the results that we consider most relevant. We will concentrate on a few sources that we judge particularly relevant because of their historic importance, as well as of their specific characteristics and behaviour, i.e., GRS 1915+105 (\cref{sec:GRS1915}), Cyg X-1 \cref{sec:CygX1}, V404 Cyg (\cref{sec:v404cyg}), GX 339$-$4 (\cref{sec:GX339}), 1E1740.7$-$2942 (\cref{sec:annihilator}). In  \cref{sec:INTdiscoveries}, we will go through the sources that were discovered by \integral, in \cref{sec:otherBHs}, we will review other BH systems that INTEGRAL observed over its 18 years of activity. Finally, we will briefly overview SS 433 and Cyg X-3, which may host a stellar-mass BH (\cref{sec:SS433} and \cref{sec:CygX3}). Table \ref{tab:transient BHBable} lists all the sources covered in this work, along with their coordinates, and the references to the most relevant works reporting findings obtained based on \integral data.


\subsection{GRS~1915+105}
\label{sec:GRS1915}

\subsubsection{A short resume of a very peculiar source}

\grs\ was discovered in 1992 with the WATCH instrument on-board \textit{Granat} \citep{castrotirado92}. It is considered a transient source, although it has always been active since its discovery. Contrary to most other sources, and until quite recently \citep[e.g.,][]{Negoro2018,Homan2019,Rodriguez2019, Motta2021}, \grs\ has always been found in a bright \xray state, and is most likely one of the brightest BHBs in our Galaxy. 

Infrared observations made with X-shooter allowed the orbital period to be refined to P$\sim$33.85 days \cite{Steeghs2013}, and to obtain a mass measurement of M$_{\mathrm{BH}}\sim$10.1~M$_\odot$ and M$_{\mathrm{Star}}\sim$0.47~M$_\odot$ for the BH and its K-M III companion star, respectively. 
A distance of $\sim$9 kpc was obtained by \citet{Reid2014} from VLBA radio parallax measurements, which allowed the mass to be re-evaluated and refined (M = 12.4$_{-1.8}^{+2.0}$~M$_\odot$), and the jet inclination and average speed to be accurately estimated - $\sim$60$^o$ and between 0.65c and 0.81c, respectively \cite{Reid2014}.  

\grs is amongst the few microquasars that do not behave like the vast majority of Galactic black hole systems (a-la GX~339$-$4)\footnote{See also V404 Cyg and Cyg X-3 as two more obvious 
'outliers', although likely for different reasons (Sec. \ref{sec:v404cyg} and \ref{sec:CygX3})}. Among its most obvious peculiarities: 
\begin{itemize} 
\item  it has been in outburst for almost $\sim$30 years, most likely because of its large accretion disc, the largest among all the known Galactic BHs, which provides an extremely large mass reservoir;
\item its outburst is characterised by extreme variability on seconds to years timescales (see Fig.~\ref{fig:1915long}), with \xray flares appearing randomly either isolated or in series, interspersed by longer phases of bright but steadier \xray emission (Fig.~\ref{fig:1915long});
\item it is highly variable at all wavelengths on short time-scales: in particular, on sub-second time scales shows both LF \& HFQPOs \citep[e.g.,][]{Morgan1997, Muno1999, rodriguez2002}, and on seconds-minutes time-scales it displays large oscillatory repetitive patterns (including the so-called heartbeats) \citep[e.g.,][]{Mirabel1998, Belloni2000, Kleinwolt2002}
\item it  does not clearly follow the common 'Q-shape' pattern in the HID, which is typical of the vast majority of BHBs (Sec. \ref{sec:states}), but shows transitions between three accretion states in many ways similar to the canonical ones \cite{Belloni2000} (see below). 
\item besides major relativistic discrete jets and the LHS compact jet, which are also observed in other BHBs, it also shows another type of jet, characterised by repeated small ejections in fast succession (tens of minutes, with fluxes of the order 30-100mJy \citep[e.g.,][]{Pooley1997}).
 \end{itemize}

  
\subsubsection{The largest display of variability patterns}

The \rxte\ satellite monitored GRS 1915+105 intensively over its 16-years long life-time, generating one of the largest soft \xrays dataset at high time resolution for this system. Based on $\approx$8 years of data, \citet{Belloni2000} identified 12 distinct variability patterns in the 1-s resolution \xray light curve of GRS~1915+105. Such variability classes repeat randomly in time, tracing specific patterns in a colour-colour diagram\footnote{A colour-colour diagram is similar to the HID, but the intensity is replaced by a second hardness ratio (or colour), obtained from intensities measured in two extra energy bands.}, and are believed to reflect transitions among three accretion states, labelled  A, B, and C. Such states are roughly equivalent to - but not coincident with - the HSS (state A), a bright SIMS (state B), and a bright HIMS (state C), with which they share common fast temporal properties, in particular in terms of the presence and behaviour of QPOs \citep[e.g][]{Muno1999,Muno2001,rodriguez2002,rodrigue04_1915,rodriguez2008}. 

Two additional variability patterns were added to the original 12 a few years after \citet{Belloni2000}, 
one thanks to the longer, quasi-uninterrupted observation possible 
with \integral \cite{hannikainen05}. 
All the variability patterns have been named using Greek letters, and are widely used to describe essentially any new \xray dataset obtained on GRS 1915+105. The 2--20 keV \xray spectral analysis also indicated that the transitions between states could be due to rapid emptying/refilling of the inner disc \citep{Belloni1997a}, which evolves on either the viscous time-scales, or on faster/(quasi-)dynamical time-scales \cite{Eardley1975}. 
With the exception of IGR J17091$-$3624 (see Sec. \ref{sec:17091}), GRS 1915+105 is the only system showing such a large number of variability patterns. No Galactic BHBs continuously shows the variability typical of GRS 1915+105, although a few other sources have shown at times a handful of classes. Such an example is the NS XRB  MXB\,1730$-$335, also known as the Rapid Burster \cite{Bagnoli2015}, which has been recently shown to display two variability classes observed in GRS\,1915+105 (namely, the $\rho$ and $\theta$ class). A second example is the ULX system 4XMM J111816.0$-$324910 in the galaxy NGC 3621, which might host a stellar mass BH or a NS, and has also clearly shown the heartbeats typical of GRS 1915+105 \cite{Motta2020}. 


\subsubsection{A lab for studying the accretion-ejection connections}

While GRS~1915+105 is in many ways unique, it also shares several properties with more conventional 
transient BHBs (e.g GX 339$-$4, \cite{Done2004,Soleri2008}), and also with some AGN (and in particular some quasars,  \cite{Marscher2002,Chatterjee2009,Punsly2013}), which are thought to display mass-scaled versions of GRS~1915+105's correlated \xray and radio behaviour close to jet ejection events. Therefore, comprehending the coupling between inflow and outflow in GRS~1915+105 is important for the understanding of XRB systems, but has also a broader relevance for studies of AGN, and potentially of all the jetted systems powered by accretion on compact objects \citep{Fender2004a}.

\grs\ is considered the archetypal galactic source of relativistic jets, and as such has been for many years the source of choice for the study of the physics of jets. Radio observations performed early after its discovery revealed the presence of large scale radio jets making \grs\ the second microquasar (see Sec. \ref{sec:history}) after 1E 1740.7$-$2942 \cite{Mirabel1992}. Such radio jets were seen to evolve on the plane of the sky with an apparent superluminal motion \citep{Mirabel1994}, and - as it was clarified later - correspond to the transient relativistic ejections observed also in other BHBs during the LHS to HSS transitions. \grs\ also shows a rather steady radio emission, associated with harder and less variable \xray emission, which has been ascribed to the presence of a compact jet of smaller extension \cite{Dhawan2000}. 
\grs\ and the high-mass BHB Cyg X-1 (Sec. \ref{sec:CygX1}) are the only sources where compact radio jets have been resolved with interferometric radio observations performed using Very Long Baseline Interferometry (VLBI) \citep{Dhawan2000, Stirling2001, Fuchs2003}. 

In addition to the radio emission associated with transient and compact, steady jets, \grs\ also shows smaller amplitude radio and infrared oscillations \citep{Pooley1997,Fender1998, Mirabel1998}, which simultaneous multi-wavelength observations in radio, IR, and \xray observations have associated with \xray oscillations on the same time-scales \citep{Eikenberry1998,Mirabel1998}. The radio/IR oscillations were interpreted as the signature of small discrete ejections of material from the system occurring on a rapid succession, possibly associated with certain disc instabilities developing in the accretion flow. This particular radio/IR and \xray correlated variability made  \grs\ the first and still in many ways the best source where the disc-jet coupling can be probed in detail. 
For this reason, \grs\ has been the target of numerous multi-wavelength observing programs, many involving \integral, over the past 19 years, and is also being currently monitored regularly, as simultanously as possible, by \integral, Insight-HXMT and Nicer in the \xrays, the Sardinia and Medicina radio telescopes in Italy, and AMI-LA in the UK\footnote{We note that monitoring observations are not restricted to these facilities, but we mention here only those that are in tentative simultaneity with the long-term \integral monitoring}.

The main questions at the core of most of the observing programs that have targeted \grs\ over the years revolve around the interplay that exists between the inner accretion flow and the generation of various types of outflows, which not only includes the jets observed in radio, but also the disc winds seen during the soft(-ish) states \cite{Neilsen2009}. The advent of \integral allowed the spectral window to be extended to energies above 20 keV, and thus to study the behaviour of the Comptonised emission, probe the presence of an additional soft $\gamma$-ray tail, and explore the  potential connections of such spectral components end their properties with the presence of jets \citep[e.g.,][]{Fuchs2003, rodrigue08_1915a,Droulans2009}.

\subsubsection{18 years of \integral observations: long-term light curve and selected results}

\begin{figure*}
    \centering
    \includegraphics[width=\textwidth]{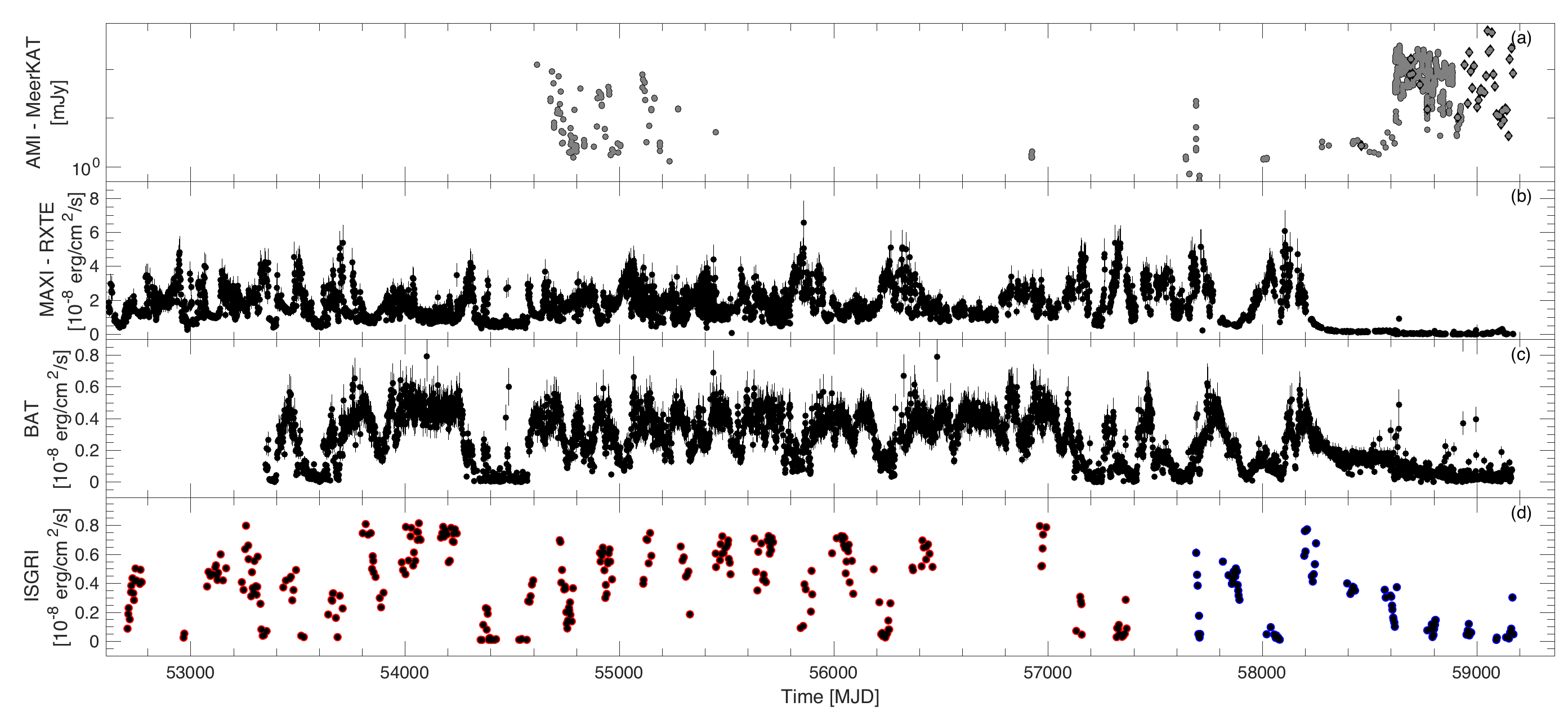}
    \caption{Long term light curves from GRS 1915+105. From the top: AMI-LA (circles) and MeerKAT (diamonds) data (panel a), MAXI data (panel b) covering the 4--20 keV energy band, BAT data (panel c) covering the 15--50 keV energy band, IBIS/ISGRI data (panel d), covering the 30--50 keV energy band. In panel (d), red- and blue-edged points mark data reduced using OSA 10.2 and OSA 11.1, respectively. Each point in panel (d) comes from one \integral revolution ($\approx$300~hr). 
    In July 2018 (~ MJD 58300) GRS 1915+105 started a slow decline towards an \xray plateau, which still continues at the time of writing. A marked change in the radio behaviour of the source is visible around MJD 58600. Figure adapted from \citet{Motta2021}.}
    \label{fig:1915long}
\end{figure*}

Figure~\ref{fig:1915long} shows the \grs's \rxte/\asm, MAXI and \integral/IBIS/ISGRI 30--80 keV light curves from 2002, October 17$^{th}$ (\integral's launch) to 2020, December
31$^{st}$. The ISGRI light curve was obtained from all the available pointings between revolution 48 (MJD 52704.523, or 2003 March, 6$^{th}$) and
2298 (MJD 59170.195 or 2020, Nov. 17$^{th}$), where we selected data with an ISGRI 
exposure greater than 500~s. We further restrained to data where \grs\ was observed with an off-axis angle of less than 10$^o$. This results in a total of 7609 good pointings.  

\begin{figure*}
    \centering
    \includegraphics[width=18cm]{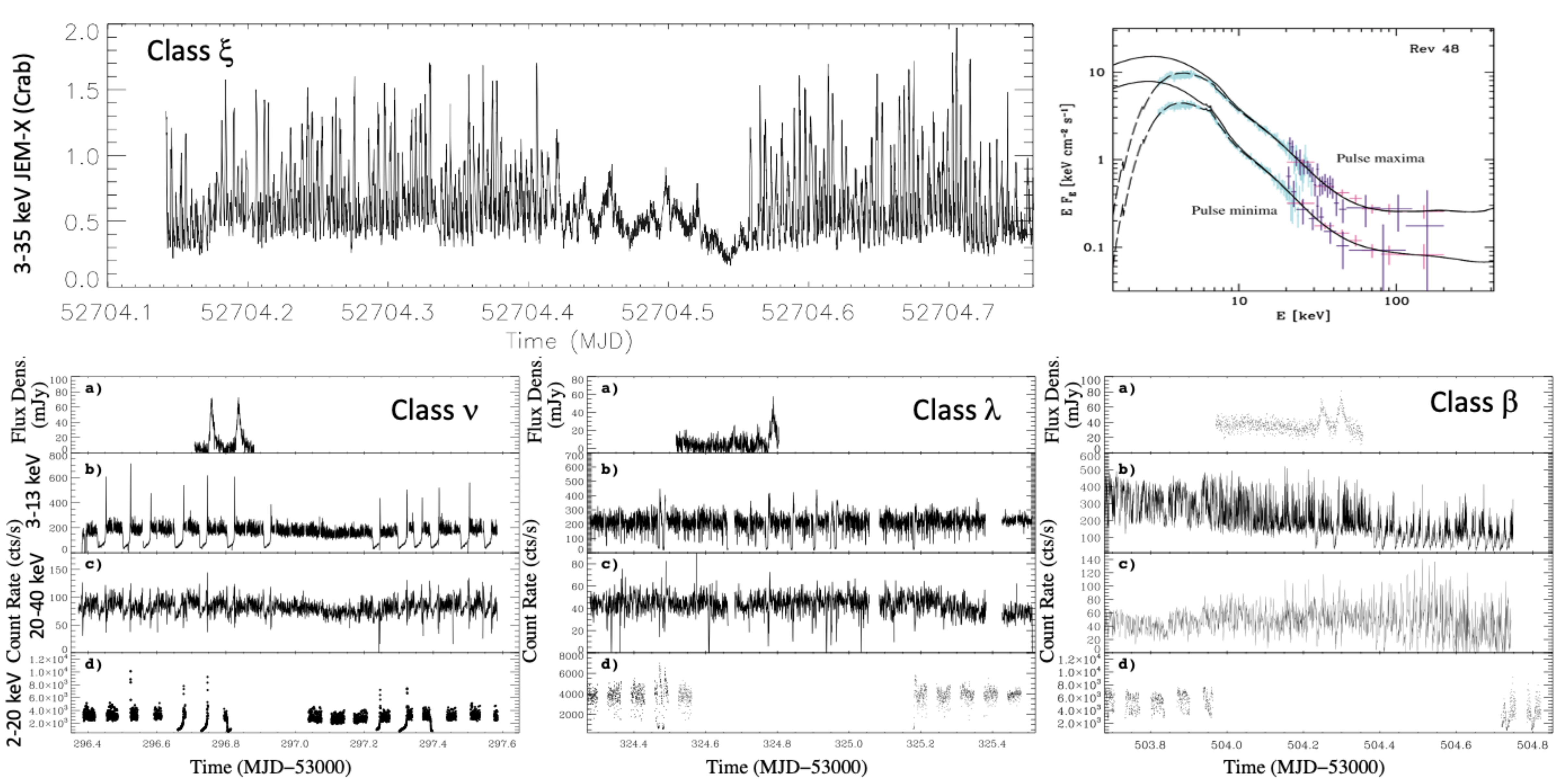}
    \caption{The upper plots show the JEM-X 3--35 keV light curve of GRS 1915+105 obtained during the first half of the rev. 48 observation (left), and  the spectra obtained from both the peak and at the minima in between flares in this observation obtained combining JEM-X, ISGRI and SPI data (right). Adapted from \citet{hannikainen05}. The lower plots represent the radio (panels (a)), \xray (panels (b) for JEM-X and (d) for \rxte/pca), and hard \xray (IBIS/ISGRI, panels (c)) 
    light curves of 3 variability classes showing radio ejections. Adapted from \citet{rodrigue08_1915a}. 
    } 
    \label{fig:1915classes}
\end{figure*}

The very first dedicated \integral observation made during revolution 48 was reported by \cite{hannikainen03}. Analysis of these data allowed a 'new' source, IGR~J19140$+$0951, to be discovered. 
The JEM-X, IBIS/ISGRI, and \rxte/PCA light curves permitted to identify a new variability class to be added to the 12 already defined by \citet{Belloni2000}. This new class, labelled $\xi$ \citep{hannikainen05}, is characterised by a flaring behaviour (Fig. ~\ref{fig:1915classes}) composed of a main, relatively hard pulse, followed by smaller amplitude softer ones, recurring on a typical time-scale of about five minutes.  The JEM-X, ISGRI, and SPI spectra obtained at the flare peaks and in correspondence with the minima between consecutive flares are well fitted by a model consisting of a bright disc and hybrid thermal/non-thermal Comptonisation above $\sim$ 10 keV, compatible with the source being  in a soft state (state A) during class $\xi$. Unlike other classes, where flares sign transitions between states A-B-C, the flares in class $\xi$ are purely isothermal and the overall variability is attributed  to limit-cycle oscillations, possibly resulting from delayed feedback in the inner accretion disc. 

\citet{Fuchs2003} reported the very first simultaneous multi-wavelength observations of \grs, which happened during revolution 57 (MJD 52731, 2003 Apr. 2$^{nd}$). Less than a month after the discovery of class $\xi$, the source was found in a 'steady' state, associated with class $\chi$ (state C). Benefiting from one of the largest frequency coverage 
ever obtained for this system, \citet{Fuchs2003} could observe (i) a strong steady optically thick radio emission, resolved with the VLBA as a strong compact jet; (ii) bright near-IR emission (probably associated with the jet); (iii) a strong QPO at 2.5 Hz in the \xrays; (iv) a power law spectrum without any cutoff in the 3–-400 keV range. 
This was the first time that all these 'hard'-state properties were observed simultaneously, thus reinforcing the suspected links among the hard accretion state, the presence of a compact jet as well 
as of a high-energy tail, and \xray QPOs.

The detection of hard \xray/soft $\gamma$-ray tails above $\sim$100 keV was studied by \citet{Droulans2009} with SPI. The analysis of 20 observations (19 of which have exposures larger than $\gtrsim$45~ks) with data covering the 20--500 keV energy range clearly shows the need for an additional power law component beside the standard thermal Comptonisation used to represent the $\sim$20--100 keV spectra. An hybrid Comptonisation model also adequately represents the spectra, and \citet{Droulans2009} argue that the 20--500 keV emission in \grs\ might be due to a combination of thermal and non-thermal Comptonisation. The comparison between the emission in the 1-12 keV (\rxte/ASM) energy range, and the 20-50 keV energy range (SPI) showed that the emission in the two bands vary in a correlated fashion. This property led the authors to conclude that the emission in these two bands arises from the same spectral component, but has no link with the hard spectral tail above 50~keV, which is rather stable and does not follow the variations of the other two components. 
The X-ray variability classes in \grs most likely reflect changes within the accretion flow (e.g., accretion rate changes, variations of the viscosity). Instead, the  emission above 10 keV reflects the properties of the electron population(s). The fact that the parameters of the population of thermalised electrons may vary, but not in a correlated fashion with the variability pattern observed in the light curves of \grs, suggest that the macroscopic properties of the Comptonising electrons evolve independently from those of the (inner) accretion flow.

\subsubsection{An \xray signal precursor to small ejections: is the corona the source of ejected material?} 

The possibility to follow \grs on long continuous observing window with \integral allowed observers to nicely monitor the variability patterns for long periods of time \cite[e.g.][]{hannikainen03}, thus accumulating data during several distinct accretion states \cite{hannikainen05, Droulans2009}. The long \xray observations possible with \integral also allowed simultaneous multi-band observing campaigns to be organised more easily, which resulted in a better understanding of the source \cite[e.g.,][]{Fuchs2003}. With the aim to study the interplay and causal relations between the main emission components in \grs (jets, disc, corona), \citet{rodrigue08_1915a} presented the results of a long term radio to X/$\gamma$-ray monitoring performed with the Ryle telescope, \rxte, and \integral. From the 11 \integral observations, these authors observed the source in 9 different variability classes. Radio observations were made during of 8 of these, and in particular allowed the intervals during which the source was displaying the variability classes $\lambda$, $\nu$, and $\beta$ to be covered. 

\citet{rodrigue08_1915a} adopted a model-independent approach, based on the inspection of light curves and colour diagrams, which allowed to witness transitions between specific multi-wavelength states on short ($\sim$100~s) time-scales (Fig. \ref{fig:1915classes}). 
This allowed \citet{rodrigue08_1915a} to discover small amplitude ejections in a new class ($\lambda$) correlated to a specific hard-dip/soft flare X-ray behaviour. This is reminiscent of what is seen during classes $\beta$, $\nu$ (confirmed in this same study), and $\alpha$.  The authors conclude that, in \grs,  small amplitude ejections systematically occur in response to a phase of a hard \xray dip ended by a short, intense, and soft \xray spike (bottom of Fig.~\ref{fig:1915classes}). This proposition is reinforced by the potential correlation between the amplitude of the radio flare and the length of the preceding hard dip. These results were later confirmed by the analysis of the entire sample of simultaneous  \rxte/Ryle observations \cite{Prat2010}.

\citet{rodriguez2008} presented both the spectral and temporal analysis of the \integral and \rxte\ data of these 4 different variability classes. In the case of the three 'cyclic' classes, they performed a state-resolved JEM-X/ISGRI spectral
analysis. For each class, all cycles were separated into 3 ($\beta$), 4 ($\nu$), or 6 ($\lambda$) spectra based on their count rates and hardness. The resultant phase-dependent spectra were modelled with a common model consisting of an absorbed thermal/disc emission and a Comptonised component. In all cases, the spectral analysis shows a comparable behaviour: the inner disc radius and temperature increase from the hard dip to the soft \xray flares, with a relative disc contribution to the 3--200 keV flux that increases monotonically over the sequence. Meanwhile the contribution of the Compton 'corona' first increases from the bottom of the dip to the rising part of the first \xray spike, and then shrinks at the spike's maximum. It should be noted that this spike is not necessarily the main obvious flare in this sequence, but is the one that is easily seen 'half' way through the dip in class $\beta$. It is interesting to note that it is also at this moment that the LFQPO observed up to this point disappears (as clearly visible in classes $\beta$ and $\nu$). The main conclusion here is that the soft spike is the true moment of the ejection (later seen in radio), and that the corona is the source of the ejected material. The same conclusion was drawn by \cite{Rodriguez2003} in the case of XTE J1550$-$564, and later discussed for other sources as well \cite{Rodrigue09_1915wkshop}. 
 
The spectral analysis of class $\chi$ revealed the presence of an additional power law component beside the disc and thermal Comptonisation component. While the low energy components (disc and thermal corona) show little variations over the course of this observation, the high-energy power law varies in apparent correlation with the radio flux. This led to the suggestion that this 'hard-tail' was somehow related to the jet emission, and thus was akin to that observed in Cyg X-1 and a handful of other sources.

\subsubsection{The recent low luminosity state: when \grs\ finally behaves like other Galactic black holes (for a while)}

On July 2018, based on MAXI/GSC data (Fig.~\ref{fig:1915long}), \citet{Negoro2018} reported the faintest \xray state observed since 1996 (i.e., the start of daily monitoring of the source with \rxte/ASM) from \grs. The observations analysed were taken between 2018 July 7 and 8 (MJDs 58306-58307), and follow a nearly exponential $\sim$100 days decrease \cite{Negoro2018}. Also the radio flux from the source appeared to be relatively low (even though not significantly so) compared with the historical data. Unfortunately, at the time of the main decline observations with \integral \ were not possible. 

On 2019 May \citet{Homan2019} and \citet{miller2019} reported a sudden, further \xray dimming as observed by  Chandra and Nicer. An active Swift+\integral monitoring of the source allowed \citet{Rodriguez2019} to show that the source was in a state very similar to the a canonical BH LHS, i.e. significantly harder and fainter that the typical C state (during class $\chi$ in particular), typically observed in \grs. 
Unlike the usual anti-correlated soft-hard \xray behaviour, a significant hard \xray   
dimming was also reported from the 2019, May, 13$^{th}$ (MJD 58616) \swift+\integral observation, 
leading to the suggestion that the source was maybe slowly decaying to quiescence. This hypothesis was refuted quickly by \citet{iwakari2019}, who observed a flare with both Maxi and \swift/BAT. 
Since then, as can be seen in the long term light curve in \cref{fig:1915long}, although the source is indeed in a long, slightly decreasing, low flux state, it still appears to have quite regular X-ray flares, and an extremely marked radio flaring, in all qualitatively similar to what has been observed over the many years of the outburst of \grs\ \cite{Motta2019_1915,Motta2021}.
This recent period certainly attracted great interest in this source as testified by the number of
Astronomer's Telegrams reporting on the changing phenomenology observed from the source (and numerous papers published over the following year), including brief transitions to faint states, sudden flares, and brief episodes during which some of the several known variability classes of \grs manifested. 
The origin of this possibly new state remained unclear for some time, although reports of obscuration episodes interleaved by short and intense flares started to highlight significant similarities between \grs and systems displaying phases of strong, variable local absorption: all such systems tend to show high-amplitude flares. Some noteworthy examples are V4641 Sgr \citep{Revnivtsev2002}, Swift J1858.6$-$0814 \citep{Hare2020,Munoz-Darias2020}, and even some Seyfert II galaxies \citep{Moran2001}, and - perhaps the clearest and most recent example - V404 Cyg. 

Radio observations of \grs during the long \xray plateau finally solved the conundrum: while during approximately half of the X-ray plateau \grs was remarkably quiet in radio, after a month-long multi-band flaring period, its radio activity resumed and continued after the multi-band flaring ceased. In this phase, absorption was a constant characteristic of the X-ray energy spectra of GRS~1915+105 \cite{Koljonen2020,Miller2020,Neilsen2020,Balakrishnan2021,Koljonen2021}, indicating the presence of an obscuring in-homogeneous absorbing medium located close to the source. The radio emission, instead, was characterised by marked flaring activity qualitatively very similar to that observed in the past. This provided evidence for two important facts: first, that the correlation between radio and X-ray emission (which for many years characterised GRS~1915+105 \cite{Fender2004a}) ceased to exist during the latest evolution of \grs; and second, that the X-ray behaviour alone---in the absence of such radio-\xray correlation---is very misleading, as it offers only a partial view of the current state of the source. It thus appears that GRS~1915+105 has not entered quiescence yet, and might not be approaching it either (but see \cite{Neilsen2020} and \cite{Koljonen2021}). Continued observations of this intriguing source will show what path the outburst evolution  will take this time.  

Based on the results obtained with the dense monitoring of this unusual phase in the life of this iconic \xray binary, it seems now clear that the decline preceding the long-lived X-ray plateau observed starting from mid-2018 was a transition to a truly dim state, very similar to the canonical LHS typical of many transient BHBs
\cite{Rodriguez2019,Motta2021}.  Such a state, characterised by low X-ray and radio flux, as well as low long-term variability in both bands, is certainly unusual for GRS~1915+105, which has never been observed before in a low-luminosity canonical LHS \citep[but see][]{Gallo2003}. 
The second half of the long \xray plateau, following the multi-band aforementioned flaring period, despite having been dubbed `the obscured state', it is not really a state, but rather a \textit{condition} directly dependent on the presence of local absorption along our the line of sight \cite{Motta2021}. 

The accretion processes that must be feeding the markedly variable jet observed in radio must be happening beneath a complex layer of material local to the source, which shields the inner portion of the accretion flow, and thus blocks a large fraction of the X-ray emission. Behind this Compton-thick curtain of material, GRS~1915+105 is most likely evolving through various states and transitions (as shown by \citet{Motta2021}), consistent with what it had been doing for 25 years until June 2018. 
This means that, perhaps, GRS~1915+105 did not really enter the outburst phase in 1992, when it was discovered, but simply emerged from an obscured phase similar to the one we are witnessing at the time of the observations presented in \citet{Motta2021}, as suggested by \citet{Miller2020}.

As to what is the reason behind the change in the behaviour of \grs, as in many other ways this systems remains a mystery, despite having been scrutinised with essentially any instrument that could ever observe it, from the radio radio to the $\gamma$-rays. A change in the accretion flow must have happened to trigger such a remarkable change in the emission from the source, but whether this was due to a change in the companion star, or in the properties of the accretion flow itself remains unclear. Certainly \grs remains one of the most important BHBs we know, and a lot has to be learned still from this system. \integral, which has greatly contributed to the understanding of \grs, will be crucial to the continuation of the studies of this system, especially if the obscured phase will continue for much longer. Instruments sensitive to the lower energies will not be able to monitor the activity of this source (currently best studied in the hard \xrays), which are much less affected by absorption, and are easily within the reach of \integral.

\subsection{Cygnus X-1}
\label{sec:CygX1}

\subsubsection{A key source for understanding black hole XRBs}

Discovered during a rocket flight in 1964 \cite{bowyer:65a}, and identified to be a massive source in orbit with the O9.7~Iab star HDE226868 \cite{Bolton1972,Webster1972}, Cyg X-1 was the first BH candidate (see \citet{oda:1977a} for a review of the early observations). As also discussed in Sect.~\ref{sec:history}, Cyg X-1 is one of only a handful of persistent stellar-mass BHs in our Galaxy. The newest VLBI parallax measurements place the source at a distance of $2.22^{+0.18}_{-0.17}$\,kpc, and yield a mass of the BH of $21.2\pm2.2 M_\odot$ \citep{Miller-Jones2021}.

\begin{figure*}\centering
  \includegraphics[width=0.95\textwidth]{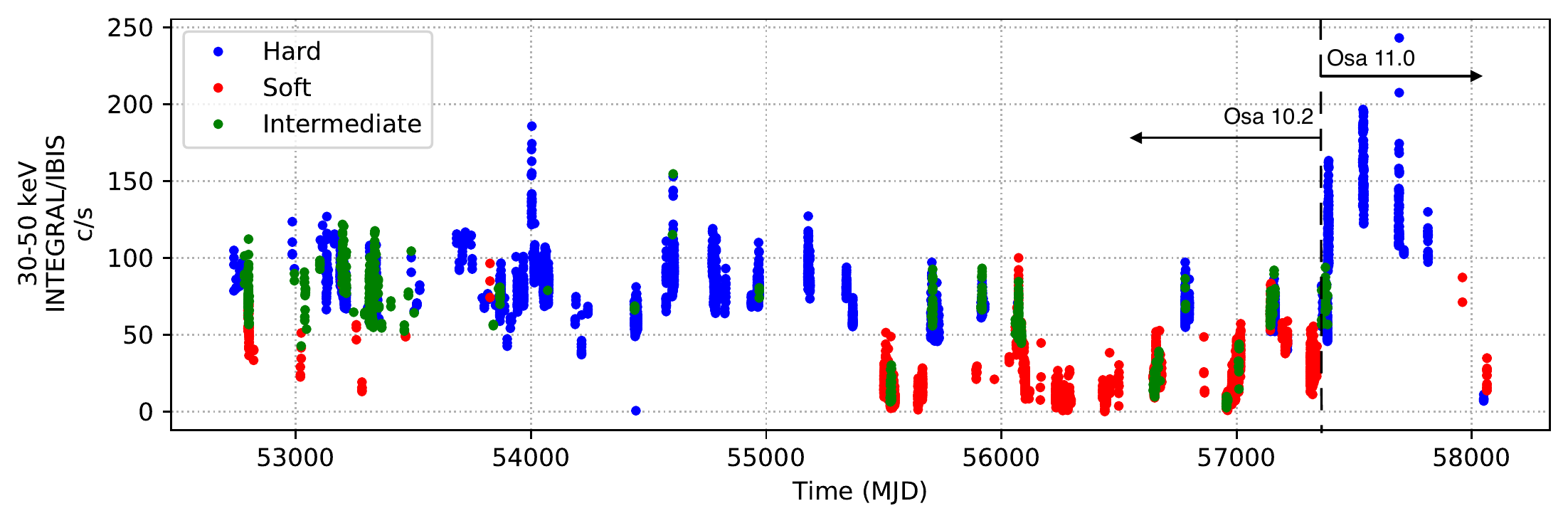}
  \caption{IBIS/ISGRI long-term light curve of Cyg X-1 since 2004
    \citep{Cangemi_2021a}. Colours indicate the state of the source
    (blue: hard, green: intermediate, red: soft). The dashed lines
    separate the time intervals included in the earlier analysis
    \citep{Rodriguez2015} and the data from newer
    programmes. Figure taken from \citet{Cangemi_2021a}. 
    }\label{fig:cygx1_longint}
\end{figure*}

With a 5.6\,d long, circular orbit and an orbital separation of
$40\,R_\odot$ between the two binary members, the BH is
close enough to the donor that the stellar wind is focused towards it
\cite{gies:1986}. Therefore, accretion occurs mainly via a wind, although the presence of a small accretion disc dampens the short term fluctuations typical of pure wind accretors. The result is
that Cyg X-1 spends a significant fraction of the time in the LHS. In fact, Cyg X-1 has been considered for long the prototype of the
BHB in the LHS, although excursions into the HSS are possible and have been often observed. In particular, 
the source has spent more time in the HSS in the last decade
\citep{Grinberg_2013a,Cangemi_2021a}. As the BH is deeply
embedded in the stellar wind, strong orbital
variability is seen in the soft \xrays; there are both an overall orbital modulation due to line of sight
effects through the wind \cite{Poutanen2008,Grinberg2015} as well as short-term
absorption dips caused by density fluctuations in the stellar wind
and/or the accretion flow passing through our line of sight
\cite[e.g.,][]{li1974,balu2000,Miller_2005a,Miskovicova_2016a,Hirsch2019}.
In addition, a strong broadened Fe K$\alpha$ fluorescence line has been
observed, which might suggest that the BH has a high to extreme intrinsic angular momentum
\cite{Fabian1989,Fabian2012,Tomsick2014,parker2015,duro2016,zhao2021}.

Cyg X-1 has been subjected to deep observations with virtually any available
\xray missions, past and present. A comprehensive review of all of its properties is
therefore not possible nor appropriate here. We will instead
concentrate on the areas where \integral contributed the most,
including the studies of the broad band \xray spectrum and of the radio
through \xray and $\gamma$-ray spectral energy distribution, the hard
\xray variability, and the question of the soft $\gamma$-ray tails and its
polarisation.

\subsubsection{Spectral behaviour and short-term variability}

In the hard state, the \xray spectrum of Cyg X-1 can be well
described by a Comptonisation continuum (see \cite{Pottschmidt_2003a,Del_Santo_2013a} for studies using
\integral), together with a Compton reflection hump \citep{duro2016}. Phenomenologically speaking this corresponds to a power law continuum with an exponential cutoff. Above a few
hundreds keV, deviations from this continuum become apparent, indicating the presence of a soft $\gamma$-ray tail. We will discuss this spectral component and its
polarisation -- which was first measured with \integral -- in
Sect.~\ref{sec:cygx1tail} below.

In order to study the broad-band long-term variability of the source, \integral has performed an extensive monitoring of the source since launch, accumulating more than 15\,Ms of exposure on this system. Figure~\ref{fig:cygx1_longint} shows the long-term light curve of the source obtained from this campaign \citep{Cangemi_2021a}, also encoding the source state, determined following the prescription of \citet{Grinberg_2013a}. These observations are part of an unique long-term observing campaign with multiple instruments that started in the late 1990s. The aim of these observations was to study the evolution of the broad-band spectral energy distribution from 1\,keV to ${\sim}$2\,MeV, the short-term \xray timing behaviour, and the relationship between the \xrays and the radio emission of the source. The campaign consisted of \integral \ key programme observations, bi-weekly 5--10\,ksec \rxte pointings, and dedicated simultaneous observing campaigns with Chandra, Suzaku, \textsl{NuSTAR} and \xmm, which have been accompanied by extended monitoring observations in the radio (AMI-LA, VLA) and in the optical bands
\citep{pottschmidt:00b,pottschmidt:03a,wilms2006,wilms:07a,Cadolle2006,Malzac2006,hanke:09a,duro:11a,Boeck_2011a,Del_Santo_2013a,Tomsick2014,duro2016,miller:12a,Grinberg_2013a,Grinberg2014,Grinberg2015,rodrigue15_cyg,Miskovicova_2016a,Tomsick2018,Hirsch2019,Lubinski2020a,Cangemi_2021a}.

The main emphasis of the spectroscopic analysis of the campaign concentrates on the statistics and interpretation of the continuum emission of the source, as other missions are better suited for studies of the soft \xray spectrum and the relativistic reflection. Statistical analysis of the hard and intermediate states of Cyg X-1 showed that the source behaviour on these states is more complex than previously assumed.  
For example, \citet{Lubinski2020a} used IBIS data in the 22--100\,keV range to constrain the correlations between hard \xray flux, spectral shape expressed by the photon index, and variability properties. The \xray flux and photon index show a multi-mode behaviour, with multiple clusters of typical parameter configurations, corresponding to different flavours of the states, especially during the LHS. These clusters show different \xray and radio variability properties and are possibly related to different accretion geometries. Similar indications of different hard state flavours had  been seen by \citet{Pottschmidt_2006a} in the short-term variability behaviour of Cyg X-1 in the $\sim$3--20\,keV energy range of \rxte\ data, and by \citet{Jourdain2012b} in the 20-200 keV energy band in the \integral SPI data. For what concerns the intermediate state, \integral \ observations also revealed similar bi-modal spectral variability \cite{Malzac2006}.

\integral also offers an unique opportunity to extend short-term variability (``timing'') studies to the hard \xrays, and to probe the short-term variability of the hard \xray emission in different spectral states. \citet{Pottschmidt_2003a} were able to obtain 15--70\,keV PDS using ISGRI, focusing on a flaring episode, while \citet{Cabanac_2011a} used SPI up to 130\,keV, concentrating on long-term averaged PDS for the LHS and the HSS. The power spectral shapes are generally consistent with those seen at lower energies with \rxte\ \citep[][and references therein]{Pottschmidt_2006a,Grinberg2015} and can be described by a sum of several Lorentzian components which show a characteristic dependency on the spectral hardness of the source \citep{Grinberg2014}. These studies showed that Cyg X-1 varies significantly up to at least 130\,keV.

\subsubsection{SED, disc-jet connection, and  polarisation}\label{sec:cygx1tail}

The major impact that \integral has had on our understanding of Cyg X-1 has been on the behaviour of its hard tail and its polarisation (see Sect.~\ref{sec:high-energy} for a general overview of the hard tail). The elusive soft $\gamma$-ray tails, i.e., excesses in the hard band above the rollover of the Comptonisation component, have first been detected in Cyg X-1 with \textit{CGRO}/COMPTEL \cite{McConnell_2000a,McConnell2002}. \integral was crucial in confirming the detection of the tail with SPI \citep{Bouchet2003,Cadolle2006,Jourdain2012} and IBIS \citep{Laurent2011,Rodriguez2015},  and in studying its possible dependency on the spectral state \citep{Walter_2017a}. \citet{Cangemi_2021a} reported the analysis of the full data set of Cyg X-1, in which the presence of a hard tail component also in the average soft state spectrum was confirmed.

The modelling of the broad band radio to MeV and even TeV spectrum of Cyg X-1 results in two equally plausible explanations for its hard emission. One explanation is that the hard tail is due to a non-thermal particle population in the Comptonising medium \citep[e.g.,][and references therein]{Romero_2014a,Chauvin2018}, while an alternative  explanation is that the hard tail is formed by synchrotron emission from a jet
\citep[e.g.,][]{Malzac2006,Del_Santo_2013a,Pepe_2015a,Kantzas2021}. The latter explanation, as already mentioned above, would provide a direct connection between the hard \xray and $\gamma$-ray emission and the jets observed in radio (see also Sect.~\ref{sec:discjet}).

Cyg~X-1 is an ideal source for tackling this question, since it is essentially always located in the region of the HID sometimes referred to as the ``jet-line''\cite{Belloni2011},  where transitions from the hard states to the soft states result in energetic, relativistic ejection events, and subsequent jet quenching.  For this reason, the coupling between the disc and the jet can be studied exceptionally well. The only other Galactic quasi-persistent hard state sources, 1E~1740.7$-$2942 (see Sec.~\ref{sec:annihilator}) and GRS~1758$-$258, both show $<$5--10\% of the typical \xray flux of \mbox{Cyg X-1}. This makes their observation more challenging, since it is much harder than in Cyg X-1 to accumulate enough photons to study the emission above $\sim$300\,keV. The other bright quasi-persistent BH, GRS 1915+105, is (almost uniquely) operating in a very different accretion regime, and thus is not particularly suited for polarisation studies (see Sec.~\ref{sec:GRS1915}).

\begin{figure}
\includegraphics[height=9cm, angle=90]{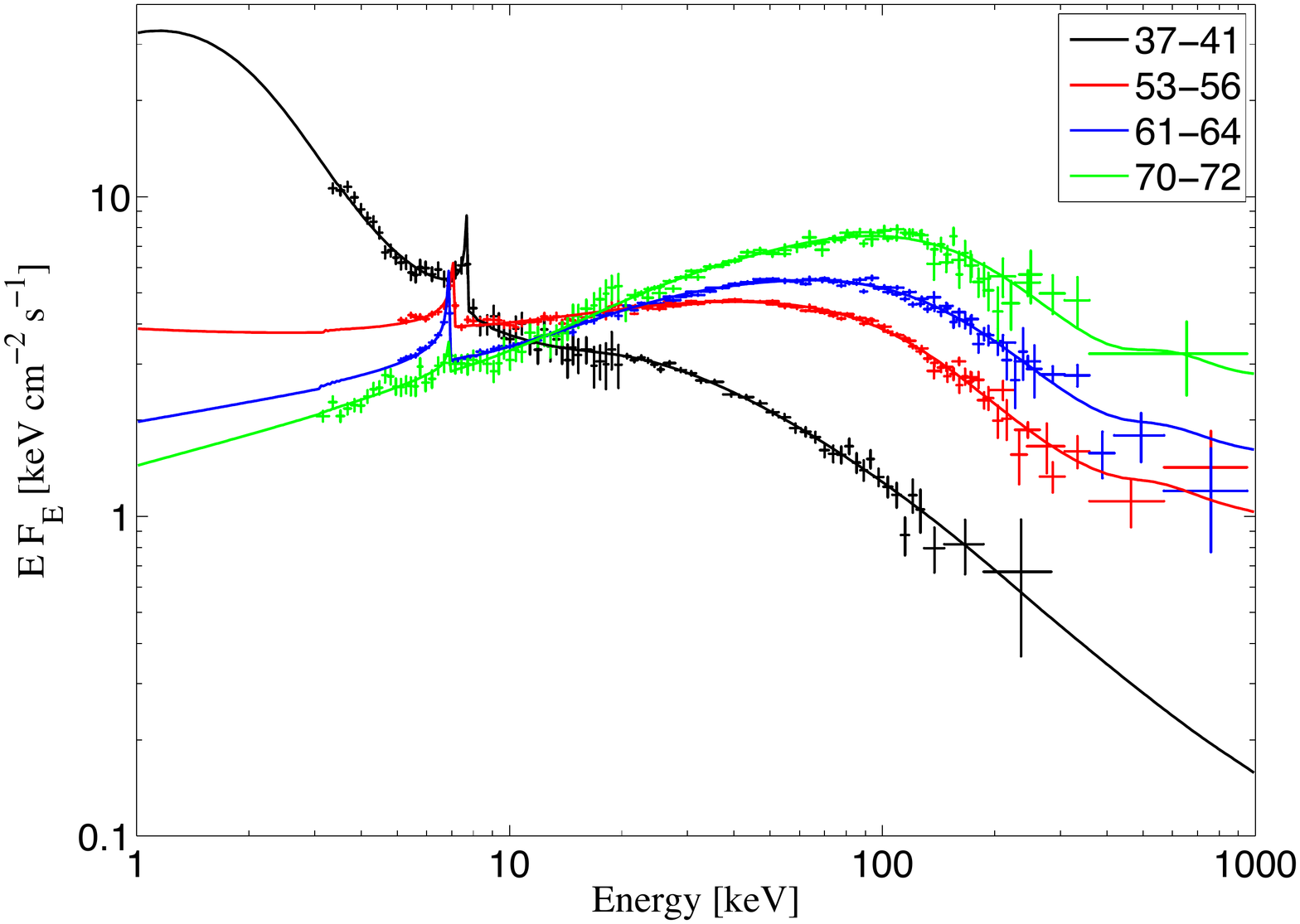}
\caption{Joint JEM-X, IBIS and SPI  energy spectra of Cyg X-1 during four different spectral states
fitted with the {\sc {belm}} model with pure non-thermal acceleration, plus {\sc {diskline}} and {\sc {reflect}} (figure taken from \citet{Del_Santo_2013a}).
}\label{fig:belm_fit}
\end{figure}

\citet{Del_Santo_2013a}  analysed 6\,years of \integral observations of Cyg X-1 using two hybrid thermal/non-thermal Comptonisation models. Besides the largely used {\sc{eqpair}}, for the first time these authors performed a spectral fits with the {\sc{belm}} model (Fig. \ref{fig:belm_fit}) which takes into account the effects of the magnetic field in the corona of BHBs \cite{Malzac2009}.  They found that while in the soft states the emission is dominated by Comptonisation of the disc photons, while in the hard states, the data are consistent with a pure synchrotron self-Compton model, which means that the synchrotron photons could be the only source of seed photons for the Comptonisation process.
Finally, \citet{Del_Santo_2013a} provided important constraints on the intensity of the magnetic field in the X-ray corona, which are in agreement with previous analytic results \cite{Wardzinski2001}. 

\begin{figure}
\includegraphics[scale=0.45, angle=-90, trim=1.cm 0cm 7cm 0cm]{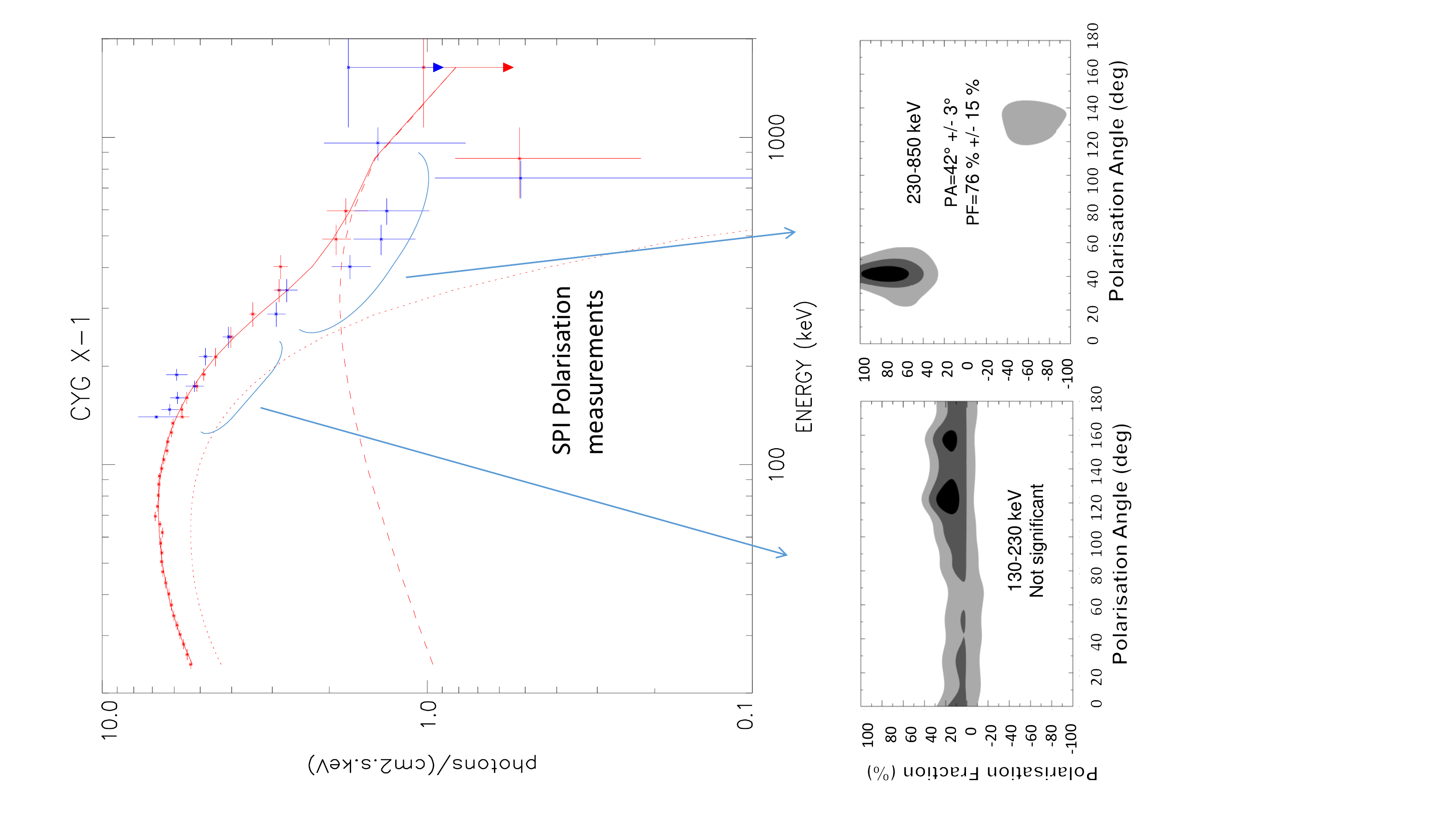}
\caption{Spectrum and polarisation results from SPI. Figure adapted from \citet{Jourdain2012}. 
}\label{fig:pol}
\end{figure}

Breaking the degeneracy between hybrid Comptonisation models and models where the tail is explained through synchrotron radiation requires additional diagnostics beyond the mere spectra analysis. \integral provides unique capabilities to help clarifying the above issue, in particular thanks to the data yielded by IBIS (ISGRI and PICsIT),  as well as by the SPI instrument, used as polarimeters (see Sect.~\ref{sec:high-energy}). 
Based on IBIS data of the first years of \integral, \citet{Laurent2011} found that above 400\,keV - 
where the hard tail dominates the emission - a high degree of polarisation is observed, with hints that the degree increases with increasing energy. These results were later independently confirmed thanks to  SPI  \citep{Jourdain2012} (Fig. \ref{fig:pol}). 

The limited availability of data required these initial analyses to use all available observations, and thus precluded a state-dependent analysis of the data. It should be noted, however, that during the initial years of the mission Cyg X-1 was predominantly observed in the LHS. The accumulation of more data allowed \citet{rodrigue15_cyg} to perform a state-resolved polarisation analysis of the data, which confirmed a high degree of polarisation in the LHS. However, only high upper limits could be placed in the HSS, mainly owing to the lower flux of source in this state in the high energies, combined with lower exposure accumulated during the HSS at the time of the analysis.

The high degree of polarisation of the hard tail in Cyg X-1 makes it very likely that the emission detected above 400\,keV is due to synchrotron radiation, as no other process is known that could produce such a high degree of polarisation. The emission below 200\,keV is not polarised, as one would expect from Comptonisation. This result therefore strongly suggests that the soft $\gamma$-ray tail emission originates in the jet, as is also suggested by the Fermi satellite detection of a very high energy component \citep{zanin:16a} and by the broad band spectral energy distribution of the source \citep{Pepe_2015a}. The fact that the degree of polarisation is so high also means that the magnetic field configuration in the source must be very simple and uniform, since only minor de-polarisation must have occurred to randomise the starting polarisation.  This is particularly important because it sets strong constraints on synchrotron models. For example, as argued by \citet{Zdziarski_2012a}, the shape of the hard tail implies a rather hard electron energy spectrum ($p\sim 1.3$--1.6,
where the electron energy distribution is $\propto E^{-p}$), i.e.,  significantly harder than what one would expect from standard electron acceleration mechanisms, which would instead give $p\sim2$. In addition, there must be highly energetic electrons relatively far away from the BH \citep{Pepe_2015a}. If the hard tail comes from the jet, then the observed Compton component may be associated to the jet base, as postulated, e.g., by \citet{Markoff2005}.

\subsection{V404 Cygni }
\label{sec:v404cyg}

\begin{figure*}
\centering
\includegraphics[width=1\textwidth]{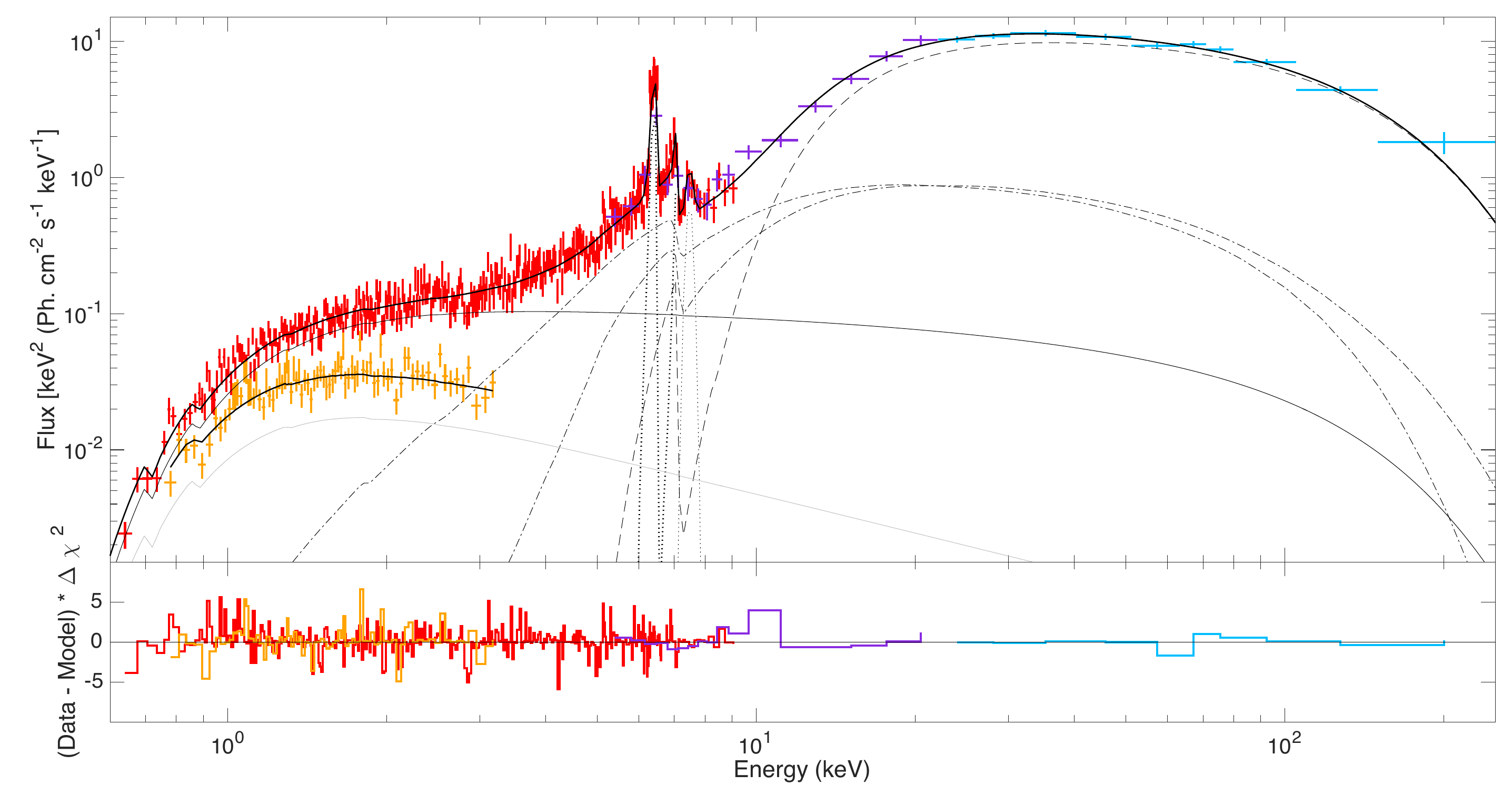}
\caption{\swift/XRT (red for the source, orange for the dust scattering halo spectrum) + \integral /JEM-X (purple) + \integral /ISGRI (clear blue) spectrum of V404 Cyg fitted with model based on \textsc{MYTORUS} \cite{Yaqoob2012}, as described in \cite{Motta2017b}. 
\textit{Upper panel}: the thick black line marks the best fit to the data. The thin black line indicates the illuminating Compton continuum, the dashed and dotted-dashed lines mark the various components of \textsc{MYTORUS}, the dotted line marks the fluorescent Fe line spectra and a Gaussian line at $\sim$7.5 keV. Finally, the solid grey line indicates the model component accounting for a dust scattering halo emission. Figure taken from \citet{Motta2017b}.
}
\label{fig:V404spectrum}
\end{figure*}

\subsubsection{A brief history of the first dynamical black hole}

The transient system V404 Cyg harbours one of the first dynamically confirmed stellar-mass BHs \cite{Casares1992} (see also \citet{Bolton1972,McClintock1986b}).  It is located at a distance d = 2.39 $\pm$ 0.14 kpc \cite{Miller-Jones2009}, and is well known by its bright and extremely variable outbursts.
The system was  detected as a powerful \xray source (GS 2023+338) by the Ginga satellite in 1989 (\cite{Makino1989}, and promptly identified with the recurrent recurrent Nova Cygni 1938 \cite{Makino1989b,Marsden1989} with recorded outbursts in 1938 and 1956 \cite{Richter1989}. The  1989 outburst was characterised by rapid and erratic flaring activity in \xrays in any time scale down to 2\,ms \cite{Tanaka1989}: in some flares, drops in flux of $\sim$\,20\,Crab in less than one hour \cite{Tanaka1989,Kitamoto1989} were observed. Despite the high fluxes exhibited by the system,  which might have reached the Eddington limit during the peak of the outburst \cite{Tanaka1989},  the source energy spectra (well-described by a power law) and  PDS (mostly consisting of steep red noise) were consistent with V404 Cyg being in the LHS during the entire outburst \cite{Tanaka1989,Oosterbroek1996}, which suggested that the source did not transition to the HSS (or intermediate states), as many other transient BHBs do.
It was also observed that the spectral parameters were strongly influenced by changes in the local absorption, which had values in the range $5 \times 10^{21}-4.5\times 10^{23}$ cm$^{-2}$ \cite{Tanaka1989, Oosterbroek1996,intZand1992}. As the outburst decayed, the average flux became less variable and the absorption episodes were less frequent, until the source reached quiescence, the level of which is notoriously higher ($L_\textrm{x} \sim 10^{34}\,\ergcms$) than for other BHBs in the case of V404 Cyg \cite{Gallo2005}.

After $\sim$26 yr in quiescence, the onset of a new outburst was detected by Swift/BAT, MAXI and Fermi/GBM on 15 June 2015 \cite{Barthelmy2015,Negoro2015,Younes2015}, although optical data suggest that the outburst  actually started one week earlier \cite{Bernardini2016}.
The peak of the outburst was reached on 26 June 2015, and the flux dropped abruptly immediately afterwards \cite{Ferrigno2015,Walton2015}. 
Following this drop, V404 Cyg slowly faded to quiescence over the subsequent weeks \cite{Sivakoff2015}.
The outburst triggered an intense multi-wavelength monitoring campaigns, which provided data over the entire electromagnetic spectrum, generating one of the richest (largely coordinated) multi-band campaign ever performed on a BHB. 
\integral played a key role in providing \xray and optical coverage as continuously as possible between June 17th and July 13th 2015 \cite{Kuulkers2015a,Kuulkers2015b}, yielding crucial pieces of information which helped piecing together the most important properties of this important system.

\subsubsection{\integral observations of V404 Cyg}\label{sec:V4042015}

Intense flaring activity was observed by \textsl{INTEGRAL} between $17-28$ June 2015 \cite{Rodriguez2015,Natalucci2015,Roques2015, SanchezFernandez2017}, as well as hardness variations on second time-scales \cite{Natalucci2015}. 
In some major flares, V404 Cyg reached peak fluxes around 50 and 40 Crab in soft and hard \xrays, respectively, and peak-to-peak flare recurrence times as short as 20 min \cite{Rodriguez2015}. The extraordinary brightness of the source allowed spectral studies with unprecedented time resolution for coded-mask instruments such as those on-board \integral. \citet{SanchezFernandez2017} analysed 602 IBIS/ISGRI spectra of V404 Cyg during the intense flaring period ($17-28$ June 2015) with integration times as short as 8 seconds in the brightest flares. All the 20--200 keV IBIS/ISGRI spectra were compatible with a Comptonised model, which in most cases required an absorption component to fit the data. The measured absorption columns were so high ($N_{\rm H} \gtrsim  10^{24}$ cm$^{-2}$) - especially during the \textit{plateaus} observed in the IBIS/ISGRI light curves -  that the effects of absorption were evident up to $\sim$ 30\,keV. 
The absorption column density and the \xray flux were anti-correlated in most of the flares, suggesting that the observed variability was partly - but not exclusively - due to variations in the absorption column.  \citet{SanchezFernandez2017} studied the correlations between the parameters derived from the fits of the IBIS/ISGRI spectra, and concluded that V404~Cyg had been in the LHS during most of the outburst, but with clear indications for softer spectra observed during some of the brightest flares. 

The detailed analysis of the broad-band spectrum of V404 Cyg during one of the June 2015 \xray plateaus showed an energy spectrum  reminiscent of that of obscured/absorbed active galactic nuclei (AGN) \cite{Motta2017b} (see Fig. \ref{fig:V404spectrum}). The joint spectral analysis of strictly simultaneous  \textsl{Swift}/XRT, JEM-X  and IBIS/ISGRI spectra revealed a Comptonised spectrum ($N_{\rm H} \sim 1-3\times10^{24}$ cm$^{-2}$)  heavily absorbed by a partial covering material, and a dominant reprocessed component, which included a narrow Iron $K_{\alpha}$ line (see Fig. \ref{fig:V404spectrum}). This finding indicated the presence of a geometrically thick accretion flow as the likely responsible of both the high intrinsic absorption and the intense reprocessed emission.

\subsubsection{Other X-ray and multi-wavelength observations}

A complete, time-resolved, spectral analysis of the available  \textsl{Swift}/XRT data \cite{Motta2017a} found that the XRT spectra could be fitted with a partially-covered, absorbed power law. A blue-shifted Iron $K_{\alpha}$ line appears  together with the signature of high column densities. 
In these fits, changes in column density relate with fast ($\sim$minutes) but moderate flux changes, while variations in the covering fraction would be responsible for the most dramatic flux and spectral changes observed during the outburst \cite{Motta2017a}. Interestingly, despite the intense flares observed (sometimes exceeding the Eddington luminosity), none of the  spectra showed the unambiguous presence of soft disc-blackbody emission. These findings were ascribed to the presence of a \textit{slim disc} in the vicinity (a few hundreds of Gravitational radii) of the central BH, which both hides the innermost (and brightest) regions of the accretion flow, and produces a cold, fragmented, high-density outflow that introduces the high-absorption and fast variability observed. Such a slim disc was possibly sustained by erratic or even continuous at Eddington/Super-Eddington accretion, and as a consequence the central BH was partly or completely obscured by the inflated disc and its outflow.

Similar extreme and variable absorption was also detected by \textsl{NuSTAR}  during observations close to the peak of the outburst \cite{Walton2017}. 
A strong relativistic reflection component was found in the \textsl{NuSTAR} flare spectra, which was interpreted as the result of illumination of the accretion disc (truncated at a small distance from the BH) by transient jet activity \cite{Walton2017}.  

\begin{figure}
\includegraphics[width=0.5\textwidth]{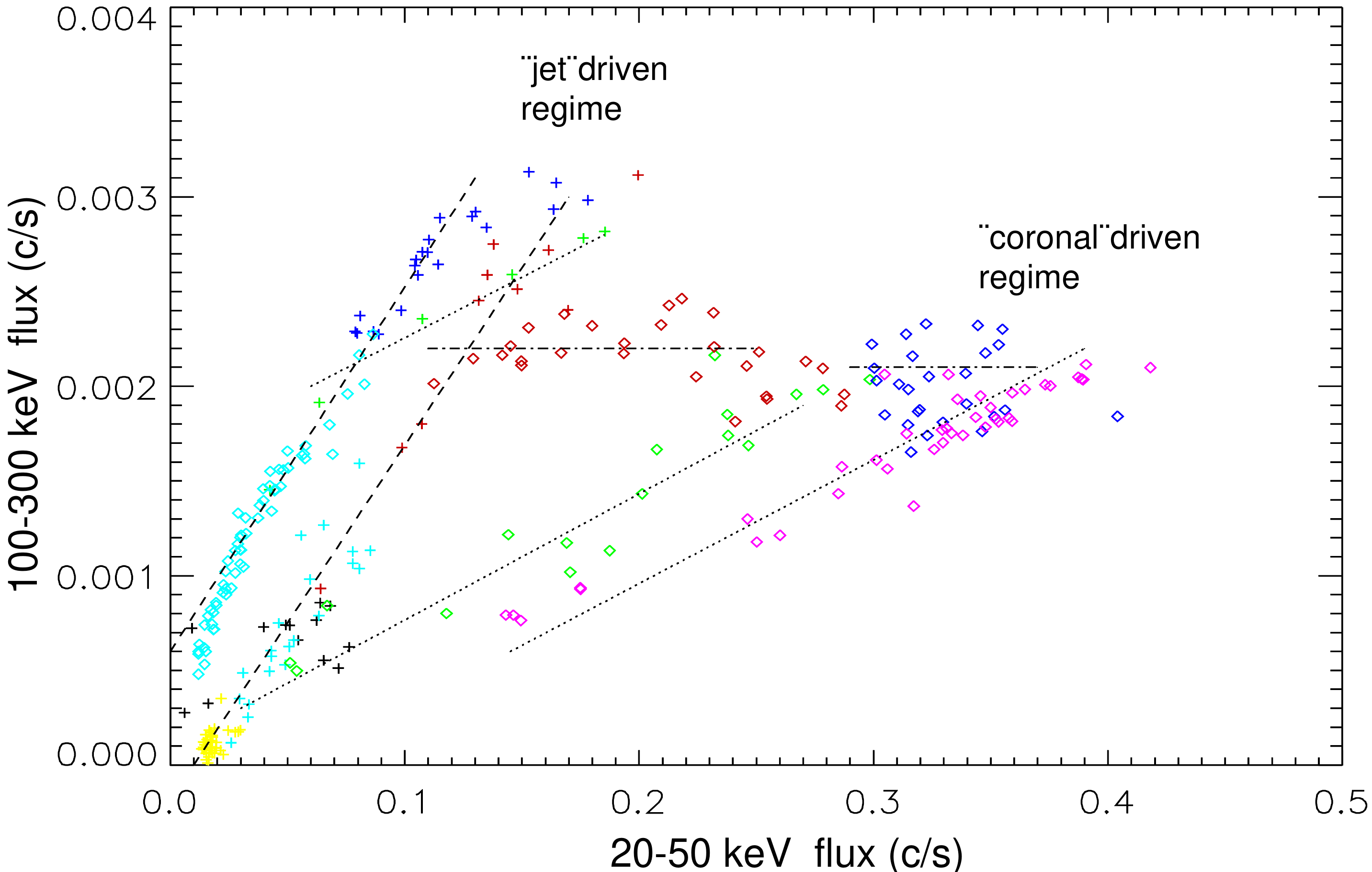}
\caption{Flux-flux relation observed in V404 Cyg during the brightest flare, as seen by \integral/SPI. Figure adapted from \citet{Jourdain2017} 
}\label{fig:V404FF}
\end{figure}

During the flares of V404 Cyg \integral/SPI detected high-energy excesses \citep{Roques2015,Jourdain2017} with respect to the thermal Comptonisation continuum that describes the ISGRI and SPI data well up to 200 keV \citep{SanchezFernandez2017, Natalucci2015}. This excess is reminiscent of the one seen in Cyg X-1 up to about 1 MeV, which has been interpreted either as the result of Comptonisation by a hybrid (i.e., thermal plus non-thermal) electron population in both the LHS and the HSS \citep{Zdziarski2017}, or as synchrotron emission from the jet based on the presence of a significant polarisation fraction (see Sec. \ref{sec:CygX1}).
Noticeably, since the GeV detection by FERMI/LAT \citep{Loh2016} came from the very same V404 Cyg flares where the 200 keV - 1 MeV excess was seen by SPI \citep{Jourdain2017}, it is possible that both originate in the jet.
Figure ~\ref{fig:V404FF} displays the behaviour of V404 Cyg along the brightest (and last) flare observed by \integral. \citet{Jourdain2017} ascribed the 20--50 keV flux to the corona and the 100--300 keV flux to the jet, and pointed out that the two components evolve independently, following two distinct evolution patterns. The flux-flux relation shows two different slopes, which can be interpreted as two regimes where the source emission is driven by the corona and the jet contributions, respectively. A third (horizontal) track, reflecting a stable 'jet' emission with a decreasing corona flux, corresponds to the transition between the two.

The soft $\gamma$-ray tail detected in V404 Cyg could be in principle ascribed to a third mechanism, the production of electron-positron pair plasma during the brightest, Eddington-limited flares. \citet{Siegert2016} reported evidence of a 511 keV annihilation line associated to a pair plasma in the SPI data. However, independent studies have failed to confirm this 511 keV line detection \citep{Jourdain2017,Roques2019}. 
Such a discrepancy was likely caused by the event selection of the SPI data, which is a particularly important issue for bright sources \cite{Roques2019}.

In the radio VLBA monitoring, \citet{Miller-Jones2019} found that the jet orientation was changing significantly in time scales ranging from minutes to hours. Their interpretation was that the Lense-Thirring precession of the slim disc that forms due to the super-Eddington accretion -- which is also likely responsible for the rapid changes in the cold absorption \citep{Motta2017a} --  is driving the changes in the jet orientation.

\subsubsection{Rapid multi-wavelength variability}
 
Rapid variability in the emission from V404 Cyg was also omnipresent at longer wavelengths.
Perhaps the most surprising feature of was the detection of heartbeat-type oscillations in the optical \citep{Kimura2016}, which are reminiscent of the \xray light curves of GRS 1915+105 (Sec. \ref{sec:GRS1915}). A particularly fascinating aspect of V404 Cyg was the complexity of how the \xray and optical emissions were correlated (see \cite{Oates2019} for a study of correlated variability in UV and \xrays, performed using \swift/UVOT and XRT data). An incredible plethora of examples of correlated behaviour was observed within the 2 central (i.e. most active) weeks of the outburst.
The detailed study by \citet{Rodriguez2015} and \citet{Alfonso2018} showed that on 1-minute and longer time scales, the \integral OMC, JEM-X and ISGRI light curves showed: (i) flares where the optical emission was delayed by minutes up to an hour following and strong \xray flare, (ii) flares where the optical and \xray emissions were practically simultaneous (i.e. almost the exact same variations were observed ), and 3) flares during which enhanced optical and soft \xray emissions preceded a hard \xray flare \citep{Kajava2020}.
Moreover, simultaneous NuSTAR and optical observations taken on sub-second time scales showed that the optical emission lagged the \xrays by  roughly 100ms \citep{Gandhi2017, Hynes2019}. This lag was interpreted as the delay between fluctuations generated in the hot corona around the BH, that propagate up to the relativistic jet base about 1000 $R_{\rm g}$ away, which is responsible for part of the optical emission \citep[e.g.,][]{Malzac2018}.
The correlated 1-minute time scale optical lags correspond to the light travel time delays from the BH to the outer disc,  were instead likely produced by the re-processing of the \xrays in the outer accretion disc, or by the illuminated binary companion \citep{Kimura2017, Alfonso2018, Hynes2019}.

Longer optical time lags of up to an hour were attributed to interactions between distinct relativistic ejecta in the form of blobs of plasma \citep{Alfonso2018}, although on these long time scales the association of a given \xray flare to its optical counterpart may be in some cases ambiguous given the stochastic nature of the light curves.
The few cases where \integral/OMC/JEM-X/ISGRI detected negative lags, i.e., where the optical (and soft \xray) emission leads the (hard) \xrays \citep{Alfonso2018,Kajava2020}, remain a puzzle yet to be solved, which could be explained in terms of mass accretion rate fluctuations propagating inwards through the accretion disc. 

Optical emission line profiles showed significant changes minutes after major \xray flaring detected by \integral \citep{Munoz-Darias2016}.
Such lines showed broad and deep P-Cyg profiles, from which outflow velocities up to 3000 km/s were estimated. The presence of an outflow in optical with such a velocity is consistent with the strong disc wind observed in \xrays by \textsl{Chandra}, likely radiation or thermally driven as V404 Cyg approached the Eddington limit \cite{King2015}. As suggested by \citet{Motta2017a}, the two outflows could be two portions of the same disc wind, which likely feature a temperature/ionisation stratification. Over the course of the 2 most active weeks of the  outburst, more than $10^{-8}\,M_{\odot}$ was lost in these variable disc winds, as well as in the Compton-thick outflow responsible for the variable partial covering, such that the suggestion was made that the outburst itself may have ceased prematurely due to excessive mass loss \citep{Munoz-Darias2016}. 

\subsubsection{The December 2015 re-flare}

On 23 December 2015, only four months after the end of the primary outburst, the \textsl{Swift}/BAT telescope detected a re-brightening of V404 Cyg \cite{Barthelmy2015b}. Follow-up \xray, optical, and radio observations confirmed the onset of a new outburst \citep{Munoz-Darias2017}. This December event was a re-flare, it was in many respects a fainter analogue of the first week in the June 2015 activity, with weaker flares and lower non-flaring emission in between them. The \integral and Swift spectra were compatible with a Comptonised continuum modified by a Compton-thick partial covering absorber \citep{Kajava2018}.
Optical photometry \citep{Kimura2017} and spectroscopy, as well as monitoring in radio \citep{Munoz-Darias2017} also showed a behaviour comparable to what was observed in June, including similar, though less pronounced, P-Cyg optical lines and radio variability. 

\subsection{Fast time variability}

The \xray timing studies by \integral \citep{Rodi2017} and NuStar \citep{Gandhi2017} did not show signs of QPOs, but instead the dominating feature was the presence of strong low-frequency (bending) power law noise, and the complete absence of high-frequency variability.
PSD featuring these properties are characteristic of heavily absorbed sources like Cyg X-3 \citep{Koljonen2018}, GRS 1915+105 in its obscured state \citep{Koljonen2021} as well as GRO J1655$-$40 in its ultrasoft state \citep{Uttley2015}. By analogy with these sources, it appeared that the high frequency noise in V404 Cyg was not intrinsically absent, but was instead smeared out or suppressed by the disc wind via heavy scattering.
Timing analyses of Chandra, FERMI/GBM and Swift on the other hand did show only weak and short-lived mHz QPOs \citep{Huppenkothen2017}. The absence of the strong type-C QPOs typically expected in a BH in its LHS or HIMS was interpreted as a consequence of the peculiar properties of V404 Cyg: the rapid evolution of either the truncation radius or the radius of a ring where the anisotropies responsible for the generation of QPOs occur \cite{Ingram2020} would make the accretion flow might be much more turbulent than it would otherwise be, so much so that the conditions for the precession of the inner accretion flow that would give origin to a Type-C QPO are not met, if not for very short times. 


\subsection{GX 339$-$4}\label{sec:GX339}

\gx\ was discovered in 1973 by {\it{OSO-7}} \cite{Markert1973}
and, based on the luminosity upper limit of the stellar companion, it was classified 28 years later as a low-mass XRB \cite{Shahbaz01}. Soon afterwards, observations in the optical band \cite{Hynes2003} revealed the motion of the companion star, based on which an orbital period of $\sim 1.7$ d was derived, as well as a mass function of  $5.8 \pm 0.5$ M$_\odot$. Later on, based on high-resolution optical spectra, a distance to the source in the range 6--15 kpc was proposed \cite{Hynes2004}, which, however, is still to be confirmed. 

Over the years after its discovery, GX 339$-$4 has shown a number of outbursts that have been extensively monitored in several multi-wavelength observations, and the properties of this system - which is considered the prototypical transient BHB - have been probed via a vast number of analysis techniques. 
The long-term spectral behaviour of the source, evidenced in the data collected from 1987 until 2004 - mostly by \rxte\, but also Ginga and \textit{CGRO} - was presented in \citet{Zdziarski2004}. This study showed that \gx\ had already undergone 15 outbursts in the period analysed, sampling several times all the known accretion states, occurring along pronounced hysteresis cycles, with the hard-to-soft state transitions occurring at much higher (and variable) luminosity than the soft-to-hard transitions. 

The outburst which occurred in 2002/2003 was studied in detail through timing and colour analysis \cite{Belloni2005}, and provided the prototype for the HID (see Sec. \ref{sec:states}). The state classification based on the HID proposed by \cite{Homan2005} has since then been  the main framework to describe the outburst evolution of transient BHBs, featuring two extra states - the HIMS and SIMS - with respect to the two historical LHS and HSS. This state classification has been  refined over the years to incorporate the increasing amount of information acquired on the more and more BHBs  that were being observed and discovered, and started to include multi-band observations, \textit{in primis} coming from the radio band.
Indeed, radio observations performed during the 2003/2004 outburst revealed  a large-scale relativistic radio jet from \gx \cite{Gallo2004}, which was then  associated with the HIMS-SIMS transition \cite{Fender2004} reported previously by \citet{Nespoli2003}.

In 2004, a new outburst occurred, and it was extensively observed by \integral as well as by \rxte \ and other observatories. 
A simultaneous \rxte\ and \integral campaign was performed on 2004 August 14-16, when the transition HIMS-to-SIMS was expected \cite{Belloni2006}.
\citet{Belloni2006} observed a state transition shorter than 10 hr, and performed the first simultaneous timing and and broad-band spectral analysis (3--200 keV) across a hard-to-soft state transition. These authors observed in the HIMS a steeper power law ($\Gamma \sim$1.9) and a lower value of the high-energy cut-off ($\sim 70$ keV) than in the LHS. They also observed the disappearance of the cut-off in the 3--200~keV energy band at the transition to the SIMS. This confirmed a behaviour  previously observed in other sources (e.g., GRO J1655$-$40 \cite{Joinet2005}), where it was argued that the cut-off might not disappear at the transition to the SIMS, but move to significantly higher energies, i.e., beyond the energy band covered by \rxte, but within the reach of \integral. 

A similar finding was reported later on by \citet{DelSanto2008} and \citet{Motta2009a}, who analysed \integral and \rxte data from the 2007 outburst of the source, and observed in greater detail the evolution of the high-energy cut-off across the long-lived ($\approx$ 10 days) hard-to-soft transition. A much denser monitoring with \rxte
of this outburst allowed \citet{Motta2009a} to confirm that the evolution of the cut-off is non-monotonic: it decreased monotonically from 120 to 60~keV during the luminosity increase in the LHS, started increasing again at the LHS-to-HIMS transition reaching approximately 100~keV just before the transition to the SIMS. The cut-off further increased to $\sim$130keV, and was no longer detected in the HSS. 

The non-monotonic evolution of the high-energy cutoff observed in GX 339$-$4  has been later reported in a number of sources (e.g. \citep{DelSanto2016}). As observed in GX 339$-$4 for the first time, and with an unprecedented clarity, the changes in the high-energy cutoff are observed to be as fast as the timing properties (in particular in the behaviour of the LFQPOs observed in the \rxte \ PDS), while other spectral properties appear to be evolving much more smoothly over the state transition. 

\begin{figure}
    \includegraphics[width=9cm]{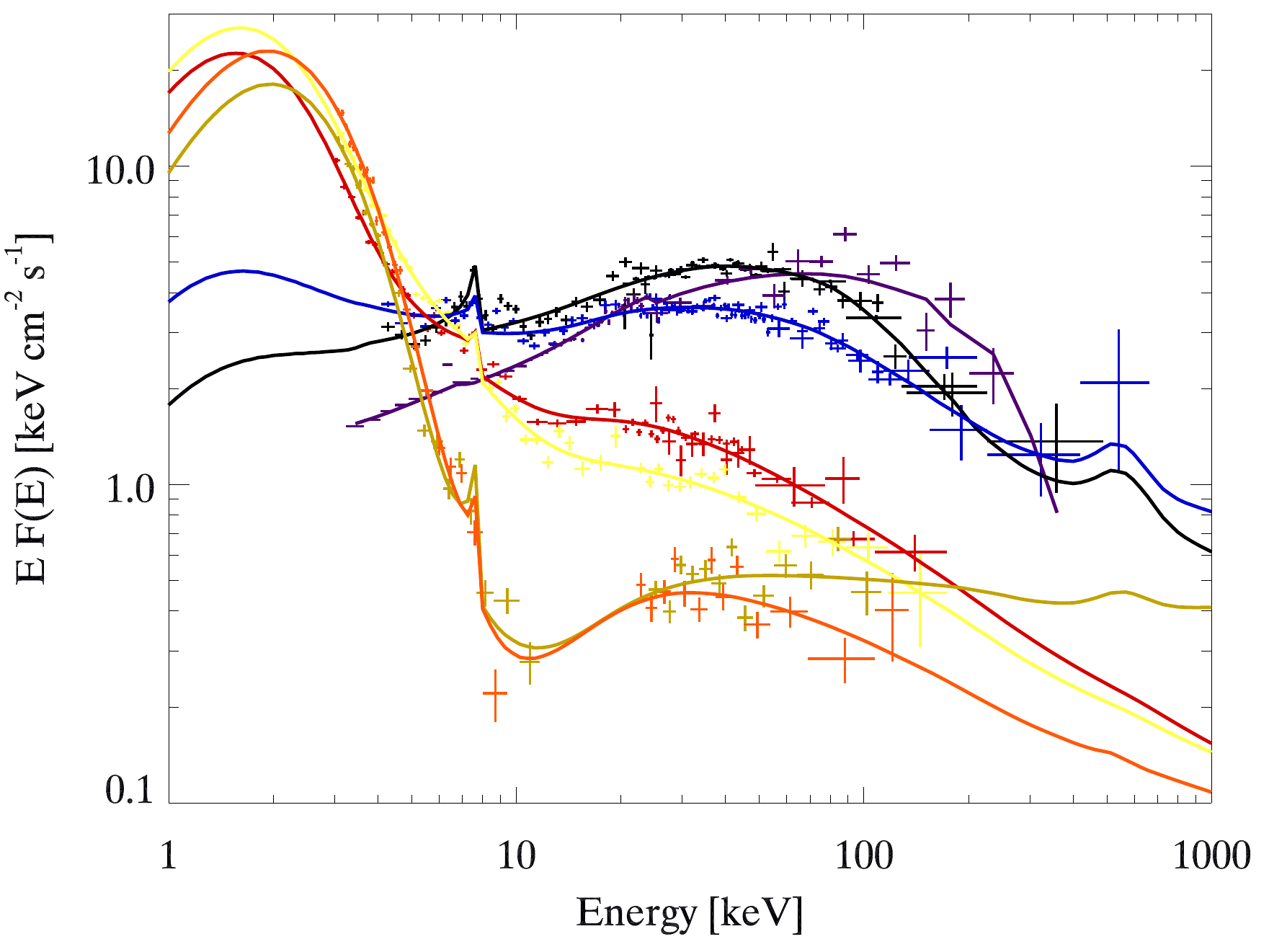}
    \caption{Joint JEM-X, IBIS and SPI energy spectra of GX 339$-$4 collected in different spectral states, from hard (violet) to soft state (orange). The data were fitted with the hybrid thermal/non-thermal  Comptonisation model {\sc eqpair} plus {\sc discline} (from \citet{DelSanto2008}). 
    }
    \label{Fig:gx339}
\end{figure}

The use of a phenomenological model such as a cut-off power law  provides a useful qualitative description of the data and of the evolution of the spectral shape at a time where the energy spectrum undergoes pronounced and relatively fast changes. However, it is worth noticing that by using such an empirical model, not much can be said on the evolution of the plasma physical parameters, as the energy cut-off is only roughly correlated to the electron temperature of the Comptonising plasma, and it may not reflect the evolution of the electron temperature of the corona.
Additional components play a role in the complex X/$\gamma$-ray spectra of BHBs (e.g., non-thermal Comptonisation and Compton reflection), which has to be investigated by means of more accurate, physically motivated spectral models. 
In this context, \citet{DelSanto2008} investigated the spectral variability of \gx\ across the different states by using the hybrid thermal/non-thermal Comptonisation model
\texttt{eqpair} \cite{Coppi1999} on the \integral spectra collected during the period August-September 2004 (see Fig. \ref{Fig:gx339}). 
These authors found that the spectral transition HIMS-SIMS-HSS was driven by changes in the soft photon flux responsible of the cooling of the hot plasma (the corona, or hot flow). Simultaneously, the increase in the disc temperature and the decrease of the inner disc radius were also observed. Such properties and behaviour are consistent with the truncated disc model \cite{Esin1997, Done2007}, although other models such as dynamic accretion disc corona models cannot be in principle ruled out.

Thanks to the high-energy coverage of the \integral telescopes, in all spectral states of \gx, a significant contribution from a (sometimes variable \cite{Droulans2010}) high-energy tail above 200 keV, (see Sec. \ref{sec:high-energy}),  was observed \citep{Joinet2007, DelSanto2008, Caballero-Garc'ia2009}.
During the 2007 outburst a peculiar spectral transition was observed by \integral and \rxte\ \cite{DelSanto2009}, with multiple transitions back and forth from/to the HSS and LHS, across the SIMS, while during the 2010 outburst a number of simultaneous multi-wavelength observations confirmed the jet quenching over the transition in many wavebands \cite{CadolleBel2011}.

\subsection{1E1740.7$-$2942: the great annihilator}
\label{sec:annihilator}

Discovered by the Einstein satellite in 1984 \cite{Hertz1984}, the bright hard \xray source 1E~1740.7$-$2942  (hereafter \GA)  is located less than one degree away from the Galactic Centre. It was classified as candidate BHB by \citet{Sunyaev1991a} because of the spectral similarities with \cyg, based on results obtained with the Franco-Russian hard \xray/soft $\gamma$-ray satellite \textit{Granat}. \GA\ is the prototype of the {\it{microquasar}} class, thanks to its double-sided radio jets reaching large angular distances from the core ($\sim 1'$), which were  discovered in 1991 by  \citet{Mirabel1992}.

This source takes its nickname from a claim that was made soon after the first observations in the soft $\gamma$-ray domain. In 1990, the \textit{SIGMA} telescope, on-board \textit{Granat}, observed \GA\ and found a spectrum characterised by a transient (less than 3 days) broad feature around the electron-positron annihilation energy at 511 keV \cite{Bouchet1991, Sunyaev1991a}. This finding attracted the attention of the community and a great deal of studies were dedicated  to investigate the possibility that the aforementioned feature could be due to the electron-positron annihilation line (the search of which is among the main aims for the \integral mission). The ``line'' was speculated to be connected with the spectacular bipolar jet seen  in radio, which could be formed by leptons rather than by baryons. Alternatively, it could be the result of productions of positrons (which would therefore annihilate the electrons) by a hot plasma with a temperature much lower than 1 MeV via photon-photon absorption, for example when the accretion flow is able to interact with a wind \cite{Oss1995}.  
\citet{Bouchet2009} performed a deep analysis based on the SPI data searching for any feature around 511 keV associated with \GA, but did not find evidence for a line at 511 keV, either narrow or broad, transient or persistent.
Later on, \citet{Decesare2011} searched for such an emission from any point sources with the IBIS telescope, and found no evidence of Galactic 511 keV point sources, but estimated a 2 $\sigma$ upper limit annihilation flux of 1.6$\times10^{-4} ph/cm^2 /s $, consistent with a diffuse electron-positron annihilation scenario, where electrons and positrons could be produced in compact objects, and propagate in the interstellar medium before annihilation. 
Thus, the annihilation scenario for \GA\  was never confirmed, but the evocative name of this source stuck to it, so that \GA\  is often still affectionately referred to (rather erroneously) as the Great annihilator. 

\citet{Smith2002a} measured a periodic modulation of $\sim 12.73 \pm 0.05$ days which, if interpreted as the orbital period of the system, would suggest  a red-giant as donor. However, the upper limit for the IR counterpart derived with the VLT seems to exclude such a nature \cite{Marti2000}. The high hydrogen column density toward the source (see e.g. \cite{Gallo2002}) does not allow any firm identification of the companion star, the characteristics of which remain unclear. 

\integral observations performed throughout several years have revealed that most of the time \GA\ is found in either one of two main states: the canonical LHS, with a mean flux of $\sim$ 50  mCrab, and a {\it{dim}} state, with a much lower flux level of a few mCrab  \cite{Delsanto2008p, Bouchet2009}. In particular, \xmm and \integral provided the arguably highest quality broadband spectrum of the source in the LHS \cite{Castro2014}.
In a few occasions, soft spectra have been observed, testifying occasional transitions to a softer state  \cite{Smith2002b}. In particular, simultaneous \integral and \rxte\ broad-band spectra taken in 2003 showed that an intermediate/soft spectral state occurred just before one of the long dim states shown by  \GA\ \cite{Delsanto2005}. This behaviour was interpreted as the effect of a drop in the accretion rate, which is expected to affect both the thin disc and the hot flow (corona). While such a change would propagate through the hot flow immediately, it might take up to several weeks to propagate through the thin accretion disc. Thus, the hot flow is temporarily drained of matter compared to the disc, therefore an accretion flow configuration very similar to that observed in the HSS of other BHBs can be temporarily observed (see also \cite{Smith2002b}). 

\citet{Bosch-ramon2006} applied a jet model to the \GA\ SED (which include \integral data) to investigate the possibility that the hard \xray emission in \GA\ was to be ascribed to the jet. They concluded that  synchrotron emission from the jet seemed in fact to be able to explain the low-hard state X-ray spectrum of the
source, but it exceeded largely the observed core radio fluxes. \citet{Bosch-ramon2006} therefore concluded that although accretion and jet phenomena are probably linked - in \GA \ (and in any other source), the dominant component in the (hard) X-rays can be either the jet or the disk/corona depending on the system considered.


\subsection{Black holes discovered by INTEGRAL}\label{sec:INTdiscoveries}

\begin{figure*}
\centering
\includegraphics[width=0.99\textwidth]{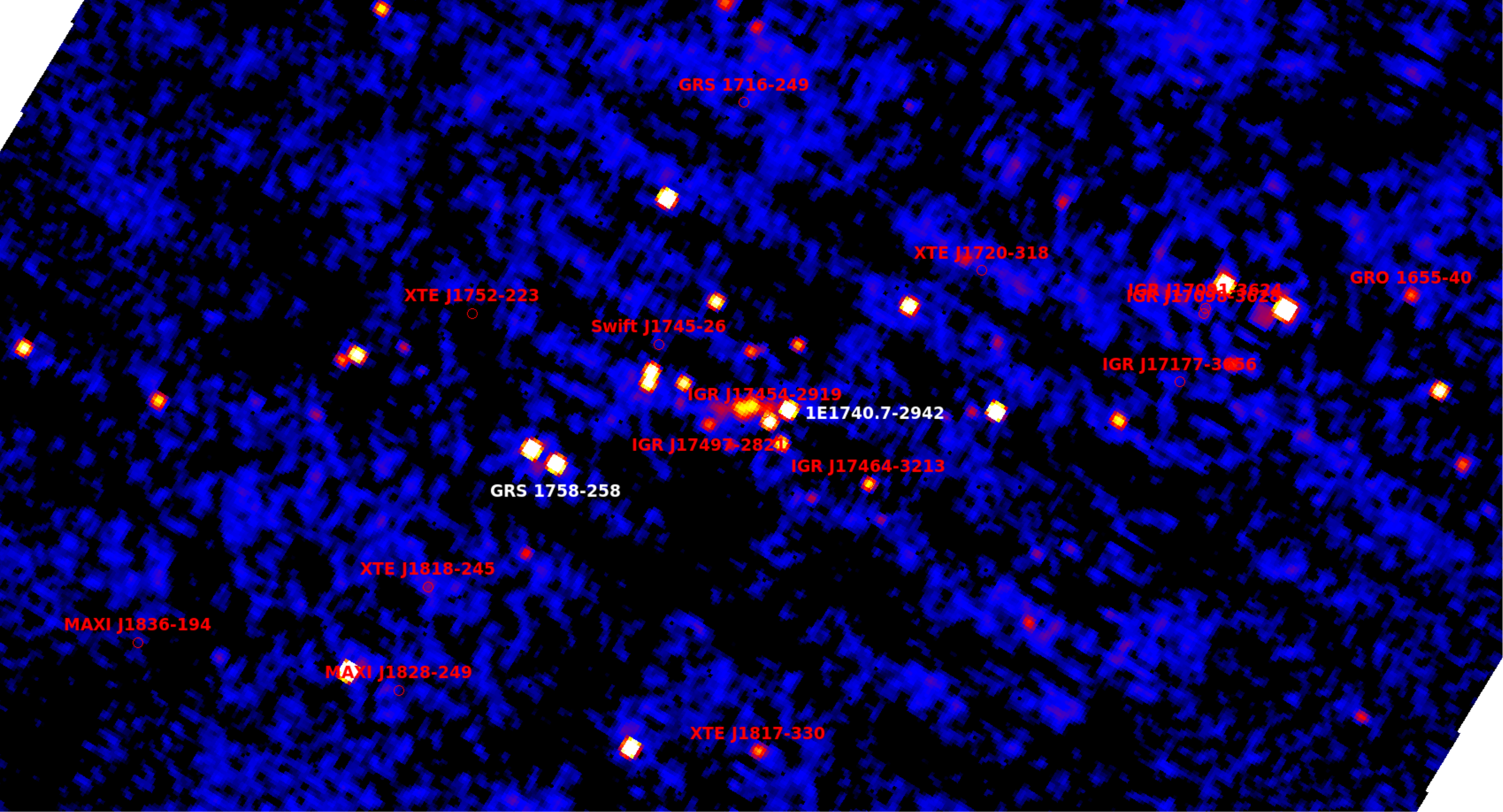}
\caption{A zoom-in to the Galactic Bulge region as seen by \integral, based on the IBIS/ISGRI (18--40\,keV) data obtained through the  Galactic bulge program (from 2005 to 2009). Transient BHBs are marked in red, persistent BHBs are marked in white. All source names are annotated above the source \xray position, except 1E1740.7$-$2942 (annotated at the right of the position), and IGR J17497$-$2821 and GRS 1758$-$258 (both annotated below their position). All the sources that are not labelled are NS \xray binaries. 
}\label{fig:GB}
\end{figure*}

The bulge of our Galaxy (Galactic Bulge) contains a variety of high-energy point sources \cite[see, e.g., ][]{Knight1985,Belanger2006,Kuulkers2007,Bird2010},
including persistent and transient (candidate) BHBs. While persistent sources offer the opportunity to probe the accretion process at play at any time, in transient BHBs the accretion rate varies by larger factors than in the persistent sources, thus offering the opportunity to study accretion in regimes not accessible in the persistent sources.
The regular and long-term monitoring of persistently bright BHBs allowed by the Galactic Bulge monitoring program enables a systematic study of the hard \xray states of BHBs (e.g., \cite{Delsanto2005,Natalucci2014} for 1E 1740.7$-$2942), which is difficult or impossible for other missions to accomplish due to a limited sensitivity at high energies and/or the lack of wide-field imaging instruments as well as of orbits wide enough to allow long, uninterrupted observations of a particular region of the sky. Furthermore, by probing the hard \xray spectra of the sources over a wide luminosity range, the Galactic Bulge monitoring allowed the dependence of the \xray spectrum on the (total) luminosity to be investigated. 

Since February 2005, whenever the Galactic Bulge region was visible, it has been monitored approximately every 2--3 days with 3.5 hours snapshots, thanks to a dedicated \integral program \cite{Kuulkers2007}. The wide-field imagers at high-energies that are key to the \integral payload allowed the monitoring of the variability of the variety of sources in this region to be performed all at once in a broad energy range over the past 15 years. 
Various bright ($\gtrsim$100 mCrab) well-known \xray transients were active during the several years of the Galactic bulge monitoring program (e.g. GRO J1655$-$40, \cite{Shaposhnikov2007a} and H1743$-$322, \cite{Prat2009}), as well as new ones
, some of which were discovered by \integral during routine observations of the Galactic Bulge region. \integral is ideally suited to detect such hard spectral state outbursts: transient and persistent BHBs account for about half of the total detections in the 150--300 keV band (e.g., \cite{Bazzano2006,Krivonos2015}).

In this section, we will briefly review the results obtained on BHBs which have been discovered by \integral, focusing in particular on the discoveries and findings made possible thanks to \integral.  

\subsubsection{IGR J17091$-$3624}\label{sec:17091}

\integral discovered IGR J17091$-$3624 on 2003 April 14, during a routine observation of the Galactic Centre region \cite{Kuulkers2003}. Follow-up observations were immediately performed by \rxte, which confirmed the presence of an hard spectrum well-fitted by an absorbed power law and fast time variability on time scales from several tenths of a second to several tens of seconds, typical of a BHB in the LHS \cite{Lutovinov2003}. 
Archival searches revealed that IGR J17091$-$3624 was also detected by the 
Roentgen observatory aboard KVANT module of the MIR space station in 1994 October, and by BeppoSAX in 1996 September and 2001 September \cite{Lutovinov2003}. \xmm observed the field of IGR J17091$-$3629 in 2006 August, and 2007 February, but while IGR J17098$-$3628 - located 9.6 arcminutes away from it was observed in a relatively bright state, IGR J17091$-$3624 was not detected. In 2011 January, Swift/BAT observed renewed activity of this system, which raised from a flux of $\sim$20 mCrab to $\sim$60 mCrab in about a week \cite{Krimm2011}.

\citet{Capitanio2012} followed the evolution of IGR J17091$-$3624 throughout its 2011 outburst with \integral and Swift. Unlike the previous outbursts of this source, which were relatively short (weeks) and faint, the 2011 outburst was significantly longer and brighter. The initial evolution of the source followed the standard state evolution of most BHBs, with a LHS followed by a HIMS and a transition to the SIMS and HSS. Radio observations of this system confirmed the presence of a flat radio spectrum in the LHS, and the detection of a radio flare during the hard to soft transition, likely associated with relativistic ejections \cite{Rodriguez2011}.
Shortly after the HSS was reached, IGR J17091$-$3624 started showing a variability pattern very reminiscent of the \textit{heartbeats} \cite{Altamirano2011,Court2017} observed in GRS 1915+105 and a handful of other sources \cite[see, e.g.,][and references therein]{Motta2020}. 

The long and bright outburst started in 2011 ended in 2013, with a return of the source to the LHS, a number of short re-flares during the LHS, and  a final descent into a very low-luminosity LHS, or a bright quiescence \cite{Pereyra2020}. In 2016 February, IGR J17091$-$3624 showed another outburst, which, however, only lasted approximately 2 months with no sign of the heartbeat variability shown during the previous outburst \cite{Pereyra2020}.

\subsubsection{IGR J17098-3628}

IGR J17098$-$3628 was discovered by \integral on 2005 March 24th during deep Open Program observations of the Galactic Centre region. The discovery was made thanks to IBIS/ISGRI data, which revealed a new \xray transient located 9.4 arcmin off IGR J17091$-$3624 \cite{Grebenev2005}. Follow-up \integral and \rxte\ observations of the Galactic centre region revealed a variable source, which was tentatively classified as a new transient BHB due to its properties typical of the sources of such a class \cite{Grebenev2005a}.
A tentative radio counterpart was identified shortly after the discovery by \citet{Rupen2005a} in VLA observations, which showed a relatively bright radio source well within the IBIS/ISGRI error circle.

\integral and \rxte\ observed the field of IGR J17098$-$3628 several times in the period between 2005 and 2007, during which the target was regularly detected \cite{Capitanio2009a,Chen2008}. Targeted observations performed with \xmm and Swift also yielded a significant detection of this source. Renewed activity from this system was reported by \citet{Prat2009a}, which reported on new observations of the Galactic Centre. The \integral data showed a hard power law spectrum compatible with a BH in the LHS. 

\subsubsection{IGR J17177$-$3656}

IGR J17177$-$3656 was discovered in 2011 March 15 \cite{Frankowski2011} in the data from IBIS/ISGRI. A subsequent \integral observation from the Galactic Bulge monitoring program was performed shortly after the discovery, confirming the detection, and revealing an increase in the source flux in comparison with previous serendipitous data. The source position was refined first thanks to a \swift  target of opportunity observation, then thanks to a \chandra  observation \cite{Zhang2011, Paizis2011}. Follow-up observations in radio with ATCA evidenced a moderately faint ($\approx$0.2 mJy at 5.5 and 9 GHz) radio counterpart to IGR J17177$-$3656, which suggested that this source could be a BHB in the LHS.

\citet{Paizis2011a} reported on multi-wavelength observations of IGR J17177$-$3656, showing that the overall behaviour of this system follows that of other BHBs, thus confirming that this system might in fact be a binary hosting a stellar-mass BH, possibly observed at high inclination.

\subsubsection{IGR J17464$-$3213 (a.k.a. H1743$-$322)}
\label{sec:H1743}

On 2003 March 21, during a routine scan of the Galactic Centre region,  \integral observed a new active \xray source - IGR J17464$-$3213 \cite{Revnivtsev2003}. Follow-up observations of the same region with \rxte\ found a source previously named XTE J1746$-$322 at a position consistent with that of IGR J17464$-$3213 \cite{Markwardt2003} with a hard spectrum. A radio counterpart in the same region of the sky was also identified by the VLA \cite{Rupen2003}. The positions of the above sources were all  compatible with the \xray transient H1743$-$322 discovered during its outburst in 1977 by Ariel V and HEAO-1 \cite{Kaluzienski1977}. 

Early time \integral observations showed a slowly rising hard \xray flux, the beginning of a long outburst ended in November 2003. Such an outburst was followed by \rxte\ and \integral, which observed significant spectral and fast-timing variations, and marked flaring activity especially in the hard X/soft $\gamma$-ray domain sampled by \integral/SPI. H1743$-$322 showed a behaviour rather typical of transient BHB, with large scale flux variations and various accretion states, based on which it was classified as a candidate BHB \cite{Parmar2003,Homan2005,Capitanio2005,Joinet2005}. The companion star in this binary is too faint to be detected by the currently available instruments, hence, the BH nature of the accretor is yet to be confirmed.

After the first outburst, to date the brightest observed for this system, H1743$-$322 underwent repeated outbursts, with a rate as high as of 1-2 outburst a year. Interestingly, H1743$-$322 showed one of the first and clearest examples of a \textit{failed} outburst (see \cite{Capitanio2009a, Motta2010a}, and \cite{Soleri2013} for the case of Swift J1753.5$-$0127), during which the source crossed the LHS-HIMS transition, but did not reach the SIMS and HSS, instead moving back to the LHS and then to quiescence in only a few weeks after the start of the outburst. The discovery of such a behaviour showed that the process that drives the evolution of a source across the various accretion states is reversible even after the LHS-HIMS transition has occurred, but still maintains its hysteresis nature, differently from the hard-only outbursts. 

\subsubsection{IGR J17454$-$2919}

IGR J17454$-$2919 was discovered in mosaic images made from JEM-X during \integral revolution 1460 (2014/09/27--30) \cite{Chenevez2014a}. 
The source was quite dim, with \xray fluxes $\sim$6.5 and $\sim$8.2 mCrab in the 
3--10 keV and 10--25 keV bands, respectively. Earlier observations of the field did not 
show any activity from this position down to about 1 mCrab, confirming that the source was a 
new transient entering the active phase. Early follow-up observations with 
\swift \cite{Chenevez2014a}, \nustar \cite{Tendulkar2014}, and \integral in the direction of IGR J17454$-$2919, confirmed that the source was indeed in a 
rather sustained outburst phase \cite{Chenevez2014b}. In particular, the \nustar spectral and temporal analysis shows a behaviour pointing towards a BHB nature.

\citet{Paizis2015} reported multi-wavelength analysis of this source. Their 
Chandra observations first allowed these authors to refine the celestial position of the source to
 RA$_{\mathrm{J2000}}$=17$^h$45$^m$27.69$^s$, Dec$_{\mathrm{J2000}}$=$-$29$^o$19$'$53.8$''$ with a 90$\%$ error box of 0.6$''$ 
\cite{Paizis2015b}, which showed that IGR J17454$-$2919 has an IR counterpart, the source 2MASS J17452768$–$2919534. The same field was observed at optical wavelengths with GROND: IGR J17454$-$2919 was not detected in the GROND $g'$, $r'$, $i'$, $z'$ bands, but it was present in the NIR (J, H K$_s$) ones at magnitudes similar to those observed in the 2MASS and UKIDSS catalogues.

\citet{Paizis2015} also reporte the broad-band analysis of \swift/XRT, BAT and \integral data along the course of the outburst. A larger absorbing column density is obtained with \swift than in the \nustar spectra, which could be indicative of variation of the intrinsic (i.e. local) absorption. The spectra appeared to be well-described by a power law, with no evidence of a high-energy cut-off. Overall the \xray behaviour of the source was reminiscent of a low-mass \xray binary, but no firm conclusion on the nature of compact object could be drawn from the available data.

\subsubsection{IGR J17497$-$2821}
\label{sec:IGRJ17497}

IGR J17497$-$2821 was discovered with IBIS/ISGRI on 2006 September 17 \cite{Soldi2006}, at a 2--100 keV unabsorbed luminosity of $10^{37}$ erg/s, obtained assuming a distance of 8 kpc \cite{Kuulkers2006}. 
Preliminary spectral analysis of the \integral data led \citet{Kuulkers2006} to conclude that the source was an XRB, either containing a BH or a NS, in a 
typical LHS. 

Very little was published based  on the  \integral data, although the source was followed with multiple facilities at various wavelengths from the radio band to the \xrays. Most results are reported in \cite{Paizis2007}, and \cite{rodrigue07_17497}. The former observed the field of the source with the High Transmission Grating Spectrometer aboard \chandra, which first allowed them to refine the position to
RA$_{\mathrm{J2000}}$=17$^h$49$^m$38.037$^s$ 
Dec$_{\mathrm{J2000}}$=$-$28$^o$21$'$17.37$''$ 
with a 90$\%$ error box of 0.6$''$. The Chandra spectral analysis shows a hard spectrum ($\Gamma \sim$1.2), significantly affected by absorption (N$_{\mathrm{H}}$=4.4$\times10^{22}$~cm$^{-2}$) \cite{Paizis2007}. Two candidate infrared counterparts could be identified, and based on the \xray and IR behaviour, \citet{Paizis2007} suggested that the source is a low-mass \xray binary with a potential red-giant K-type companion. 

A 50~ks Suzaku observation showed that the source spectrum was well described by a power law continuum up to 70 keV, which then rolled-over due to a high-energy cut-off around 100 keV \cite{Itoh2006}. 
At lower energies, structures in the 5--8 keV range were ascribed to an Iron absorption edge. The results of (quasi-)simultaneous radio (ATCA) and \xray (\rxte) observations were reported by 
\citet{rodrigue07_17497}. These authors showed that the 3--200 keV spectra are indeed well 
represented by a power law with a marked cut-off at high-energy, or by a thermal Comptonisation model, similarly to what is typically observed for BHB in the LHS. 
While the power density spectra are also typical of those of a BH LHS, with about 36\% RMS variability (0.1--100 Hz), no QPO were detected down to about 2\% RMS.

The radio observations allowed \citet{rodrigue07_17497} to show that the source lies below  the track traced by GX 339$-$4 and V404 Cyg in the so-called radio-\xray flux plane \citep[e.g.,][]{Gallo2003, Corbel2003, Plotkin2012}, and is instead located on the lower track, which is followed by a growing number of BH systems, including H1743$-$322 \cite[see e.g.][]{Coriat2011,Corbel2013}.

\subsection{Other black holes}\label{sec:otherBHs}

\subsubsection{GRO J1655$-$40} 
\label{sec:GRO1655}

GRO J1655$-$40 was discovered on 1994 July 27 by the {\it{CGRO}} satellite, and was originally named Nova Scorpii 1994 \cite{Zhang1994}. Soon after its discovery, radio observations revealed apparently super-luminal relativistic jets moving in opposite directions almost perpendicular to the line of sight ($\sim$85$^{\circ}$) with a relativistic velocity of 0.92c \cite{Hjellming1995, Tingay1995}. 
The optical counterpart was discovered soon after by \citet{Bailyn1995}, and subsequent optical studies showed that the system is a low-mass \xray binary hosting a blue sub-giant secondary and a BH primary with a mass of $\sim$6\msun \cite{Orosz1997, Greene2001}. At a distance of $\sim$3.2kpc, GRO J1655$-$40 is one of the closest BHBs we know. The orbital inclination inferred from optical observations and from the detection of absorption dips ($\sim$70$^{\circ}$, e.g., \cite{Bailyn1995a} \cite{Kuulkers1998}) suggests that the inner disc inclination (which could coincide with the direction of the jet or of the BH spin axis) may differ from the inclination of the binary system, implying a misalignment of up to 15$^{\circ}$.
GRO~J1655$-$40 has been subject of a considerable number of \xray timing studies, which revealed the presence of several unique features in its PDS, including the only confirmed QPO triplet (i.e. a type-C QPO and a pair of HFQPOs detected simultaneously, see Sec. \ref{sec:timing}, and \cite{Strohmayer2001a, Motta2014}) in a BHB. 

During its first outburst, started in 1994, GRO J1655$-$40 showed a power law spectrum extending up to 600~keV in the \textit{CGRO}/OSSE data, and which softened and subsequently hardened over the course of the outburst \cite{Greiner1995}.
After the first, relatively short outburst (a few weeks), GRO J1655$-$40 underwent two other major outbursts, one started in 1996 April \cite{Remillard1996} and the other started in 2005 February \cite{Markwardt2005}. The former was regularly covered by \rxte, which confirmed that the overall behaviour of the system resembled that of other BHBs \cite{Mendez1998}. The latter was covered by \rxte, as well as by \integral and \xmm, which provided key information on the properties of this system. 

Thanks to the \integral sensitivity to the high-energy photons from GRO J1655$-$40, the broad-band spectrum of the system could be probed in detail as it evolved over the several accretion states sampled by the source (including the ULS, of which GRO J1655-40 to date provided the best example \cite{Motta2012}). Clear state transitions were observed, and a high-energy cut-off associated with the power law-like continuum was clearly detected during the LHS (see \cite{Joinet2005, Shaposhnikov2007}, but also \cite{Caballero-Garcia2007}). The high-energy cut-off evolved significantly as the source moved away from the LHS and approached the soft states, rising to values $>$ 300 keV and then disappearing completely near the transition to the HSS. As already mentioned in Sec. \ref{sec:GX339}, a similar behaviour was later observed - in greater detail - in the BHB GX 339$-$4. \citet{Joinet2005} also reported on the  behaviour of the radio jet, which was quenched around the time of the disappearance of the cut-off. The combined use of \integral and \xmm data allowed \citet{DiazTrigo2007} to constrain the shape of the broad-band continuum, thus evidencing both the presence of a relativistically broadened Iron K-$\alpha$ line and a variable photo-ionised absorber, which induced the presence of strong K absorption Fe XXV and Fe XXVI lines, and blue-shifted Ne X and Fe XXIV features. 

\subsubsection{GRS 1716$-$249}
\label{sec:GRS1716}

GRS 1716$-$249, also called Nova Oph 1993, was discovered in 1993 by {\it CGRO}/BATSE \cite{Harmon1993} and \textit{Granat}/SIGMA \cite{Ballet1993}. 
The spectral type K (or possibly later) star V2293 Oph was identified as possible donor
and a distance of $2.4 \pm 0.4$\,kpc was derived \citep{Dellavalle1994}. 
\citet{Masetti1996} estimated a lower limit on the mass of the compact object as  4.9\,\msun, which confirmed its BH nature, and an orbital period of the
system of 14.7\,hr.

The brightest outburst of the transient BHB GRS 1716$-$249 occurred in 2016 after more than 20 years of quiescence. {\it MAXI} first discovered the start of a new outburst from this source, and \integral observed it with the aim to study and characterise its soft $\gamma$-ray
emission. Simultaneous multi-wavelength observations were also triggered with the ATCA
radio telescope and the REM infrared camera \cite{Delsanto2017}. This outburst lasted more than one year and was followed almost continuously by \swift. XRT and BAT results of this monitoring campaign
\cite{Bassi2019} showed that GRS 1716$-$24 belongs a group of transient BHBs 
that show failed outbursts. 

\citet{Bassi2020} presented the SPI spectral analysis of the \integral observations, which showed a strong non-thermal component above 200 keV. The SED of GRS 1716--24 was modelled with the internal shock emission model \citep[{\sc ishem},][]{Malzac2013} to investigate on the possible jet nature of the soft $\gamma$-ray emission. The authors measured a spectral cooling break at $\sim$10\,keV and 
found that the soft $\gamma$-ray emission of GRS 1716--24 cannot be explained by
synchrotron emission in the radio jet, unless the index of the electron energy distribution responsible for the emission is lower than 2. However, such a value is difficult to reconcile with the known shock acceleration mechanisms, which suggests that the jet might not contribute much to the soft $\gamma$-ray emission in this source.

\subsubsection{GRS 1758$-$258} 
\label{sec:GRS1758}

GRS 1758$-$258 was discovered in 1990 during \textit{Granat} observations of the Galactic Centre region \cite{Mandrou1990,Sunyaev1991}. This system shows hard \xray properties typical of Galactic BHBs  in the
LHS, i.e., a comparatively hard power law shaped spectrum with a cut-off around $\sim$100 keV, and marked fast time-variability \cite{Lin2000}. Owing to such properties, together with the detection of a double-sided radio counterpart \cite{Rodriguez1992}, GRS 1758$-$258 is considered to be a BHB. Despite the rare occurrence of dim soft states with a typical duration of several months \cite{Hirsch2020}, GRS 1758$-$258 is a persistent source, and one of the very few persistent BHBs known to date. 

GRS 1758$-$258 was observed extensively within the \integral Galactic Centre
Deep Exposure program in 2003 and 2004, which was complemented by numerous \rxte\ observations \cite{Pottschmidt2006}. During this period, the system was initially found in one of its rare dim soft states, and underwent a full transition to the LHS. \citet{Pottschmidt2006} showed that the dim soft state in GRS 1758$-$258 is different from the soft states typical of other persistent high-mass XRBs such as Cyg X-1 (see Sec.~\ref{sec:CygX1}) - where softening is connected  with an increase in luminosity - and is more similar to those typical of Low-mass BHBs. A soft $\gamma$-ray tail extending above 300 keV has possibly been seen as well in the PICsIT data \cite{Pottschmidt2008}. 

Recently, \citet{Tetarenko2020} used ALMA data to explore the molecular gas emission at the location of jet-ISM interaction zones near GRS 1758$-$258. They identified molecular structures that might trace
jet-blown cavities in the gas surrounding this system, confirming previous radio results from the GRS 1758$-$258, which led \citet{Marti2015} to postulate the presence of a jet-blown cocoon structure in deep radio continuum maps of the region surrounding GRS 1758$-$258 \cite{Marti2017}. Such a result also confirms that suggestion that the lobes around GRS 1758$-$258 are being powered by a variable ``dark'' relativistic jet, which in turn has to be fed by high-rate accretion \cite{Tetarenko2020}.

\subsubsection{GS 1354$-$64}
\label{sec:GS1354}

GS 1354$-$64 (also known as BW Cir) is a transient BHB which showed three \xray outbursts over the past 28 yr. During its 1997 outburst, \rxte\ detected a moderate flux increase (up to $\sim$30--40 mCrab), although the source remained in the LHS. GS 1354$-$64 is close to the poorly located Cen X-2, a soft \xray transient discovered in 1967 with a peak \xray flux of $\sim$8 Crab, which might mean that the two sources could be coincident \cite{Kitamoto1990}.

This system is located at a distance of $\sim$25 kpc, and it is observed at an inclination of $\sim$79\degrees\  \citep{Casares2004,Casares2009}. Optical studies allowed a dynamical BH mass of $7.9\pm0.5$\msun to be derived, as well as an orbital period of 2.54 d. The donor star in the system is a type G0-5 III with a mass of $1.1\pm0.1$\msun.  
In quiescence, GS 1354$-$64 shows strong optical variability in the R filter, and a dynamic range of  4 mag ($\sim$17 to $\sim$21) over a period of 14 yr \citep{Casares2009}.

\integral observed GS 1354$-$64 during its 2015 outburst in a coordinated campaign with the \swift \xray observatory and the South African Large Telescope (SALT). 
The results were presented in \citet{Pahari2017}, who showed evidence for optical cyclo-synchrotron emission from the hot accretion flow. In particular, during the rise phase, a significant optical-\xray lag was reported, with the \xray photons lagging the optical ones \cite{Pahari2017}. In addition,  both the optical and \xray PDS exhibited a $\sim$18 mHz frequency peak. Simultaneous fitting of the \swift/XRT and \integral spectra showed a non-thermal, power law-dominated (by over 90 per cent) spectrum with a power law index of $1.48 \pm0.03$, an inner disc temperature of $0.12\pm0.01$ keV, and an inner disc radius of $\sim$3000 km. The above is consistent with synchrotron radiation in a non-thermal cloud of hot electrons, extending to $\sim$100 Schwarzschild radii, being also a major physical process for the origin of optical photons. 

At the outburst peak, about one month later, when the \xray flux rose and the optical dropped, the features in the optical/\xray corss-correlation vanished. A $\sim$0.19 Hz QPO was still observed in the  \xray PDS, although the optical variability was dominated by broad-band noise.
These observations suggest a change in the dominant optical emission source between outburst rise and peak, consistent with a progressive shrinking (and possibly cooling) of the hot flow as the disc moves in during a transition to a softer state.


\subsubsection{MAXI J1828$-$249}
\label{sec:MAXIJ1828}

MAXI J1828$-$249 was discovered at the beginning of an outburst on 2013, 15 October \cite{Nakahira2013} by MAXI/GSC, and was soon afterwards detected at higher energies by \integral \cite{Filippova2013}. 
The flux at discovery was 93$\pm$9 mCrab (4--10 keV) in MAXI \citep{Nakahira2013}, and 45$\pm$2 mCrab (48$\pm$2 mCrab) at 20--40 keV (40--80 keV) in \integral/IBIS \cite{Filippova2013}.
Follow-up observations with \swift/XRT provided a more accurate position \citep{Kennea2013a}, which permitted the identification of the UV counterpart \citep{Kennea2013b}, as well as of the optical/IR counterpart \cite{Rau2013}.
Radio observations carried out about two days after the onset of the outburst did not reveal significant emission from the source \citep{Miller-Jones2013}.

During the first two days of the outburst, the source was observed by MAXI to undergo a very rapid transition from a hard to a softer state \citep{Negoro2013}, so that observations of the source in the LHS were limited  because of the swiftness of the transition. 
The rest of the outburst (essentially the HSS) was covered by a joint observation campaign over a wide energy range from 0.6--150 keV by the instruments on \swift and \integral, although MAXI J1828$-$249 did not show significant spectral evolution \cite{Filippova2014}, reaching, however, a  peak luminosity L $\approx10^{38}$\ergpersec assuming a distance of 8~kpc.

The broad-band energy spectrum comprised of an accretion disc component with temperature kT$\sim$0.7 keV, and a power law with a photon index $\Gamma\approx2.2$ extending with no measurable break up to 200 keV. 
The observational evidences, supported by both the spectral and timing analyses, suggest that MAXI J1828$-$249 is a BHB that evolved rapidly in the first two days of the outburst from the LHS to the HSS, where it remained during the entire \swift-\integral observation campaign. 

Data collected on 2014 February 14--16 found that the source transitioned back to a LHS, and the energy spectrum featured a power law with a photon index of $\Gamma\approx1.7$, and a 0.5--10 keV flux of $4\times10^{-11}$\ergpersec \cite{Tomsick-Corbel2014}. Significant radio
emission was also detected at flux densities of $1.38\pm0.05$ mJy at 5.5 GHz and 1.28$\pm$0.06 mJy at 9 GHz \citep{Corbel2014}, further strengthening the case that the source was indeed a BHB.

Two additional works were dedicated to the 2013 outburst of this source, which reported results consistent with those mentioned above \citep{Grebenev2016,Oda2019}.

\subsubsection{MAXI J1836$-$194}
\label{sec:MAXIJ1836}

MAXI J1836$-$194 was discovered in outburst by MAXI/GSC on 30 August 2011 (55803 MJD), and detected simultaneously by the \swift\ BAT instrument  with a flux of 25 mCrab in the 4--10 keV band, and 30 mCrab in the 15--50 keV band \cite{Negoro2011}.
The discovery of this system triggered a multi-wavelength observing campaign led by \integral, which carried out observations for 60 days, from MJD 55804 to 55864.
The analysis of all the data collected by \integral, \swift/XRT, and \rxte/PCA over this period were presented in \citet{Ferrigno2012}.

\swift/XRT and UVOT provided an accurate source position, which allowed the optical counterpart to be identified \cite{Kennea2011}. Based on \rxte/PCA data, \citet{Strohmayer2011b} suggested that MAXI J1836$-$194 could be a BHB undergoing a LHS to HIMS transition.
The 2.2-m ESO/MPI telescope showed the source to have a flat broadband spectrum, consistent with the emission from the companion star or the accretion disc \cite{Rau2011}, but also akin to that of synchrotron radiation from relativistic jets. Radio observations found emission consistent with the presence of a compact jet (\cite{Miller-Jones2011b, Trushkin2011}), and the optical monitoring of the target revealed  variability correlated with the radio emission, which further supported the jet origin of the optical emission  \cite{Pozanenko2011}.

MAXI J1836$-$194 was detected by IBIS/ISGRI from MJD 55816.9 to 55850.4.
The evidence gathered during the campaign suggested that the source was first observed during the transition from the LHS to HIMS, then it remained in the HIMS for a few days, and finally slowly moved back to the LHS before fading out \cite{Ferrigno2012}. 
The timing and spectral properties evidenced in the \rxte, \swift, and \integral data showed that MAXI J1836$-$194 was one of the few candidate BHBs, together with, e.g., H 1743$-$322 in its 2008 outburst, and SAX J1711.6$-$3808 in its 2001 outburst (see \cite{Capitanio2009} and references therein), that made the transition to the HIMS, but never reached the HSS before going back to quiescence, in what is referred to as a failed outburst.

A second outburst from this system was observed a few months later in March 2012. Although significantly weaker than the first, it was followed by \swift and \integral. The resulting spectrum was consistent with a LHS, with no clear evidence of the presence of a soft component due to the accretion disc \citep{Grebenev2013}. MAXI J1836$-$194 has been the subject of many studies since these initial observations, \citep{Russell2014b,Russell2014c,Russell2015,Jana2016,Peault2019,Jana2020,Dong2020,Lucchini2021}, but all were based on the data collected during the powerful first outburst, which demonstrates the importance of the role played by the wide field of view high-energy instruments like \integral.

\subsubsection{Swift J1745$-$26} 
\label{sec:SwiftJ1745}

On 2012 September 16 the \swift/BAT telescope discovered a new bright \xray source, i.e. Swift J174510.8$-$262411 (hereafter \swJ) \cite{Cummings2012}.
Soon after its discovery, \swift\ and \integral monitoring campaigns were triggered. Based on the spectral and timing properties revealed by these two missions, \swJ\ appeared immediately to be a new bright Galactic transient BHB \cite{Tomsick2012} with a low-mass companion, which was detected in the IR \cite{Rau2012}).
Later on, the BH nature of the source was further supported by optical observations, which displayed a broadened H$\alpha$ emission line \cite{Munoz2013}.

Results from two \integral target-of-opportunity campaigns have been reported by \citet{DelSanto2016}, which covered the first part of the outburst, and \citet{Kalemci2014}, which report on the outburst decay.
The former observed that, in spite of the high luminosity of \swJ\ (up to 1 Crab in hard \xrays), the source never reached the HSS, but moved back to quiescence after reaching the HIMS, thus showing another example of failed outburst. 
The spectral analysis performed by \citet{DelSanto2016} showed that the evolution of the disc spectral parameters and the properties of the fast-time variability were consistent with the truncated disc scenario, and that a significant  non-thermal contribution to the total Comptonisation emission was strongly requested by the data in the HIMS.
\citet{Kalemci2014} found that the \xray spectra during the decay phase of the outburst, and during a 50 days-long re-brightening are consistent with thermal Comptonisation emission, and that the physical origin of the re-brightening (visible both in the \xray and in optical) was likely due to enhanced mass accretion in response to an earlier heating event. 

Interestingly, \citet{Curran2014}, who observed a radio flare when the source was thought to be still in the HIMS. Such flares are typically associated to discrete ejection events occurring near the HIMS-SIMS transition. Therefore, it is possible that the source made a very short transition to the SIMS before moving back to the HIMS, and then decaying towards quiescence. It is worth noticing that this is to date the only case of a possible jet ejection occurred during a failed outburst.

\subsubsection{Swift J1753.5$-$0127}
\label{sec:Swift J1753}

Swift J1753.5$-$0127 was discovered on 2005, May 30, by \swift/BAT \cite{Palmer2005}, and it was subsequently monitored by several facilities.
The initial multi-wavelength coverage -- including results from an \integral ToO observations -- of the outburst was reported by \citet{Cadolle2007}, who found that the properties of Swift J1753.5$-$0127 were typical of a transient BHB in the LHS.
The low extinction towards Swift J1753.5$-$0127 allowed the detection of a soft excess in \swift/XRT observations to be made, which may be due to a cool 0.2 keV disc \citep{Chiang2010, Kajava2016}.

Swift J1753.5$-$0127 is unique in its outburst evolution: after the initial rise to a very bright LHS, it did not make the transition ot the HIMS, nor return to quiescence, but instead decayed to a low-luminosity LHS and lingered for over ten years. In 2012, Swift J1753.5$-$0127 showed a failed outburst, which was followed by \swift \citep{Soleri2013}. In 2015, after spending another 4 years in a low-luminosity LHS, Swift J1753.5$-$0127 started showing  peculiar low flux soft state episodes between 2009--2012, who were observed in the MAXI and \swift/XRT data \citep{Yoshikawa2015, Shaw2016, Kajava2016}, where the emission was for the first time dominated by a disc component, perhaps suggesting that some sort of peculiar soft state was occurring. After that, the source showed a number of hard-only mini outbursts, finally ending its activity phase in 2017 \citep{ZhangBernardini2019}. 

At higher energies, the long duration of the outburst allowed \integral to accumulate very high S/N spectra both with SPI \citep{Rodi2015} and ISGRI \citep{Kajava2016}.
By combining the \integral data with Swift/UVOT (optical and UV) and XRT (soft \xray) observations, \citet{Kajava2016} was able to deduce that during the outburst peak in 2016 the SED was consistent with the disc extending close to the ISCO, with the soft \xray and UV excess consistent of being from an irradiated disc with a modest reflection hump. During the prolonged outburst tail, instead, the entire optical-\xray SED was consistent with a pure synchrotron-self-Compton spectrum.

Interestingly, the changes in the peculiar optical-\xray cross correlation functions \cite{Durant2008} varied hand-in-hand with the spectral changes over the outburst \citep{Veledina2017}, such that both the SED and the timing variability could be attributed to an expanding hot flow, and a change between synchotron self-Compton and disc Comptonisation between the outburst peak and the tail.
This is significant because Swift J1753.5$-$0127 is also known to move along the radio-quiet track of the radio/\xray correlation \citep{Gallo2012}, making it a good example of a case where jets are not required to explain the complex optical-\xray cross-correlation functions \cite{Veledina2017}.

\subsubsection{XTE J1550$-$564}
\label{sec:XTEJ1550}

XTE J1550$-$564 was discovered in 1998 by \rxte\ \cite{Smith1998}, at the time of its first observed outburst, which lasted approximately 200 days and displayed the typical behavior of a BHB in outburst \cite{Sobczak2000, Homan2001, Remillard2002}. \citet{Orosz2002} performed optical spectroscopic observations of the companion star in XTE J1550$-$564, confirming that the primary object should be a BH of $\approx$ 9.4 M$_{Sun}$. With this mass, and at a distance of $\approx$5.3~kpc, XTE J1550$-$564 reached and possibly exceeded its Eddington luminosity during this first outburst \cite{Orosz2002}). 
Another major, although less luminous, outburst occurred in 2000, which, however, did not show any obvious super-Eddington event.

Radio observations obtained during a subsequent outburst, occurred in 2003 outburst evidenced a strong radio flare, and subsequent VLBI radio observations taken soon after it revealed the presence of super-luminal jet ejection event \cite{Hannikainen2001}. Such ejections were later  observed also in the \xrays with Chandra \cite{Corbel2002, Tomsick2003, Kaaret2003}, months and years after the inferred ejection time, offering one of the best examples of long-lived ejections interacting with the ISM. 

 After the 2003 outburst, XTE J1550$-$564 has remained relatively active, exhibiting sporadic, low-amplitude, hard-only flaring episodes with a tentative recurrent time of 300--500 days. One such under-luminous event was observed with \integral, which allowed the broad spectrum of the source to be probed \cite{Sturner2005}.  The energy spectrum of XTE J1550$-$564 was fully consistent with a LHS observed both in the same system in outbursts, and in other BHBs. The spectral variability was limited, but a mild softening was observed as the source approached the quiescence level. The broad band \integral spectrum could be equally well fit with a standard Comptonisation model, and with a model requiring Comptonisation in an outflow, suggesting a possible contribution of a jet to the LHS emission (see Sec. \ref{sec:states}.


\subsubsection{XTE J1720$-$318}
\label{sec:XTEJ1720}


XTE J1720$-$318 was discovered on 2003 January 9 with the ASM on-board \rxte,  as a new transient source \citep{Remillard2003}.
The 1.2--12 keV flux rose to 430 mCrab in 2 days before starting to decay.
The radio counterpart of this system was identified with the VLA and ATCA \citep{Rupen2003a,O'Brien2003}, and the IR counterpart was affected by large extinction, suggesting that the source was located at a large distance \citep{Nagata2003}. An extensive campaign was organised to observe the new candidate BHB with \integral, \xmm, and \rxte, the results of which were presented in \citet{Cadolle2004}.

\xmm observed about 40 days after the outburst peak, and found the source clearly in a HSS characterised by a strong thermal component, well-modelled by a disc component with an inner disc temperature of kT$\sim$0.7 keV and a weak power law tail.
The high equivalent absorption column density derived from the \xmm data confirmed that XTE J1720$-$318 is  likely at a large distance, possibly located at the Galactic Centre or further, most probably within the Galactic Bulge.

Simultaneous observations with \integral and \rxte at the end of February 2003 found the source still in the HSS, and allowed to measure with higher precision the slope of the spectral continuum, characterised by a power law-like shape with photon index $\sim$2.7. The disc component accounted for more than 85\% of the 2--100 keV source luminosity, estimated to be $2.5\times10^{37}$\ergpersec with a distance of 8 kpc.

At the end of 2003 March , \integral observed a dramatic change in the source behaviour.
Following the decay that brought the source emission below detection level, a rise of the high-energy component was observed approximately 75 days after the main outburst peak. Over the course of 10 days, the 20--200 keV unabsorbed luminosity increased from below the \integral detection level, to a value of $7\times10^{36}$\ergpersec. This re-brightening was observed by \integral for an additional 25 days.
The increase in flux was not seen at low energies (below 10 keV) in the \rxte/ASM and \jemx data, suggesting that the source was undergoing a hard-only outburst, with an energy spectrum well-described by a power law with a photon index of 1.9, or by a thermal Comptonisation model with a (poorly constrained) plasma temperature of $\sim$40 keV. 

The observational characteristics of this source---high peak luminosity, fast rise and slow decay time scales, HSS followed by a reflare in the LHS, and spectral parameters --- all suggested that XTE J1720$-$318 is a new transient candidate BHB. Only the long, continuous, and high-energy sensitive \integral observations allowed the study of this very distant and absorbed source, which for a large fraction of its activity phase could only be detected in the hard \xrays. 

\subsubsection{XTE J1752$-$223}
\label{sec:XTE J1752} 


XTE J1752$-$223 was discovered by \rxte in the Galactic Bulge in 2009 \citep{Markwardt2009,Markwardt2009b}. A distance to the source of 3.5$\pm$0.4 kpc was reported in \citet{Ratti2012}. The source position was  accurately determined using the astrometric optical observations, and very long baseline interferometry radio imaging \citep{Miller-Jones2011}, which also revealed decelerating relativistic jets \citep{Yang2011}.
XTE J1752$-$223 showed a typical outburst evolution for transient BHBs as observed by \rxte\ \citep{Munoz-Darias2010,Shaposhnikov2010a}, MAXI \citep{Nakahira2010}, and Swift \citep{Curran2011}. Observations with \suzaku and \xmm avidenced a strong, relativistic Iron emission lines in the energy spectra extracted in the LHS \cite{Reis2011}, and 

\integral observed XTE J1752$-$223 during its 2010 outburst, as part of a multi-wavelength campaign that included \rxte, \swift, SMARTS in O/NIR, and the Australia Telescope Compact Array (ATCA) in the radio band.
\citet{Chun2013} reported on the results of such a campaign and showed that a radio core flare observed in the VLBI images corresponded to a small flare in the I-band flux, which was followed by a larger flare in the I- and H-band flux; they also showed that a third radio flare seen by ATCA - which observed the radio spectrum becoming from optically thick to optically thin - coincided with a flare detected in the I and the H bands.  Observations in the X-rays indicated that  the short term \xray timing properties of XTE J1752$-$223 did not show a significant change during flares in the O/NIR.
\citet{Chun2013} also analysed the broad-band \xray spectrum of XTE J1752$-$223, which included IBIS/ISGRI data, and found that it was consistent with thermal Comptonisation, with a clear high energy cut-off. 

\subsubsection{XTE J1817$-$330}
\label{sec:XTEJ1817}

The galactic BH candidate XTE J1817$-$330 was discovered by \rxte\ in January 2006 at the beginning of an outburst \citep{Remillard2006}.
The discovery of counterparts at different wavelengths followed closely in the radio \citep{Rupen2006a,Rupen2006b}, NIR \cite{D'Avanzo2006}, and optical \citep{Torres2006}.

\integral and \xmm observed XTE J1817$-$330 in 2006 February and March, respectively, 18 and 44 days after the outburst peak. The results of these observations were reported by \citet{Sala2007}. The source was clearly detected by both observatories, both in the X-rays and in UV (V and UVW1 filters with \integral/OMC and \xmm/OM, respectively) thanks to the Optical monitors on-board both missions. 
The combined \xmm and \integral spectrum was fit with a two-component model consisting of a thermal accretion disc component,  and a Comptonised continuum. The soft \xray spectrum was dominated by an accretion disc component, with a maximum temperature decreasing from 0.96$\pm$0.04 keV at the time of the \integral observation, to 0.70$\pm$0.01 keV on 13 March at the time of the \xmm observation. A low equivalent hydrogen column density affected the source, and a number of absorption lines, most likely of interstellar origin, were detected in the \xmm/RGS spectrum.
A limit on the absolute magnitude of M$_{v} >$6 mag implies that the secondary star must be a K-M star, and a giant could be excluded, even assuming a distance of 10 kpc distance.

\subsubsection{XTE J1818-245}
\label{sec:XTEJ1818}

\begin{figure}
    \includegraphics[width=0.49\textwidth]{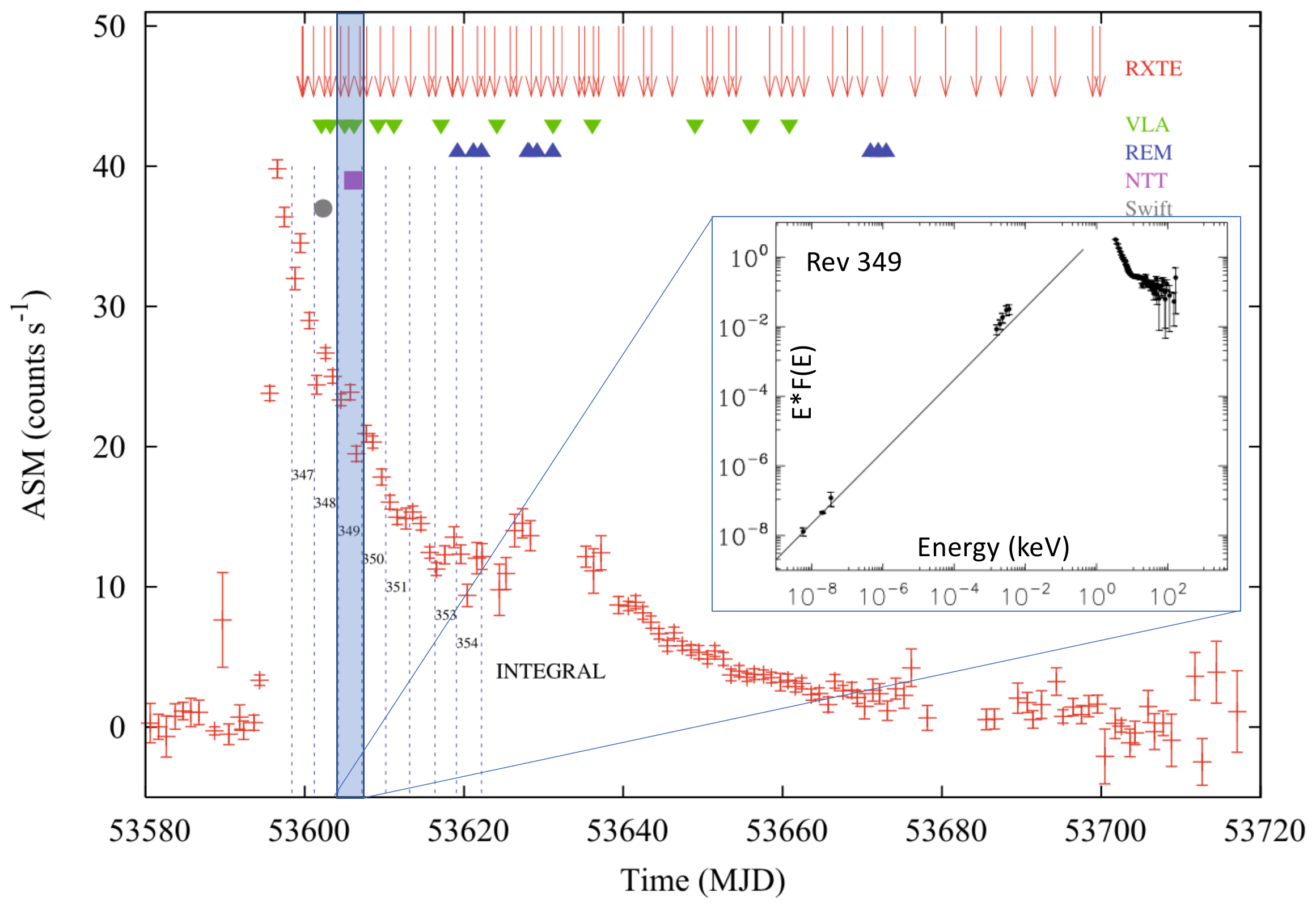}
    \caption{\rxte/ASM light curve of XTE~J1818$-$245 indicating (with various symbols) the dates of the multi-wavelength observations obtained by \cite{Cadolle2009}. The insert shows a broad band radio to soft $\gamma$-rays SED obtained during the observation performed in \integral revolution 349. It includes data from the VLA in radio, ESO/NTT in Optical, Swift, \rxte, and \integral at high-energies. Adapted from \citet{Cadolle2009}.
    }
    \label{fig:1818Multi}
\end{figure}

XTE J1818$-$245 was discovered with the ASM onboard \rxte\ on 2005, August 12$^{th}$ (MJD 53594) \cite{Levine2005}. Its flux rapidly increased by a factor of 10 in a couple of days (Fig.~\ref{fig:1818Multi}), reaching a  2--12 keV peak flux slightly exceeding 500 mCrab. This and the very soft spectrum of the source led \citet{Levine2005} to suggest the source was a new transient BHB. Prompt multi-wavelength follow-up observations involving \rxte, \integral and  \swift in the X-rays, a optical 1m telescope in Chile, and the VLA in radio \cite{Still2005,Shaw2005,Steeghs2005,Rupen2005b} allowed the position of the source to be refined, and both an optical and a radio counterparts to be identified. All the results obtained from this campaign supported the probable  BH nature of the primary component in this XRB \cite{Cadolle2009}. A rough  distance estimate of 2.8--4.3 kpc was obtained based on the overall behaviour of the source. 

During  \integral/revolution 349 (MJD 53605--53607, highlighted in Fig.~\ref{fig:1818Multi}), shortly before the state transition to the HSS, the source shows the presence of a 
compact jet, with a ``flat'' radio spectrum, which may significantly contribute to the NIR/Optical band. The NIR spectrum, however, showed a steeper index than in radio (see the inset in Fig.~\ref{fig:1818Multi}), and which would result in a radio excess if extrapolated down to a few GHz, thus indicating that the jet is only one of the components contributing to the NIR emission.

Two radio flares are reported after the source's transition to the HSS, both ascribed to discrete ejection events, which, however, occurred about 5 days after the transition to the HSS, casting doubts on the existence of a direct link between the transition to the soft states and the launch of relativistic jets (see also \cite{Fender2009}). This behaviour, however, could be in principle explained assuming that a shock happened within previously ejected material, far from the launching site, perhaps due to the interaction with the ISM.

\subsubsection{MAXI J1348$-$630}
\label{sec:MAXIJ1348}

MAXI J1348-630 was discovered during its outburst on 2019, January 26th  by MAXI/GSC \citep{Yatabe2019} and was monitored by a number of \xray telescopes: \textit{Swift}/XRT \cite{Bassi2019}, \textit{Nicer} \cite{Sanna2019} and \integral \cite{Cangemi2019a, Cangemi2019b, Lepingwell2019}. A few candidate radio counterparts were reported by ATCA \citep{Russell2019}, the MWA \citep{Chauhan2019} and MeerKAT \cite{Carotenuto2019}.  
The source showed clear state transitions and a multi-wavelength behaviour consistent with that of many BHBs, thus suggesting that MAXI J1348$-$630 is a new candidate BHB \citep{Tominaga2020}. The fast time variability properties of the source were also investigated, leading to the detection of different types of QPOs typical of BHBs \citep{Belloni2020, Zhang2020}.

\integral followed the source evolution during its entire outburst. The light curve showed a spectacular rising phase from January 26th to January 29th. During the four following days, \citet{Cangemi2019a} observed the flux decreasing slightly and then reaching its maximum of $\simeq 5$\,Crabs in the 30--50\,keV ISGRI energy band on February 2nd \citep{Cangemi2019b}. During the following decay, the light curve exhibited three weak rebrightenings (on February 6th, 8th and 10th) before the source reached quiescence.

An initial spectral study with \integral revealed the presence of a high-energy tail above the Componised continuum in the LHS. This component, detected by both IBIS/ISGRI and SPI is characterised by a photon index $\Gamma \simeq 2$. Further analysis suggested that this component is also strongly polarised with a polarisation fraction consistent with a synchrotron emission coming from the base of the jet (Cangemi et al. in prep).

\subsubsection{MAXI J1631$-$479}
\label{sec:MAXIJ1631}

MAXI J1631$-$479 was discovered as a new \xray transient on 2018 December 21$^{st}$ when the MAXI/GSC nova alert system reported a bright \xray outburst from a source in the Norma region, which was initially -  erroneously - associated with the \xray pulsar AX J1631$-$4752 \cite{Kobayashi2018}. Observations performed with \swift and \nustar refined the position of the new transient, and confirmed that MAXI J1631$-$479 was a previously unknown source \cite{Miyasaka2018}. A radio counterpart to this system was found by ATCA \cite{Russell2019}, while no clear optical counterpart was identified, consistently with the high column density in the direction of the source, found in the \nustar observation.

\integral started observing the Norma region as part of the Galactic Plane Scan on 2019 January 21th, and MAXI J1631$-$479 was clearly detected by IBIS/ISGRI \cite{Onori2019}. Subsequent \integral observations, as well as data from the MAXI/CSG and \swift/BAT monitors showed that the source underwent a few state transitions during which the energy spectrum evolved coherently with what expected from a BHB in outburst \cite{Fiocchi2020}. Since 2019 January 12 (MJD 58495), MAXI J1631$-$479 was monitored weekly in radio with \textit{MeerKAT} as part of the ThunderKAT Large Survey Project \cite{Fender2016} and \swift. Both \textit{MeerKAT} and \swift confirmed that MAXI J1631$-$479 exhibited a number of \xray states, and in particular that both high- and low-luminosity hard states were interleaved by extended soft states \cite{Monageng2021}. MeerKAT data revealed radio flaring observed shortly after a transition from the HIMS to the SIMS, which is broadly in agreement with existing empirical models describing the disc-jet coupling in BHBs (see Sec. \ref{sec:discjet}). However, the extended duration of the flare hints at multiple unresolved flares, or at jet-ISM interactions.

\subsubsection{MAXI J1820+070}
\label{sec:MAXIJ1820}

MAXI J1820+070 was discovered in March 2018, first in the optical range by the AllSky Automated Survey for SuperNovae (ASAS-SN) as ASASSN-18ey \cite{Tucker2018}, and soon after in the \xray data from the MAXI mission \cite{Kawamuro2018}. MAXI J1820+070 is located at a distance d = 3.0 $\pm$ 0.3 kpc, a value obtained from a radio parallax measurement \cite{Atri2020}, and consistent with determinations based on the Gaia Data Release 2 parallax, d = 3.5 $^{+2.2}_{-1.0}$ kpc \cite{Bailer-Jones2018, Gandhi2019,Atri2020}. Based on optical spectroscopic measurements, MAXI J1820+070 hosts a stellar mass BH of $\approx$ 6 \msun and is observed at a relatively high inclination to the line of sight (66$^{\circ}$ to 81$^{\circ}$), as confirmed by the presence of clear absorption dips in the light curve \cite{Kajava2019}. 

MAXI J1820+070 was the target of an intense multi-wavelength observing campaign, which yielded a large amount of data on the source. Particularly noteworthy are the radio data obtained from a number of radio facilities (MeerKAT, eMERLIN, AMI-LA, VLBA) thanks to which extended super-luminal jets were discovered in the radio band \cite{Bright2020}, and which were also detected in the \xrays by Chandra \cite{Espinasse2020}.

\integral observed MAXI J1820+070 around the time of the outburst peak, as well as during part of its decay to quiescence, revealing intriguing properties of the hard X/soft $\gamma$-ray emission of this BHB. \citet{Roques2019} reported on the spectral analysis of the data collected by \integral, and in particular from the SPI instrument, and showed that the high energy emission consists of two components evolving independently. While the emission below 50 keV (consistent with a Comptonised radiation plus reflection) evolved rapidly, reaching a peak in approximately 12 days after the start of the outburst, the emission in the 100--300 keV energy band (best-described by a cut-off power law) evolved much more slowly, with a later peak and a slower decay. The decoupled evolution of such components indicates that they originate from two distinct emission regions, with the power law component possibly arising from close to or within the jet. 

\subsubsection{A 0620-00 }
\label{sec:A0620-00}

The X-ray nova A 0620-00 was discovered in 1975 with the Ariel V Sky Survey Experiment \cite{Elvis1975}. For two months in 1975 this object was the brightest celestial X-ray source with a peak luminosity in the 1-10 keV range of L $\sim$ 10$^{38} \ergs$. Photometric observations of A0620--00 in quiescence yielded the first estimate of the orbital period and of the mass function \citet{McClintock1986b}, which showed that the primary in this system could be a BH, located at the small distance of  0.87 kpc (see also \cite{Gandhi2019}).
Beside being one of the first confirmed BH XRBs, quiescence observations of A0620-00 were also instrumental to determine that the radio-\xray correlation that holds for BH XRBs \cite{Gallo2003,Corbel2003} in the LHS extends down to quiescence \cite{Gallo2006}. 
As far as we are aware, INTEGRAL never targeted A0620--00, and no INTEGRAL-specific results have been reported so far on this systems. 


\subsection{Possible black holes?}\label{sec:notbh}


\subsubsection{SS433}
\label{sec:SS433}

The microquasar SS433 is not only a remarkable, but also a highly intriguing and peculiar XRB whose nature as a BH or NS system is still not fully clear (but see \cite{vandenheuvel2017}). \integral discovered hard-\xray emission from the source up to 100 keV soon after launch, and contributed significantly over the years to our understanding of SS433 \citep{Cherepashchuk2003,Cherepashchuk2005,Cherepashchuk2009,Cherepashchuk2013}. Considered a highly interesting high-energy source, SS~433 is the subject of a recent and extensive review paper by \citet{Cherepashchuk2020} to which we direct the reader for a thorough discussion of its properties and characteristics.

\subsubsection{Cygnus X-3}
\label{sec:CygX3}

Cygnus X-3 (Cyg X-3) is one of the first discovered \xrbs \cite{Giacconi1967}. In this system, the compact object  is very close to its Wolf-Rayet companion \cite{Koljonen2017, VanKerkwijk1992}, which makes this system peculiar in many ways if compared to other \xrbs with low mass companions. It is located at a distance of $7.4 \pm 1.1$\,kpc \citep{McCollough2016} with an orbital period of 4.8 hours \citep{Parsignault1972}. 

The nature of the accretor in Cyg X-3 remains a mystery as it has not yet been possible to obtain an estimate of the mass function of the system \citep{Hanson2000, Vilhu2009}. However, its spectral behaviour, the various accretion states it shows, and the overall properties of this system seem to indicate that a BH rather than a NS is powering Cyg X-3 \cite{Cangemi2021, Hjalmarsdotter2009, Szostek2008, Zdziarski2013}. Cyg X-3 is also one of the brightest radio sources among the known \xray binaries, \citep{McCollough1999} and shows both compact radio jets and discrete ejections. 

While the global shape of the spectra from Cyg X-3  is very similar to those of other BHBs, the value of the spectral parameters can be markedly different: the exponential cutoff is at a lower energy of $\sim$20 keV in the hardest states, whereas the disc is very strong in the softest states \citep{Cangemi2021, Hjalmarsdotter2009, Szostek2008}. Those peculiarities and their correlation with the radio behaviour led to a definition of six different spectral states; five states defined both  according to the \xray spectral shapes and levels of radio fluxes \citep{Szostek2008}, an a  \textquotedblleft hypersoft\textquotedblright\ state, defined based on the HID of the source \cite{Koljonen2010}. This hypersoft state is dominated by a strong presence of the disc and precedes the emission of powerful radio ejections \citep{Beckmann2007, Koljonen2010, Zdziarski2018}.

\citet{Hjalmarsdotter2009} followed the evolution of the complex behaviour of Cyg X-3 with \textit{RXTE}, and showed that its spectra could be well-explained with a hybrid thermal/non-thermal corona model. The hard state, in particular, was studied using \integral observations (JEM-X, ISGRI and SPI data, \cite{Hjalmarsdotter2008}). In this study, the authors found that this state could be well described by a model consisting of a power law with a cut-off, and an accretion disc with a (variable) inner truncation radius. \citet{Hjalmarsdotter2008} suggested that the low electron temperature observed in the data could be due to down-scattering in a dense plasma surrounding the compact object. The physical origin of this plasma is likely related to the strong stellar wind of the Wolf-Rayet star, which interacts with a small accretion disc formed close to the compact object from wind accretion \citep{Zdziarski2010}. This scenario is consistent with the energy dependency of the orbital modulation (0.2 days) measured using \swift, \integral and \rxte\ data \cite{Zdziarski2012}.

Although the flux of the source is very weak in the \xrays, Cyg X-3 was recently detected up to 200\,keV thanks to the observations of more than sixteen years of \integral data \citep{Cangemi2021}. Using JEM-X, ISGRI and SPI, the authors created six accumulated spectra according to each spectral state and detected a non-thermal component above 50 keV in the hardest states. Following the previous results of \citep{Hjalmarsdotter2009}, they applied a hybrid thermal/non-thermal corona model and found a more efficient acceleration of electrons in states where major discrete ejections are observed. In these states, they also found a slight dependence of the photon index of the power law as a function of the strong orbital modulation suggesting a higher absorption when Cyg X-3 is behind its companion.

\section{Conclusions and future perspectives}

As emphasised multiple times in this review, the contribution of the \integral observatory to the investigation of the hard X-ray sky has been significant and multi-fold. Its unique combination of instruments sensitive to both the hard X-rays and the soft $\gamma$-rays, a large field of view, and the ability of performing long, uninterrupted observations are the distinctive characteristics of \integral, which have led to dramatic advancements in our understanding of the high-energy emission of black hole X-ray binaries.
\integral brought to the next level the exploration of the hard X-ray/soft $\gamma$-ray sky, thus extraordinarily deepening our understanding of it. The importance of the discoveries in the field of the  high-energy emission from accreting systems which \integral made possible can be largely ascribed to the unique payload it carries, which has no comparable counterpart in the current and past missions.  

\integral allowed the discovery of several new sources to be made, as well as the observations of the very early stages of the outbursts of BHBs, which often prove to be elusive. The \integral sensitivity to the high-energy \xray and $\gamma$-ray photons makes it capable of registering the early activity of accreting source, often well before they become detectable with the currently operational all-sky monitors, which are sensitive to significantly  lower energies.

\integral massively increased the understanding of the emission above 100 keV from BHBs, enabling the systematic study of all the systems observed, through various accretion states. 
\integral contributed to building the vast and detailed phenomenological picture that framed over the past two decades, which is the foundation of our understanding of accreting systems.  The complementarity of the instruments on-board \integral with those of other past and last generation X-ray facilities, such as \rxte until 2012, \xmm, Chandra, {\it{NuSTAR}}, {\it{NICER}}, and \swift still in flight, proved fundamental to dig deeply inside the properties of BHBs, disentangling the contributions of the several spectral components in a way that was only possible thanks to the \integral's contribution in the hardest X-ray energy band. 
\integral helped shedding light on the (complex) interplay between the physical processes at play during the active phases of such sources. In particular: \integral confirmed the presence of a soft $\gamma$-ray-energy tail in the spectra of several BHBs not only in the soft states, but also in a hard or hard intermediate state. This definitely proved that an additional radiative process (aside from  thermal Comptonisation) is at play and has to be taken into account in explaining the high energy emission from BHBs. Furthermore, polarisation measurements of this  soft $\gamma$-rays emission performed by \integral in Cyg X-1 brought to the forefront the hypothesis of a link between this emission and the (polarised) jets. 

\integral's legacy is a huge and still growing data base, largely still to be explored, which nicely complements the growing amount of data available at lower energies, and  constitutes a fundamental benchmark for the study and understanding of the high-energy sky, particularly for accreting systems (see also \cite{Sazonov2020}). As new models for accretion flow emission become available, the data from \integral will continue to be exploited to help understanding these complex systems.

The \integral mission has been operating for over 18 years, and is expected to remain fully operational at least until 2022, or until 2029, when it is scheduled to reenter the Earth atmosphere. 
Observations of the hard X-ray sky with \integral in the coming years will continue to contribute to our understanding of BHBs, especially from a multi-wavelength context, in an era where coordinated, multi-band observations are becoming the standard approach to probe the accretion/ejection physics around compact objects.

\section{Acknowledgements}

All the authors acknowledge the scientists and engineers who have participated to the great adventure of the \integral project. We also acknowledge all the authors that over the years have contributed to the field of high-energy astrophysics making use of data provided by \integral.

SEM acknowledges financial support from the Violette and Samuel Glasstone Research Fellowship programme, the UK Science and Technology Facilities Council (STFC), and Oxford Centre for Astrophysical Surveys, which is funded through generous support from the Hintze Family Charitable Foundation. 
SEM and MDS acknowledge financial contribution from the agreement ASI-INAF n.2017-14-H.0 and the INAF mainstream grant. SEM acknowledges support from PRIN-INAF 2019 n.15.
JR acknowledges partial fundings from the CNES, and the French Programme National des Hautes \'Energies.  
JJEK acknowledges support from the Academy of Finland grant 333112. 
JW acknowledges support from the Bundesministerium f\"ur Wirtschaft und Technologie through Deutsches Zentrum f\"ur Luft- und Raumfahrt grant 50\,OR\,1909.
KP acknowledges support by NASA under award number 80GSFC17M0002.

This work is based on observations with INTEGRAL, an ESA project with instruments and science data centre funded by ESA member states (especially the PI countries: Denmark, France, Germany, Italy, Switzerland, Spain) and with the participation of Russia and the USA.
The INTEGRAL SPI project has been completed under the responsibility and leadership of CNES. The SPI team is grateful to ASI, CEA, CNES, DLR, ESA, INTA, NASA, and OSTC for their support.

SEM sincerely thanks Edward van den Heuvel for the immense patience he has shown for the past many months. 
JR is infinitely grateful to Marion Cadolle Bel, Philippe Ferrando, Andrea Goldwurm, Diana Hannikainen, Philippe Laurent, Fran{\c c}ois Lebrun, Jacques Paul, Christian Gouiff\`es. All the authors wish to thank the anonymous referee who reviewed this work, for providing useful comments, which help improving this paper. 

SEM ``thanks'' SARS-COV-2 for causing the COVID-19 global pandemic in 2020-2021, thus prompting her and consequently all the co-authors to finally finish this review. SEM also acknowledges Miel and Sesamo, for being very patient office mates over the past many months of lock-down, despite their habit of sitting on her keyboard while she was using it. 
MDS acknowledges \integral,  which has ensured she will have a full plate for the rest of her life.

\bibliographystyle{model2-names-astronomy.bst}
\bibliography{15years-bh.bib}


\end{document}